
%
\input phyzzx
\def\NPrefmark#1{\attach{[ #1 ]}}
\Pubnum={ INS-Rep.-1017 }
\date={   December 1993 }
\titlepage
\vskip 32pt
\title{ Construction of Non-critical String Field Theory  \nextline
        by Transfer Matrix Formalism in Dynamical Triangulation
      }
\author{ Yoshiyuki WATABIKI }
\vskip 16pt
\address{ Institute for Nuclear Study, University of Tokyo,
          \nextline
          Tanashi, Tokyo 188, Japan }
\vskip 32pt
%
%
\font\BIGr=cmr10 scaled \magstep2
\def\NextPage{\vskip 30pt}
\def\sp(#1){ \noalign{\vskip #1pt} }
\def\mbar#1{\overline {#1} \hskip 1pt{}}
\def\der{\partial}
\def\e{\varepsilon}
\def\Distance#1#2{ d ( #1 , #2) }
\def\dim#1{ { \hbox{dim} [ #1 ] } }
\def\bra#1{ \langle #1 | }
\def\ket#1{ | #1 \rangle }
\def\vac{ \bra{{\rm vac}} }
\def\cuum{ \ket{{\rm vac}} }
\def\k{\kappa}
\def\inv#1{ {1 \over #1} }
\def\Real{ {\Re e} }

\def\define{ \mathrel{\mathop\equiv^{\rm def}} }
\def\intdz#1{ { \oint\limits_{|z|=#1} \! \! {d z \over 2 \pi i z} } }
\def\intdzeta{ { \int\limits_{- i \infty}^{+ i \infty}
               {d \zeta \over 2 \pi i} } }
\def\Surf{{\cal S\/}}
\def\Loop{{\cal C}}
\def\void{\hskip 10pt}
\def\H{ {\cal H} }

\def\Hone{\mbar \H}
\def\cc{t} 
\def\t{d}
\def\T{D}
%
%
\abstract{
We propose a new method which analyzes the dynamical triangulation
from the viewpoint of the non-critical string field theory.
By using the transfer matrix formalism,
we construct the non-critical string field theory
(including $c>1$ cases) at the discrete level.
For pure quantum gravity,
we succeed in taking the continuum limit
and obtain the $c=0$ non-critical string field theory
at the continuous level.
We also study about the universality of the non-critical string field theory.
}
\endpage
%
%
%
\REF\KPZ{
             V.G. Knizhnik, A.M. Polyakov and A.A. Zamolodchikov,
                Mod.\ Phys.\ Lett.\ {\bf A3} 819 (1988).
             }
\REF\DDK{
             F. David,
                Mod.\ Phys.\ Lett.\ {\bf A3} 1651 (1988);
\nextline
             J. Distler and H. Kawai,
                Nucl.\ Phys.\ {\bf B321} 509 (1989).
             }
\REF\DT{
             F. David,
                Nucl.\ Phys.\ {\bf B257} 45 (1985);
\nextline
             V. A. Kazakov,
                Phys.\ Lett.\ {\bf B150} 282 (1985);
\nextline
             J. Ambj\o rn, B. Durhuus and J. Fr\"ohlich,
                Nucl.\ Phys.\ {\bf B257} 433 (1985).
             }
\REF\BKKM{
             V. A. Kazakov, I. K. Kostov and A. A. Migdal,
                Phys.\ Lett.\ {\bf B157} 295 (1985);
\nextline
             D. Boulatov, V. A. Kazakov, I. K. Kostov and A. A. Migdal,
                Phys.\ Lett.\ {\bf B174} 87 (1986);
                Nucl.\ Phys.\ {\bf B275} 641 (1986).
             }
\REF\Brezin{
             M. Douglas and S. Shenker,
                Nucl.\ Phys.\ {\bf B335} 635 (1990);
\nextline
             E. Br\'ezin and V.A. Kazakov,
                Phys.\ Lett.\ {\bf B236} 144 (1990);
\nextline
             D.J. Gross and A.A. Migdal,
                Phys.\ Rev.\ Lett.\ {\bf 64} 127 (1990);
                Nucl.\ Phys.\ {\bf B340} 333 (1990).
             }
\REF\BIPZ{
             E. Br\'ezin, C. Itzykson, G. Parisi and J.B. Zuber,
                Commun.\ Math.\ Phys.\ {\bf 59} 35 (1978).
             }
\REF\Migdal{
             M. E. Agishtein and A. A. Migdal,
                Int.\ J.\ Mod.\ Phys.\ {\bf C1} 165 (1990);
                Nucl.\ Phys.\ {\bf B350} 690 (1991).
             }
\REF\FRone{
             N. Kawamoto, V.A. Kazakov, Y. Saeki and Y. Watabiki,
                Phys.\ Rev.\ Lett.\ {\bf 68} 2113 (1992);
                Nucl.\ Phys.\ {\bf B}(Proc. Suppl.){\bf 26} 584 (1992);
\nextline
             N. Kawamoto, Y. Saeki and Y. Watabiki,
                in preparation.
             }
\REF\KKMW{
             H. Kawai, N. Kawamoto, T. Mogami and Y. Watabiki,
                Phys.\ Lett.\ {\bf B306} 19 (1993).
             }
\REF\Klebanov{
             S.S. Gubser and I.R. Klebanov,
                Princeton Preprint PUPT-1422.
             }
\REF\Kazakov{
             V.A. Kazakov,
                Mod.\ Phys.\ Lett.\ {\bf A4} 2125 (1989).
             }
\REF\Staudacher{
             M. Staudacher,
                Nucl.\ Phys.\ {\bf B336} 349 (1990).
             }
\REF\IK{
             N. Ishibashi and H. Kawai,
                Phys.\ Lett.\ {\bf B314} 190 (1993).
             }
\REF\IKb{
             N. Ishibashi and H. Kawai,
                KEK-TH-378 December (1993).
             }
\REF\MSS{
             G. Moore, N. Seiberg and M. Staudacher,
                Nucl.\ Phys.\ {\bf B362} 665 (1991).
             }
\REF\KakuKikkawa{
             M. Kaku and K. Kikkawa,
                Phys.\ Rev.\ {\bf D10} 1110 (1974);
                Phys.\ Rev.\ {\bf D10} 1823 (1974).
             }
\def\REFmark#1{\PRrefmark{#1}$\!$}

%
%
%
\def\FigGenAmp{1}
\def\FigGenTM{2}
\def\FigCompLaw{3}
\def\FigSlice{4a}
\def\FigPeel{4b}
\def\FigPeelTriI{5a}
\def\FigPeelTriII{5b}
\def\FigPeelSkinI{6a}
\def\FigPeelSkinII{6b}
\def\FigPeelSkinIII{6c}
\def\FigPeelSkinIV{7}
\def\FigPeelTriA{8a}
\def\FigPeelTriB{8b}
\def\FigPeelTriC{8c}
\def\FigPeelTriD{8d}
\def\FigPeelTriE{8e}
\def\FigPeelTriF{8f}
\def\FigPeelTriG{8g}
%
%
\topskip 30pt
\chapter{\ Introduction }
\vskip 10pt
\par
The quantization of gravity is one of the most serious problems
which have been unsolved yet.
When one considers gravity together with the elementary particle field theory,
one is confronted with the problem of the quantization of gravity.
At present, the critical string theories are
only hopeful candidates of quantum gravity.
However, no one has extracted any phenomenological predictions
from the critical string theories,
though the theories incorporate
gravity as well as Yang-Mills gauge fields at the quantum level.
\par
Recently, there has been successful progress in understanding
the two-dimensional quantum gravity coupled to $c \le 1$ matter
from both viewpoints of
the Liouville theory\REFmark{\KPZ,\DDK}
and the dynamical triangulation.\REFmark{\DT,\BKKM,\Brezin}
The two-dimensional quantum gravity coupled to $c \le 1$ matter
is equivalent to the $c \le 1$ non-critical string theory.
Since the critical bosonic string theory is equivalent to
the $c=25$ non-critical string theory,
the $c \le 1$ non-critical string theories have been investigated
as the toy models of
not only the four-dimensional quantum gravity
but also the critical bosonic string theory.
In the Liouville theory
the path integration of metric is performed,
while
in the dynamical triangulation
all possible triangulated surfaces are summed up
where each triangle is a regular triangle
with the same size.
In order to investigate the dynamical triangulation,
we have, at present, two effective methods:
matrix models\REFmark{\BIPZ,\Brezin} and
numerical simulation.\REFmark{\Migdal,\FRone}
In the present paper we propose the third method
to analyze the dynamical triangulation.
So, no knowledge about the matrix models as well as numerical simulation
is necessary in this paper.
\par
As was shown in ref.\ [{\KKMW}],
the transfer matrix formalism for two-dimensional pure quantum gravity
is powerful to analyze
the fractal structure of quantized surface.
They have obtained a \lq\lq Hamiltonian formalism''
in which the geodesic distance plays the role of time.
In ref.\ [{\Klebanov}]
this analysis was applied to
$m$-th multicritical one matrix model\NPrefmark{\Kazakov}
which is identified\NPrefmark{\Staudacher} with $(2,2m-1)$ minimal model.
In ref.\ [{\IK}] the authors have push the transfer matrix formalism forward
and proposed the non-critical string field theory
for two-dimensional pure gravity.
Recently, they have also constructed
$c \le 1$ non-critical string field theory.\NPrefmark{\IKb}
\par
In this paper we propose a new method
to analyze the dynamical triangulation
from the viewpoint of the non-critical string field theory,
which is constructed by using the transfer matrix formalism.
The proper time plays an important role in this string field theory.
By using the minimal-step decompositions which are less than
one-step decomposition used in ref.\ [{\KKMW}],
we construct the non-critical string field theories (including $c>1$ cases)
at the discrete level.
In the case of pure gravity we succeed in taking the continuum limit,
and then obtain the $c=0$ non-critical string field theory
at the continuous level.
Though the Hamiltonian obtained in this paper is slightly different from
that in ref.\ [{\IK}],
both theory lead to the same amplitudes.
We also study about the universality of
the $c=0$ non-critical string field theory.
\par
The organization of the present paper is as follows:
In section 2 we give the definition of transfer matrices
as well as amplitudes
in the framework of the dynamical triangulation.
We also define the \lq peeling decomposition'
which is one of the transfer matrices and plays an essential role
in the construction of discretized string field theories.
In section 3
we construct the discretized $c=0$ non-critical string field theory
by using the \lq peeling decomposition'.
In section 4 we take the continuum limit
and obtain the $c=0$ non-critical string field theory at the continuous level.
In section 5 we investigate the universality of
the $c=0$ non-critical string field theory.
The same Hamiltonian is always obtained after taking the continuum limit,
in spite of the modification of
\lq peeling decomposition' at the discrete level.
In section 6 we study the fractal structure of quantized surface by using
the number operator of universes.
In section 7 we extend our formalism to
the string field theory with matter fields on surface.
The matter fields are introduced
by replacing triangles with some kinds of regular polygons
like $m$-th multicritical one matrix model
or by putting matter fields on each link naively.
The last section is devoted to the conclusion.
In appendix A we give the definition and the properties of
the discrete Laplace transformation,
which plays an important role when one takes the continuum limit.
The usual Laplace transformation is obtained
after taking the continuum limit of the discrete Laplace transformation.
In appendix B we summarize the notations and the properties
of the transfer matrices and the amplitudes.
In appendices C and D we derive the Schwinger-Dyson equations
at the discrete level and at the continuous level respectively
by using the non-critical string field theory.
We also calculate the explicit forms of some amplitudes
at the continuous level
in appendix D.
\NextPage
\topskip 30pt
\chapter{\ Transfer Matrix Formalism in the Dynamical Triangulation }
\vskip 10pt
\par
In this section we explain the foundation of
the transfer matrix formalism in the framework of the dynamical triangulation
for pure gravity.
The extension to the gravity theory coupled to matter fields
is straightforward.
The transfer matrix plays an essential role in
the construction of non-critical string field theory in this paper.
\par
The dynamical triangulation is the two-dimensional lattice gravity
whose space--time is regularized by regular triangles with the same size.
The curvature on a site $i$ is expressed by
$R_i = \pi ( 6 - q_i ) / q_i$,
where $q_i$ is the number of triangles concentrated on the site $i$.
The path integration of the metric is performed
by summing up all possible triangulated two-dimensional surfaces.
\par
The amplitude of a connected surface
with $h$ ($\ge 0$) handles and $N$ ($\ge 0$) boundaries
is defined by
$$ \eqalign{ \sp(2.0)
& F_N^{(h)} ( l_1, \ldots, l_N ; \k )  \ = \
  \sum_{a=0}^\infty \, \sum_{\Surf_N^{(h)}} \, \k^a
\cr\sp(4.0)
& \hbox{with} \qquad
  \Surf_N^{(h)}  \ = \  \Surf_N^{(h)} ( l_1, \ldots, l_N ; a ) \, ,
\cr\sp(3.0)} \eqn\EqGenAmplitude $$
where
$\Surf_N^{(h)}$ is one of the triangulated connected lattice surfaces
with $h$ handles and
$N$ boundaries denoted by
$\Loop_1$, $\ldots$, $\Loop_N$.
We also fix
the number of triangles on $\Surf_N^{(h)}$ as $a$
(which corresponds to the volume of surface $\Surf_N^{(h)}$)
and
the number of links on each $\Loop_i$ as $l_i$
(which corresponds to the length of each string $\Loop_i$).
In Fig.\ {\FigGenAmp} one of surfaces $\Surf_N^{(h)}$ is illustrated.
For later convenience we mark one of links on each boundary $\Loop_i$
as is shown in Fig.\ {\FigGenAmp}.
The parameter $\k$ is related to
the cosmological constant on lattice, $\tilde\cc$, by $\k = e^{-\tilde\cc}$.
In this paper, we use $\k$ instead of $\tilde\cc$.
$\k$ is considered to be put on each triangle.
\par
Next let us consider a connected transfer matrix
which makes
$N$ ($\ge 1$) initial closed strings merge and split into
$M$ ($\ge 0$) final closed strings
during $\t$-step lapse of time.  
Similarly to {\EqGenAmplitude},
the connected transfer matrix is defined by
$$ \eqalign{ \sp(2.0)
& T_{M,N}^{(h)} ( l'_1, \ldots, l'_M ; l_1, \ldots, l_N ; \k ; \t )
  \ = \
  \sum_{a=0}^\infty \, \sum_{\Surf_{M,N}^{(h)}} \, \k^a
\cr\sp(4.0)
& \hbox{with} \qquad
  \Surf_{M,N}^{(h)}
  \ = \
  \Surf_{M,N}^{(h)} ( l'_1, \ldots, l'_M ; l_1, \ldots, l_N ; a ; \t ) \, ,
\cr\sp(3.0)} \eqn\EqGenTM $$
where
$\Surf_{M,N}^{(h)}$ is one of the triangulated connected lattice surfaces
with $h$ handles and
$N$ initial string boundaries denoted by $\Loop_1$, $\ldots$, $\Loop_N$ and
$M$ final string boundaries denoted by $\Loop'_1$, $\ldots$, $\Loop'_M$.
We also fix
the number of triangles of $\Surf_{M,N}^{(h)}$ as $a$ and
the number of links of $\Loop_i$ and $\Loop'_j$
as $l_i$ and $l'_j$, respectively.
The geodesic distance $\t$ on a lattice surface is introduced
in order to fix the shape of $\Surf_{M,N}^{(h)}$
by the following two conditions:
\item{  i) }{
for any link $p$ ($p \in \Loop'$), \quad
$\min_{q \in \Loop} \Distance{p}{q} = \t$,
}
\item{ ii) }{
for any link $p$ ($p \in \Surf_{M,N}^{(h)}$ and $p \not\in \Loop'$), \quad
$\min_{q \in \Loop} \Distance{p}{q} < \t$,
}
\par{\noindent}where
$\Loop   =  \bigcup_{i=1}^N  \Loop_i$
and
$\Loop'  =  \bigcup_{i=1}^M \Loop'_i$.
The geodesic distance $\Distance{p}{q}$ is
defined as how many centers of triangles one can minimally pass through
on the dual link of the triangulated surface between two links $p$ and $q$.
The conditions {\it i}) and {\it ii}) define
the $C'$ as a set of all links $p$'s
each of which satisfies $\min_{q \in \Loop} \Distance{p}{q} = \t$.
Thus, the connected transfer matrix,
$T_{M,N}^{(h)}$, is defined by
the summation of all triangulated connected lattice surfaces,
$\Surf_{M,N}^{(h)}$,
which has $h$ handles and $M+N$ boundaries,
and at the same time satisfies the above conditions {\it i}) and {\it ii}).
In Fig.\ {\FigGenTM} one of surfaces $\Surf_{M,N}^{(h)}$ is illustrated.
For later convenience we mark one of links
on each initial string $\Loop_i$ as is shown in Fig.\ {\FigGenTM}.
As is manifest from the definition,
the transfer matrix is not invariant under the time reversal,
$\t \rightarrow -\t$ and $\Loop \leftrightarrow \Loop'$.
Since the number of triangles on surface is finite,
we find for $N \ge 1$ that
$$ \eqalign{ \sp(2.0)
& \lim_{\t \rightarrow \infty}
  T_{0,N}^{(h)} ( \void ; l_1, \ldots, l_N ; \k ; \t )
  \ = \
  F_N^{(h)} ( l_1, \ldots, l_N ; \k ) \, ,
\cr\sp(6.0)
& \lim_{\t \rightarrow \infty}
  T_{M>0,N}^{(h)} ( l'_1, \ldots, l'_M ; l_1, \ldots, l_N ; \k ; \t )
  \ = \
  0 \, .
\cr\sp(3.0)} \eqn\EqAmpAndTM $$
We also find that
$$ \eqalign{ \sp(2.0)
  \lim_{\t \rightarrow 0}
  T_{M,N}^{(h)} ( l'_1, \ldots, l'_M ; l_1, \ldots, l_N ; \k ; \t )
  \ = \
  \delta_{h,0} \delta_{M,1} \delta_{N,1} \delta_{l'_1,l_1} \, ,
\cr\sp(3.0)} \eqn\EqTMandDelta $$
because $T_{M,N}^{(h)}$ is the summation of the connected surfaces.
\par
The important property that the transfer matrix satisfies
is the composition law.
For example,
$T_{M=0,N=1}^{(h=0)} ( \void ; l ; \k ; \t_2 + \t_1 )$
is decomposed as
$$ \eqalign{ \sp(2.0)
& T_{0,1}^{(0)} ( \void ; l ; \k ; \t_2 + \t_1 ) \
\cr\sp(6.0)
& = \
  \sum_{m=0}^\infty \, {1 \over m!}
  \sum_{l_1=1}^\infty \! \cdots \! \sum_{l_m=1}^\infty
  T_{0,1}^{(0)} ( \void ; l_1 ; \k ; \t_2 )
  \cdots
  T_{0,1}^{(0)} ( \void ; l_m ; \k ; \t_2 ) \,
\cr\sp(3.0)} \eqn\EqCompoLaw $$
where the blanks between \lq$($' and \lq$;$'
mean that there are no final string states.
$1/m!$ is the symmetric factor.
The right-hand side of eq.\ {\EqCompoLaw} is illustrated
in Fig.\ {\FigCompLaw}.
Thus,
any $\t$-step transfer matrix is decomposed into
a product of connected minimal-step transfer matrices.
\par
Next, we introduce the discrete Laplace transformation.
Its properties are explained in detail in appendix A.
The discrete Laplace transformations of
the transfer matrices and the amplitudes are
$$ \eqalign{ \sp(2.0)
& T_{M,N}^{(h)} ( y_1, \ldots, y_M ; x_1, \ldots, x_N ; \k ; \t )
\cr\sp(6.0)
& = \
  \sum_{l'_1,\ldots,l'_M,l_1,\ldots,l_N = 1}^\infty
  y_1^{l'_1} \cdots y_M^{l'_M} \,
  x_1^{l_1}  \cdots x_N^{l_N} \,
  T_{M,N}^{(h)} ( l'_1, \ldots, l'_M ; l_1, \ldots, l_N ; \k ; \t ) \, ,
\cr\sp(3.0)} \eqn\EqLTransTMDis $$
and
$$ \eqalign{ \sp(2.0)
  F_N^{(h)} ( x_1, \ldots, x_N ; \k ) \
  = \
  \sum_{l_1,\ldots,l_N = 1}^\infty
  x_1^{l_1}  \cdots x_N^{l_N} \,
  F_N^{(h)} ( l_1, \ldots, l_N ; \k ) \, .
\cr\sp(3.0)} \eqn\EqLTransAmpDis $$
Let $x_c$, $y_c$ and $\k_c$ be fixed real numbers and assume that
\item{ a) }{
the convergence radii of
$x_i$ ($1 \le i \le N$) are equal to $x_c$,
}
\item{ b) }{
the convergence radii of
$y_j$ ($1 \le j \le M$) are equal to $y_c$,
}
\item{ c) }{
the convergence radius of
$\k$ is equal to $\k_c$,
}
\par{\noindent}for any transfer matrix $T_{M,N}^{(h)}$ in {\EqLTransTMDis} and
any amplitude $F_N^{(h)}$ in {\EqLTransAmpDis}.
The assumptions $a$), $b$) and $c$) lead to the fact that
all transfer matrices $T_{M,N}^{(h)}$ as well as
all amplitudes $F_N^{(h)}$ are analytic in the region
$|x_i| < x_c$ ($1 \le i \le N$), $|y_j| < y_c$ ($1 \le j \le M$)
and $|\k| < \k_c$.
In the next section, we will extend the above analytic region to
$|x_i| \le x_c$ ($1 \le i \le N$), $|y_j| \le y_c$ ($1 \le j \le M$)
and $|\k| < \k_c$ in order to define inner products
in the Laplace transformed representation.
The above assumptions are very natural
because the local structure of surface is independent of the global structure.
\par
In the following we will propose new minimal-step transfer matrices,
which plays an essential role
in the construction of the discretized non-critical string field theory
in the present paper.
In ref.\ [{\KKMW}] the authors have considered
one-step decomposition as a minimal-step one.
In stead of the one-step decomposition
like Fig.\ {\FigSlice},
we consider in the present paper the decomposition
like \lq peeling an apple' in Fig.\ {\FigPeel}.
In the case of \lq peeling decomposition',
we need a marked link which indicates the present peeling point
and has been already introduced on each initial string
in the definitions of $F_N^{(h)}$ and $T_{M,N}^{(h)}$.
A minimal-step \lq peeling decomposition' removes
one triangle with a marked link from the triangulated surface.
Therefore, the minimal-step \lq peeling decomposition'
corresponds to $(1/l)$-step one
if the length of the initial string is $l$.
Since the surface is triangulated,
there are three different types of minimal-step \lq peeling decompositions'
illustrated in Fig.\ {\FigPeelTriI}.
One of them adds the length of string by one,
while others reduce it by one.
In order to identify these three decompositions with one decomposition,
we introduce two-folded loop parts
like $\alpha$ and $\beta$ in Fig.\ {\FigPeelTriII}.
Then, owing to the two-folded parts,
only one decomposition Fig.\ {\FigPeelSkinI} is necessary and sufficient
when one removes a triangle.
However, in addition to Fig.\ {\FigPeelSkinI},
we need two decompositions illustrated
in Figs.\ {{\FigPeelSkinII} and {\FigPeelSkinIII}},
because we have to remove the two-folded parts.
Thus, we find three different types of
minimal-step \lq peeling decompositions'
illustrated
in Figs.\ {{\FigPeelSkinI}, {\FigPeelSkinII} and {\FigPeelSkinIII}}.
The first decomposition Fig.\ {{\FigPeelSkinI} removes
a triangle with a marked link
and adds the length of string by one.
The second decomposition Fig.\ {{\FigPeelSkinII} removes
a two-folded part with a marked link
and splits a string with length $l$
into two strings with lengths $l'$ and $l-l'-2$ respectively.
The third decomposition Fig.\ {{\FigPeelSkinIII} removes
a two-folded part with a marked link
and merges two strings with lengths $l$ and $l'$ respectively
into one string with length $l+l'-2$.
The number of decompositions for
Figs.\ {{\FigPeelSkinI} and {\FigPeelSkinII} are only one respectively
if one fixes the length of strings.
On the other hand, the number of decompositions for
Fig.\ {{\FigPeelSkinIII} is $l'$
because of the location of a marked point on the string with length $l'$.
These three decompositions are considered to be $(1/l)$-step decompositions
if the length of the initial string is $l$.
By performing three minimal-step decompositions
Figs.\ {{\FigPeelSkinI}, {\FigPeelSkinII} and {\FigPeelSkinIII}}
over and over again,
we construct the string field theory
which produces any kinds of genus topology.
By using only two decompositions
Figs.\ {{\FigPeelSkinI} and {\FigPeelSkinII}},
one can construct the string field theory for disk amplitude,
because the decomposition Fig.\ {\FigPeelSkinIII} makes genus higher.
\NextPage
\topskip 30pt
\chapter{\ Discretized String Field Theory }
\vskip 10pt
\par
In this section we construct
the discretized string field theory of the dynamical triangulation
with no matter fields on surface.
In order to construct the string field theory
which produces the amplitude of any genus topology,
we perform three minimal-step \lq peeling decompositions'
in Figs.\ {{\FigPeelSkinI}, {\FigPeelSkinII} and {\FigPeelSkinIII}}
over and over again.
\par
Let $\Psi^\dagger (l)$ and $\Psi (l)$ be operators,
which creates and annihilates one closed string with length $l$ ($\ge 1$),
respectively,
where a string has one marked link
in order to indicate the present peeling point.
Their commutation relations are
$$ \eqalign{ \sp(2.0)
& [ \, \Psi (l) \, , \, \Psi^\dagger (l') \, ]  \ = \  \delta_{l,l'} \, ,
\cr\sp(6.0)
& [ \, \Psi^\dagger (l) \, , \, \Psi^\dagger (l') \, ]  \ = \
  [ \, \Psi (l) \, , \, \Psi (l') \, ]  \ = \  0 \, ,
\cr\sp(3.0)} \eqn\EqCommutePsiDisI $$
where $\delta_{l,l'}$ is the Kronecker's delta.
The vacuum states, $\cuum$ and $\vac$, satisfy
$$ \eqalign{ \sp(2.0)
  \Psi (l) \cuum  \ = \  \vac \Psi^\dagger (l)  \ = \  0 ,
  \qquad \hbox{ ( for any $l \ge 1$ ) }
\cr\sp(3.0)} \eqn\EqVacuumDisI $$
where $\langle {\rm vac} | {\rm vac} \rangle  = 1$.
Thus, for example,
$\Psi^\dagger (l) \cuum$
represents one closed string state with length $l$.
$\Psi^\dagger (l_1) \Psi^\dagger (l_2) \cdots \Psi^\dagger (l_N) \cuum$
represents $N$ closed strings
with lengths $l_1$, $l_2$, $\ldots$, and $l_N$, respectively.
As typical examples of physical observables,
we have
$$ \eqalign{ \sp(2.0)
  v^{(n)}  \ = \
  \sum_{l=1}^\infty \, l^n \, \Psi^\dagger (l) \, \Psi (l) \, ,  \qquad
  \hbox{ ( for $n = 0, 1, \ldots$ ) }
\cr\sp(3.0)} \eqn\EqUnivNoDisI $$
which satisfy
$$ \eqalign{ \sp(2.0)
  [ \, v^{(n)} \, , \, \Psi^\dagger (l) \, ] \ = \ l^n \, \Psi^\dagger (l) \, ,
\qquad
  [ \, v^{(n)} \, , \, \Psi (l) \, ] \ = \ - \, l^n \, \Psi (l) \, .
\cr\sp(3.0)} \eqn\EqUnivNoDisII $$
The physical meanings of $v^{(n)}$ are, for example, as follows:
$v^{(0)}$ counts the number of strings
(the number of one-dimensional universes),
and
$v^{(1)}$ estimates the total length of all strings
(the total volume of one-dimensional universes).
\par
Now, let us consider the Hamiltonian,
$\H (g,\k)$,
which generates one-step decomposition of surface
or equivalently one-step deformation of wave function.
The coupling constant $g$ counts the number of handles
and will be explained in detail later.
The $(1/l)$-step deformation of the wave function,
$\Psi^\dagger(l) \rightarrow \Psi^\dagger(l) + \delta_{1/l} \Psi^\dagger(l)$,
is derived from the Hamiltonian $\H (g,\k)$ as
$$ \eqalign{ \sp(2.0)
\delta_{1/l} \, \Psi^\dagger (l)
\ = \
- \, [ \, {1 \over l} \, \H (g,\k) \, , \, \Psi^\dagger (l) \, ]  .
\cr\sp(3.0)} \eqn\EqOneOverLHamilton $$
The minimal-step \lq peeling decompositions'
illustrated
in Figs.\ {{\FigPeelSkinI}, {\FigPeelSkinII} and {\FigPeelSkinIII}}
deform the wave function $\Psi^\dagger (l)$ as
$$ \eqalign{ \sp(2.0)
& \Psi^\dagger (l)  \ \longrightarrow \
  \k \Psi^\dagger (l+1) \,  ,
  \hskip 90pt \qquad \hbox{ ( for Fig.\ {\FigPeelSkinI} ) }
\cr\sp(6.0)
& \Psi^\dagger (l)  \ \longrightarrow \
  \sum_{l'=0}^{l-2} \Psi^\dagger (l') \Psi^\dagger (l-l'-2) \, ,
  \hskip 2pt \qquad \qquad \hbox{ ( for Fig.\ {\FigPeelSkinII} ) }
\cr\sp(6.0)
& \Psi^\dagger (l)  \ \longrightarrow \
  \sum_{l'=1}^\infty l' \Psi^\dagger (l+l'-2) \Psi (l') \, ,
  \qquad \qquad \hbox{ ( for Fig.\ {\FigPeelSkinIII} ) }
\cr\sp(3.0)} \eqn\EqOneOverLDeformI $$
where we have introduced
$$ \eqalign{ \sp(2.0)
  \Psi^\dagger (l=0)  \ = \  1 \, ,
\cr\sp(3.0)} \eqn\EqNormalize $$
in order to simplify the second and the third deformations
in {\EqOneOverLDeformI}.
Note that we do not introduce the operator $\Psi (l=0)$.
The factor $l'$ is necessary
in the right-hand side of the last deformation in {\EqOneOverLDeformI}
because there are $l'$ types of figures for Fig.\ {\FigPeelSkinIII}
owing to the location of the marked link on $\Psi(l')$.
Note that three deformations in {\EqOneOverLDeformI}
do not depend on the location of the next peeling point on the boundary,
while they depend on that of the present peeling one.
Therefore, we can arbitrarily mark one of links of next string
after the deformations {\EqOneOverLDeformI}.
In the dynamical triangulation
all possible triangulated surfaces are summed up.
Therefore, we have to sum up all three kinds of $(1/l)$-step decompositions
in {\EqOneOverLDeformI}
in order to obtain the $(1/l)$-step deformed wave function,
$\Psi^\dagger(l) + \delta_{1/l} \Psi^\dagger(l)$.
Then, we find
$$ \eqalign{ \sp(2.0)
\delta_{1/l} \, \Psi^\dagger (l) \ = \
&  - \, \Psi^\dagger (l)
\, + \, \k \Psi^\dagger (l+1)
\, + \, g' \, ( 1 - \delta_{l,1} ) \,
        \sum_{l'=0}^{l-2} \Psi^\dagger (l') \Psi^\dagger (l-l'-2)
\cr\sp(4.0)
&  + \, 2 g \, \sum_{l'=1}^\infty \Psi^\dagger (l+l'-2) l' \Psi (l') \, ,
\cr\sp(3.0)} \eqn\EqOneOverLDeformII $$
where $g'$ and $g$ are the coupling constants of string interaction.
By the scaling field redefinition,
$\Psi^\dagger \rightarrow (1/g') \Psi^\dagger$
and
$\Psi \rightarrow g' \Psi$,
one of the coupling constants is removed.
Therefore we set $g' = 1$ in {\EqOneOverLDeformII}.
Another string coupling constant $g$ cannot be removed
by the field redefinition,
because this coupling constant distinguishes
the surfaces with different number of handles.
{}From {\EqOneOverLHamilton} and {\EqOneOverLDeformII},
we obtain the Hamiltonian which generates one-step deformation as
$$ \eqalign{ \sp(2.0)
\H (g,\k)  \ = \
&\sum_{l=1}^\infty \bigl\{ \,
 \Psi^\dagger (l) \, - \, \k \Psi^\dagger (l+1)
 \, \bigr\} \, l \, \Psi (l)
 \, - \,
 \sum_{l=2}^\infty \,
 \sum_{l'=0}^{l-2} \Psi^\dagger (l') \Psi^\dagger (l-l'-2)
 \, l \, \Psi (l)
\cr\sp(6.0)
&- \, g \, \sum_{l=1}^\infty \sum_{l'=1}^\infty
 \Psi^\dagger (l+l'-2) \, l \, \Psi (l) \, l' \, \Psi (l') \, .
\cr\sp(3.0)} \eqn\EqH $$
Thus, we have obtained the Hamiltonian
of the discretized $c=0$ string field theory.
Note that the Hamiltonian {\EqH} has tadpole diagrams,
$-2 \Psi(2)$ and $-g \Psi(1) \Psi(1)$,
because of {\EqNormalize}.
\par
The transfer matrix operator for $\t$-step is
$e^{- \t \H (g,\k)}$.
The coupling constant $g$ in the Hamiltonian {\EqH}
appears when two strings merge together into one string.
Therefore,
expanding the amplitudes in terms of $g$,
the contribution from each surface
is proportional to $g^{ h + N - 1 }$.
Thus, the transfer matrix $T_{M,N}^{(h)}$ is obtained
from the transfer matrix operator $e^{- \t \H (g,\k)}$ as
$$ \eqalign{ \sp(2.0)
& \sum_{h=0}^\infty g^{h+N-1} \,
  T_{M,N}^{(h)} ( l'_1, \ldots, l'_M ; l_1, \ldots, l_N ; \k ; \t )
\cr\sp(6.0)
& = \
  \vac \,  \Psi (l'_1) \, \cdots \, \Psi (l'_M) \,
           e^{- \t \H (g,\k)} \,
           \Psi^\dagger (l_1) \, \cdots \, \Psi^\dagger (l_N) \,
           \cuum_{\rm connected} \, ,
\cr\sp(3.0)} \eqn\EqGenTMDisI $$
where the suffix \lq connected' means that
only connected Feynman diagrams are estimated.
Note that
the Hamiltonian $\H (g,\k)$ is not invariant under time reversal,
because
$\H (g,\k) \cuum = 0$,
on the other hand,
$\vac \H (g,\k) \neq 0$.
This leads to the fact that
the transfer matrices are not invariant under time reversal,
which was manifest in the definition of the transfer matrices {\EqGenTM}
with the conditions {\it i}) and {\it ii}).
{}From eqs.\ {\EqAmpAndTM} and {\EqGenTMDisI}
we also obtain the amplitude $F_N^{(h)}$ as
$$ \eqalign{ \sp(2.0)
& \sum_{h=0}^\infty g^{h+N-1} \,
  F_N^{(h)} ( l_1, \ldots, l_N ; \k )
\cr\sp(6.0)
& = \
  \lim_{\t\rightarrow\infty}
  \vac \, e^{- \t \H (g,\k)} \,
          \Psi^\dagger (l_1) \, \cdots \, \Psi^\dagger (l_N) \,
          \cuum_{\rm connected} \, .
\cr\sp(3.0)} \eqn\EqGenAmpDisI $$
Especially, the disk amplitude, $F_{N=1}^{(h=0)} (l;\k)$, is
$$ \eqalign{ \sp(2.0)
  F_1^{(0)} ( l ; \k )  \ = \
  \lim_{\t\rightarrow\infty}
  \vac \, e^{- \t \H (g=0,\k)} \,
          \Psi^\dagger (l) \, \cuum \, ,
\cr\sp(3.0)} \eqn\EqAmpDiskDisI $$
where we do not need the suffix \lq connected' in this case.
The Hamiltonian $\H (g=0,\k)$ is considered to be
the Hamiltonian for the disk amplitude.
Here note that
from the viewpoint of the string field theory,
eqs.\ {\EqAmpAndTM} are rederived from the fact that
any state goes to the vacuum state for $\t\rightarrow\infty$,
i.e.,
$$ \eqalign{ \sp(2.0)
  \lim_{\t\rightarrow\infty} e^{- \t \H (g,\k)} \,
  \Psi^\dagger (l_1) \, \cdots \, \Psi^\dagger (l_N) \, \cuum
  \  \propto  \
  \cuum \, ,
\cr\sp(3.0)} \eqn\EqToVacuum $$
because the number of triangles is finite.
Thus, we have completed the construction of
the discretized $c=0$ non-critical string field theory.
\par
Next, we introduce the discrete Laplace transformation.
In the discrete Laplace transformed representation
one can let the lattice spacing constant $\e$ go to zero continuously.
Namely, there are no ambiguities
when one takes the continuum limit in this representation.
We explain in detail about the discrete Laplace transformation
and its continuum limit in appendix A.
Since any transfer matrix and any amplitude are written
by the wave functions and the transfer matrix operator,
for example, as {\EqGenTMDisI} or {\EqGenAmpDisI},
the assumptions {\it a}), {\it b}) and {\it c})
about the convergence radii
make us possible
to apply the discrete Laplace transformation to the wave functions,
$\Psi^\dagger$ and $\Psi$.
Thus, the wave functions $\Psi^\dagger(x)$ and $\Psi(y)$ are considered to be
analytic in the region $|x| < x_c$ and $|y| < y_c$,
where $x_c$ and $y_c$ are the convergence radii of
$\Psi^\dagger(x)$ and $\Psi(y)$, respectively.
The discrete Laplace transformation
of the wave functions are defined by
$$ \eqalign{ \sp(2.0)
  \Psi^\dagger (x)  \ \define \
  \Psi^\dagger(l=0) \, + \,
  \sum_{l=1}^\infty \, x^l \, \Psi^\dagger (l) \, ,
  \qquad
  \Psi (y)  \ \define \
  \sum_{l=1}^\infty \, y^l \, \Psi (l) \, ,
\cr\sp(3.0)} \eqn\EqLaplacePsiDis $$
where we have added $\Psi^\dagger(l=0)$
in order to simplify the form of the Hamiltonian.
Especially, we have
$$ \eqalign{ \sp(2.0)
  \Psi^\dagger (x=0)  \ = \  1 \, , \qquad
  \Psi         (y=0)  \ = \  0 \, ,
\cr\sp(3.0)} \eqn\EqNormalizeDis $$
because of {\EqNormalize} and {\EqLaplacePsiDis}.
The commutation relations {\EqCommutePsiDisI} become
$$ \eqalign{ \sp(2.0)
& [ \, \Psi (y) \, , \, \Psi^\dagger (x) \, ]  \ = \  \delta ( y , x ) \, ,
\cr\sp(6.0)
& [ \, \Psi^\dagger (x_1) \, , \, \Psi^\dagger (x_2) \, ]  \ = \
  [ \, \Psi (y_1) \, , \, \Psi (y_2) \, ]  \ = \  0 \, ,
\cr\sp(3.0)} \eqn\EqCommutePsiDisII $$
where
$\delta ( y , x ) = y x / ( 1 - y x )$
is the discrete Laplace transformation of $\delta_{l',l}$
and is convergent if $|x y| < 1$.
The discrete Laplace transformation of the transfer matrix {\EqGenTMDisI}
has the form,
$$ \eqalign{ \sp(2.0)
& \sum_{h=0}^\infty g^{h+N-1} \,
  T_{M,N}^{(h)} ( y_1, \ldots, y_M ; x_1, \ldots, x_N ; \k ; \t )
\cr\sp(6.0)
& = \
  \vac \,  \Psi (y_1) \, \cdots \, \Psi (y_M) \,
           e^{- \t \H (g,\k)} \,
           \Psi^\dagger (x_1) \, \cdots \, \Psi^\dagger (x_N) \,
           \cuum_{\rm connected} \, ,
\cr\sp(3.0)} \eqn\EqGenTMDisII $$
while that of the amplitude {\EqGenAmpDisI} has the form,
$$ \eqalign{ \sp(2.0)
& \sum_{h=0}^\infty g^{h+N-1} \,
  F_N^{(h)} ( x_1, \ldots, x_N ; \k )
\cr\sp(6.0)
& = \
  \lim_{\t\rightarrow\infty}
  \vac \, e^{- \t \H (g,\k)} \,
          \Psi^\dagger (x_1) \, \cdots \, \Psi^\dagger (x_N) \,
          \cuum_{\rm connected} \, .
\cr\sp(3.0)} \eqn\EqGenAmpDisII $$
Especially, the disk amplitude {\EqAmpDiskDisI} becomes
$$ \eqalign{ \sp(2.0)
  F_1^{(0)} ( x ; \k )  \ = \
  \lim_{\t\rightarrow\infty}
  \vac \, e^{- \t \H (g=0,\k)} \, \Psi^\dagger (x) \, \cuum \, .
\cr\sp(3.0)} \eqn\EqAmpDiskDisII $$
\par
Next, we consider to express inner products,
like $v^{(n)}$ in {\EqUnivNoDisI} or $\H(g,\k)$ in {\EqH},
in terms of $\Psi^\dagger(x)$ and $\Psi(y)$.
As a simplest example, let us consider $v^{(0)}$ at first.
{}From {\EqUnivNoDisI} and {\EqLaplacePsiDis} we obtain
$$ \eqalign{ \sp(2.0)
  v^{(0)}  \ = \
  \oint {d z \over 2 \pi i z} \, \Psi^\dagger (z) \Psi (\inv{z}) \, ,
\cr\sp(3.0)} \eqn\EqUnivNoDisIII $$
where the loop-integration contour should satisfy
$1/y_c < |z| < x_c$.
The Laplace transformation of {\EqUnivNoDisII} is
$$ \eqalign{ \sp(2.0)
  [ \, v^{(0)} \, , \, \Psi^\dagger (x) \, ]  \ = \  \Psi^\dagger (x) \, ,
\qquad
  [ \, v^{(0)} \, , \, \Psi (y) \, ]  \ = \  - \, \Psi (y) \, .
\cr\sp(3.0)} \eqn\EqUnivNoDisIV $$
In order that $v^{(0)}$ in {\EqUnivNoDisIII} satisfies eqs.\ {\EqUnivNoDisIV}
for any values of $x$ ($|x| < x_c$) and $y$ ($|y| < y_c$),
we need $x_c y_c = 1$ and
the analyticity of $\Psi^\dagger(x)$ and $\Psi(y)$
on the region $|x| = x_c$ and $|y| = y_c$.
Namely, we have to assume that
\item{ d) }{
$y_c \, = \, 1 / x_c$, $0 < x_c$, $0 < y_c$,
}
\item{ e) }{
$T_{M,N}^{(h)} ( y_1, \ldots, y_M ; x_1, \ldots, x_N ; \k ; \t )$
and
$F_N^{(h)} ( x_1, \ldots, x_N ; \k )$
are analytic in the region
$|x_i| \le x_c$ ($1 \le i \le N$), $|y_j| \le y_c$ ($1 \le j \le M$)
and $|\k| < \k_c$
for any values of $h$, $M$, $N$ and $\t$.
}
\par{\noindent}Note that the assumption $e$)
extends the initial analytic region of
$T_{M,N}^{(h)}$ and $F_N^{(h)}$ in section 2,
and is consistent with the assumptions $a$) and $b$).
Under the assumptions $d$) and $e$), we can construct
not only $v^{(0)}$ but also $v^{(n)}$ as
$$ \eqalign{ \sp(2.0)
  v^{(n)}  \ = \
  \intdz{x_c} \, \Psi^\dagger (z)
                 \bigl( - z {\der \over \der z} \bigr)^n \Psi (\inv{z}) ,
  \qquad
  \hbox{ ( for $n = 0, 1, \ldots$ ) }
\cr\sp(3.0)} \eqn\EqUnivNoDisVI $$
where the loop integration contour is counterclockwise satisfying $|z| = x_c$.
{}From {\EqUnivNoDisVI} and {\EqCommutePsiDisII}, we find
$$ \eqalign{ \sp(2.0)
  [ \, v^{(n)} \, , \, \Psi^\dagger (x) \, ]
  \ = \
  \bigl( x {\der \over \der x} \bigr)^n \, \Psi^\dagger (x) ,
\quad
  [ \, v^{(n)} \, , \, \Psi (y) \, ]
  \ = \
  - \, \bigl( y {\der \over \der y} \bigr)^n \, \Psi (y) ,
\cr\sp(3.0)} \eqn\EqUnivNoDisV $$
which are also derived from
the discrete Laplace transformation of {\EqUnivNoDisII}.
\par
The definition of the vacuum state in {\EqVacuumDisI} is rewritten as
$$ \eqalign{ \sp(2.0)
& \Psi (y) \cuum  \ = \  0 \, , \,
  \qquad \hbox{ ( for any $y \le y_c = 1/x_c$ ) }
\cr\sp(4.0)
& \vac \Psi^\dagger (x)  \ = \  0 \, .
  \qquad \hbox{ ( for any $x \le x_c$ ) }
\cr\sp(3.0)} \eqn\EqVacuumDisII $$
The Hamiltonian in {\EqH} is expressed by
$$ \eqalign{ \sp(2.0)
  \H (g,\k)  \ = \  \intdz{x_c}
  \, \Bigl\{ \,
&    \bigl\{ \,
     ( 1 - {\k \over z} ) \, \Psi^\dagger (z) \,
       - \, z^2 \, ( \Psi^\dagger (z) )^2 \,
     \bigr\} \, ( - z {\der \over \der z} \, \Psi (\inv{z}) )
\cr\sp(6.0)
&    - \, g \, z^2 \, \Psi^\dagger (z) \,
     ( - z {\der \over \der z} \, \Psi (\inv{z}) )^2
  \, \Bigr\} \, .
\cr\sp(3.0)} \eqn\EqHDis $$
In the derivation of {\EqHDis}, the assumptions $d$) and $e$) are crucial
as the same as before.
In ref.\ [{\KKMW}] the authors have required the assumption $d$)
in order to take the continuum limit.
As a matter of fact, the assumption $d$) is already essential
to the construction of the Hamiltonian {\EqHDis} at the discrete level.
By using {\EqAmpDiskDisII} with {\EqHDis},
we have derived the Schwinger-Dyson equation in appendix C,
which agrees with the result by ref.\ [{\BIPZ}].
We have also derived the Schwinger-Dyson equations
for general genus amplitudes.
\par
Next we consider the time evolution of string states
for later convenience.
The explicit form of the Hamiltonian $\H (g,\k)$ is determined uniquely from
$$ \eqalign{ \sp(2.0)
  \H (g,\k) \, \Psi^\dagger (x_1) \, \cdots \, \Psi^\dagger (x_N) \,
  \cuum  \, , \quad
  \hbox{( for any $N > 0$ )}
\cr\sp(3.0)} \eqn\EqHDetI $$
or equivalently,
$$ \eqalign{ \sp(2.0)
  [ \, \cdots \, [ \, \H (g,\k) \, , \, \Psi^\dagger (x_1) \, ] \, ,
  \cdots \, , \, \Psi^\dagger (x_N) \, ] \,   \cuum  \, .  \quad
  \hbox{( for any $N > 0$ )}
\cr\sp(3.0)} \eqn\EqHDetII $$
{}From {\EqHDis} and {\EqCommutePsiDisII},
we find
$$ \eqalign{ \sp(2.0)
& [ \, \H (g,\k) \, , \, \Psi^\dagger (x) \, ]  \, \cuum
\cr\sp(6.0)
& = \
  x {\der \over \der x} \bigl\{ \,
  \Psi^\dagger (x) \, - \, {\k \over x} ( \Psi^\dagger (x) - 1 )
  \, - \, x^2 ( \Psi^\dagger (x) )^2  \, \bigr\} \, \cuum ,
\cr\sp(3.0)} \eqn\EqOneDeformPsiI $$
$$ \eqalign{ \sp(2.0)
& [ \, [ \, \H (g,\k) \, , \, \Psi^\dagger (x_1) \, ]
                      \, , \, \Psi^\dagger (x_2) \, ] \, \cuum
\cr\sp(6.0)
& = \
  - \, 2 g \, x_1 x_2 \, {\der \over \der x_1} \, {\der \over \der x_2} \,
  \intdz{x_c} \, z^2 \Psi^\dagger (z)
         \, \delta( \inv{z} , x_1 ) \, \delta( \inv{z} , x_2 )
  \, \cuum ,
\cr\sp(3.0)} \eqn\EqOneDeformPsiII $$
and $\hbox{otherwise = 0}$.
\par
Before taking the continuum limit,
we redefine the wave function as
$$ \eqalign{ \sp(2.0)
\Phi^\dagger (x,\k)
\  \define \
\Psi^\dagger (x) \, - \, \lambda ( x , \k ) ,
\cr\sp(3.0)} \eqn\EqDefPhiDagger $$
where
$$ \eqalign{ \sp(2.0)
\lambda ( x , \k )  \ \define \
{1 \over 2 x^2} \bigl( 1 - {\k \over x} \bigr)  .
\cr\sp(3.0)} \eqn\EqDefLambda $$
Since $\Psi^\dagger(x)$ is analytic in the region $|x| \le x_c$,
$\Phi^\dagger(x,\k)$ is analytic in the region $0 < |x| \le x_c$.
Substituting {\EqDefPhiDagger} and {\EqDefLambda} into
eqs.\ {\EqOneDeformPsiI} and {\EqOneDeformPsiII},
we obtain
$$ \eqalign{ \sp(2.0)
& [ \, \H (g,\k) \, , \, \Phi^\dagger (x,\k) \, ]  \, \cuum
\cr\sp(6.0)
& = \  \Bigl\{ \,
  - \, \omega ( x , \k ) \, - \,
  x {\der \over \der x} \bigl\{ \, x^2 ( \Phi^\dagger (x,\k) )^2 \, \bigr\}
  \, \Bigr\} \, \cuum ,
\cr\sp(3.0)} \eqn\EqOneDeformPhiI $$
$$ \eqalign{ \sp(2.0)
& [ \, [ \, \H (g,\k) \, , \, \Phi^\dagger (x_1,\k) \, ]
                      \, , \, \Phi^\dagger (x_2,\k) \, ] \, \cuum
\cr\sp(6.0)
& = \
  - \, 2 g \, x_1 x_2 \, {\der \over \der x_1} \, {\der \over \der x_2} \,
  \intdz{x_c} \, z^2 \Phi^\dagger (z,\k)
         \, \delta( \inv{z} , x_1 ) \, \delta( \inv{z} , x_2 )
  \, \cuum ,
\cr\sp(3.0)} \eqn\EqOneDeformPhiII $$
and $\hbox{otherwise = 0}$,
where
$$ \eqalign{ \sp(2.0)
\omega ( x , \k ) \ \define \
- \, x {\der \over \der x} \bigl\{ \,
{1 \over 4 x^2} \bigl( 1 - {\k \over x} \bigr)^2 \, + \, {\k \over x}
\, \bigr\} .
\cr\sp(3.0)} \eqn\EqOmegaDis $$
Thus, the linear term of the wave function
in the right-hand side of {\EqOneDeformPsiI} vanishes
because of the field redefinition.
The commutation relations of the wave function
are still unchanged, i.e.,
$$ \eqalign{ \sp(2.0)
& [ \, \Psi (y) \, , \, \Phi^\dagger (x,\k) \, ]  \ = \  \delta ( y , x ) \, ,
\cr\sp(6.0)
& [ \, \Phi^\dagger (x_1,\k) \, , \, \Phi^\dagger (x_2,\k) \, ]  \ = \
  [ \, \Psi (y_1) \, , \, \Psi (y_2) \, ]  \ = \  0 \, .
\cr\sp(3.0)} \eqn\EqCommutePhiDis $$
Substituting {\EqDefPhiDagger} into {\EqGenTMDisII} and {\EqGenAmpDisII},
we find
$$ \eqalign{ \sp(2.0)
& \sum_{h=0}^\infty g^{h+N-1} \,
  T_{M,N}^{(h)} ( y_1, \ldots, y_M ; x_1, \ldots, x_N ; \k ; \t )
\cr\sp(6.0)
& = \
  \vac \,  \Psi (y_1) \, \cdots \, \Psi (y_M) \,
           e^{- \t \H (g,\k)} \,
           \Phi^\dagger (x_1,\k) \, \cdots \, \Phi^\dagger (x_N,\k) \,
           \cuum_{\rm connected}
\cr\sp(6.0)
&\phantom{ = \ } \ \
  +  \, \delta_{M,0} \, \delta_{N,1} \, \lambda (x_1,\k) \, ,
\cr\sp(3.0)} \eqn\EqGenTMDisIII $$
and
$$ \eqalign{ \sp(2.0)
& \sum_{h=0}^\infty g^{h+N-1} \,
  F_N^{(h)} ( x_1, \ldots, x_N ; \k )
\cr\sp(6.0)
& = \
  \lim_{\t\rightarrow\infty}
  \vac \, e^{- \t \H (g,\k)} \,
          \Phi^\dagger (x_1,\k) \, \cdots \, \Phi^\dagger (x_N,\k) \,
          \cuum_{\rm connected} \,
  +  \, \delta_{N,1} \, \lambda (x_1,\k) \, .
\cr\sp(3.0)} \eqn\EqGenAmpDisIII $$
Thus, the field redefinition {\EqDefPhiDagger} contributes only to
the amplitudes of disk topology, $T_{0,1}^{(0)}$ and $F_1^{(0)}$.
Especially, we find
$$ \eqalign{ \sp(2.0)
  F_1^{(0)} (x;\k)  \ = \  \hat F_1^{(0)} (x;\k) \, + \, \lambda (x,\k) ,
\cr\sp(3.0)} \eqn\EqAmpDiskDisIII $$
where
$$ \eqalign{ \sp(2.0)
  \hat F_1^{(0)} (x;\k)  \ \define \
  \lim_{\t\rightarrow\infty}
  \vac \, e^{- \t \H (g=0,\k)} \,
          \Phi^\dagger (x,\k) \, \cuum \, .
\cr\sp(3.0)} \eqn\EqAmpDiskDisHatF $$
As was discussed before,
$T_{M,N}^{(h)}$ and $F_N^{(h)}$ in {\EqGenTMDisIII} and {\EqGenAmpDisIII}
are assumed to be analytic in the region
$|x_i| \le x_c$, $|y_j| \le y_c$ and $|\k| < \k_c$ for any $i$ and $j$.
\par
Note that the inverse Laplace transformation of $\Phi^\dagger (x,\k)$ is
$$ \eqalign{ \sp(2.0)
  \Phi^\dagger (l,\k)
  \ = \
  \Psi^\dagger (l) \,
  - \, {1 \over 2}  \delta_{l,-2} \,
  + \, {\k \over 2} \delta_{l,-3} ,
\cr\sp(3.0)} \eqn\EqDefPhiDaggerL $$
where we have defined that $\Psi^\dagger (l<0) = 0$.
Thus, at this stage, the introduction of the new wave function
$\Phi^\dagger$ seems to be meaningless,
because there is no physical interpretation for operators,
$\Phi^\dagger (l=-2,\k) = -1/2$
and
$\Phi^\dagger (l=-3,\k) = \k/2$.
We will show in the next section that
this field redefinition
extracts the non-universal contribution
from the amplitudes at the continuous level.
In the right-hand sides of {\EqGenTMDisIII} and {\EqGenAmpDisIII},
only $\lambda$ depends on the cut-off parameter in the continuum limit.
\NextPage
\topskip 30pt
\chapter{\ Taking the Continuum Limit }
\vskip 10pt
\par
In this section
we take the continuum limit of the discretized $c=0$ string field theory.
The continuum limit is taken by
$\e \rightarrow 0$ and $l \rightarrow \infty$
while $L = \e l$ is fixed to be finite,
where $L$ is the length of a string at the continuous level.
{}From the viewpoint of the Laplace transformation,
the continuum limit is taken by
$\e \rightarrow 0$ with
$$ \eqalign{ \sp(2.0)
x  \ &= \ x_c \, e^{- \e \xi} ,
\qquad
y  \  = \ y_c \, e^{- \e \eta} ,
\cr\sp(4.0)
\k \ &= \ \k_c \, e^{- \e^2 c_0 \cc}  ,
\cr\sp(3.0)} \eqn\EqConLimitxyk $$
where the value of $c_0$ is positive real and will be chosen later
so as to make the forms of equations simple.
Under the continuum limit with {\EqConLimitxyk},
the factors in the integrand of the discrete Laplace transformation become
those of the continuous Laplace transformation,
i.e.,
$(x / x_c)^{l_i}  = e^{- L_i  \xi}$,
$(y / y_c)^{l'_j} = e^{- L'_j \eta}$, and
$(\k / \k_c)^a  = e^{- A \cc}$,
where
$L_i = \e l_i$, $L'_j = \e l'_j$, and $A = c_0 \e^2 a$ are considered to be
the length of the initial string, the length of the final string,
and the area of surface at the continuous level.
The $\cc$ is the cosmological constant at the continuous level.
Thus, we obtain the usual continuous Laplace transformation
from the discrete Laplace transformation,
which is explained in detail in appendix A.
The continuum limit of the wave functions is assumed to be
$$ \eqalign{ \sp(2.0)
&\Psi^\dagger (\xi)  \ = \  \lim_{\e \rightarrow 0}
 c_1 \, \e^\dim{\Psi^\dagger} \, \Psi^\dagger (x) ,
\qquad
 \Psi (\eta)        \ = \   \lim_{\e \rightarrow 0}
 c_2 \, \e^\dim{\Psi} \, \Psi (y)  ,
\cr\sp(4.0)
&\Phi^\dagger (\xi,\cc)  \ = \  \lim_{\e \rightarrow 0}
 c_3 \, \e^\dim{\Phi^\dagger} \, \Phi^\dagger (x,\k) ,
\cr\sp(3.0)} \eqn\EqConLimitWave $$
where $\dim{P}$ is the dimension of a function $P(\zeta)$
in the unit of $\dim{\e} = 1$.
For example, $\dim{L} = \dim{L'} = 1$ and $\dim{A} = 2$,
because $L = \e l$, $L' = \e l'$ and $A = c_0 \e^2 a$.
The coefficients $c_1$, $c_2$ and $c_3$ are non-zero real numbers and
will be chosen later so as to make the forms of equations simple.
$\Psi^\dagger(\xi)$, $\Psi(\eta)$ and $\Phi^\dagger(\xi',\cc)$
are considered to be analytic in the region
$0 \le \Real(\xi)$, $0 \le \Real(\eta)$ and $0 \le \Real(\xi') < \infty$,
because
$\Psi^\dagger(x)$, $\Psi(y)$ and $\Phi^\dagger(x',\k)$
are analytic in the region
$|x| \le x_c$, $|y| \le y_c$ and $0 < |x'| \le x_c$.
\par
By substituting {\EqConLimitxyk} and {\EqConLimitWave}
into {\EqDefPhiDagger} and {\EqDefLambda},
we find
$\dim{\Psi^\dagger} = \dim{\Phi^\dagger}$.
If we set $c_1 = c_3$, we obtain
$$ \eqalign{ \sp(2.0)
  \Phi^\dagger (\xi,\cc)
  \ = \
  \Psi^\dagger (\xi) \, - \, \lambda(\xi,\cc)
\cr\sp(3.0)} \eqn\EqDefPhiDaggerCon $$
with
$$ \eqalign{ \sp(2.0)
  \lambda (\xi,\cc)
  \ \define \
  \lim_{\e \rightarrow 0} c_1 \, \e^\dim{\Psi^\dagger} \, \lambda ( x , \k )
  \, .
\cr\sp(3.0)} \eqn\EqDefLambdaConI $$
When
$\dim{\Psi^\dagger} + \dim{\Psi} = 1$,
we obtain the continuum limit of the commutation relations
from {\EqCommutePsiDisII} and {\EqCommutePhiDis} as
$$ \eqalign{ \sp(2.0)
& [ \, \Psi (\eta) \, , \, \Psi^\dagger (\xi) \, ]  \ = \
  [ \, \Psi (\eta) \, , \, \Phi^\dagger (\xi,\cc) \, ]  \ = \
  \delta ( \eta , \xi ) \, ,
\cr\sp(6.0)
& [ \, \Psi^\dagger (\xi_1) \, , \, \Psi^\dagger (\xi_2) \, ]  \ = \
  [ \, \Phi^\dagger (\xi_1,\cc) \, , \, \Phi^\dagger (\xi_2,\cc) \, ]  \ = \
  [ \, \Psi (\eta_1) \, , \, \Psi (\eta_2) \, ]  \ = \  0 \, ,
\cr\sp(3.0)} \eqn\EqCommutePsiPhiCon $$
where we have set $c_1 c_2 = 1$.
$\delta(\eta,\xi) = 1/(\eta+\xi)$
is the continuous Laplace transformation of
$\delta(L-L')$.
\par
We also assume that the continuum limit of
the Hamiltonian and the coupling constant are
$$ \eqalign{ \sp(2.0)
  \H (G,\cc)  \ = \
  \lim_{\e \rightarrow 0} c_4 \, \e^\dim{\H} \, \H (g,\k) \, ,
\cr\sp(3.0)} \eqn\EqConLimitH $$
and
$$ \eqalign{ \sp(2.0)
  G  \ = \  \lim_{\e \rightarrow 0} c_5 \, \e^{\dim{G}} \, g  \, ,
\cr\sp(3.0)} \eqn\EqConLimitG $$
where $G$ is the coupling constant for string interaction
at the continuous level.
Here we have introduced the non-zero real numbers $c_4$ and $c_5$,
which are also determined later so as to make the forms of equations simple.
The continuum limit of the $\t$-step transfer matrix operator,
$e^{- \t \H (g,\k)}$, will be
$$ \eqalign{ \sp(2.0)
  e^{- \T \H (G,\cc)}  \ = \
  \lim_{\e \rightarrow 0} e^{- \t \H (g,\k)}  ,
\cr\sp(3.0)} \eqn\EqConLimitDeform $$
where $\T$ is considered to be a proper time on surface
at the continuous level.
{}From {\EqConLimitH} and {\EqConLimitDeform}, we find
$$ \eqalign{ \sp(2.0)
  \T  \ = \  \lim_{\e \rightarrow 0} {\t \over c_4} \, \e^{-\dim{\H}}  .
\cr\sp(3.0)} \eqn\EqConLimitIV $$
Then, we find that the dimension of $\T$ is $\dim{\T} = - \dim{\H}$.
In order to take the continuum limit,
we let $\e \rightarrow \infty$ and $\t \rightarrow \infty$
while $\T$ is fixed.
Therefore, we can take the continuum limit
if and only if the following condition is satisfied:
$$ \eqalign{ \sp(2.0)
\dim{\H}  \ = \ - \, \dim{\T}  \ < \  0  .
\cr\sp(3.0)} \eqn\EqConLimitCondition $$
\par
Now let us consider the continuum limit of the Hamiltonian $\H (g,\k)$
given in {\EqHDis}.
However, we need careful treatment for the integration on the complex plain
in the Hamiltonian {\EqHDis}.
As was introduced in the last part of the previous section,
we consider the continuum limit of
the time evolution of string states
{\EqOneDeformPsiI} and {\EqOneDeformPsiII}
instead of the Hamiltonian {\EqHDis}.
\par
Firstly, we consider the continuum limit of {\EqOneDeformPsiI},
from which one can derive the explicit form of the Hamiltonian $\H (G=0,\cc)$.
By substituting {\EqConLimitxyk}, {\EqConLimitWave} and {\EqConLimitH}
into {\EqOneDeformPsiI},
we obtain the time evolution of $\Psi^\dagger(\xi) \cuum$.
The naive calculation leads to
$\dim{\Psi^\dagger} = 0$ and $\dim{\H} = 1$,
which does not satisfy the condition {\EqConLimitCondition}.
So, it seems impossible to take the continuum limit.
We will see in the following that the continuum limit can be easily taken
by using the redefined wave function $\Phi^\dagger$.
The time evolution of $\Phi^\dagger(\xi,\cc) \cuum$ is obtained
by substituting {\EqConLimitxyk}, {\EqConLimitWave} and {\EqConLimitH}
into {\EqOneDeformPhiI},
$$ \eqalign{ \sp(2.0)
& [ \, \H (G,\cc) \, , \, \Phi^\dagger (\xi,\cc) \, ] \, \cuum
\cr\sp(6.0)
& = \
  \Bigl\{ \,
  - \, c_3 c_4 \, \e^{\dim{\H}+\dim{\Phi^\dagger}} \,
       \omega ( x , \k )
\cr\sp(6.0)
& \phantom{= \  \Bigl\{ \, } \,
  + \, {c_4 \over c_3} \, \e^{\dim{\H}-\dim{\Phi^\dagger}-1} \, x_c^2 \,
       {\der \over \der \xi} \,
       \bigl\{ \, e^{-2 \e \xi} ( \Phi^\dagger (\xi,\cc) )^2 \, \bigr\} \,
  \Bigr\} \, \cuum \, ,
\cr\sp(3.0)} \eqn\EqOneDeformConI $$
where
$$ \eqalign{ \sp(2.0)
\omega ( x , \k ) \ = \
{1 \over 2 x_c^2} \, \bigl\{
& 1 - {3 \k_c \over x_c} + {2 \k_c^2 \over x_c^2} + 2 \k_c x_c
\cr\sp(4.0)
\, &+ \,
( \, 2 - {9 \k_c \over x_c} + {8 \k_c^2 \over x_c^2} + 2 \k_c x_c \, )
\, \e \xi
\cr\sp(4.0)
\, &+ \,
( \, 2 - {27 \k_c \over 2 x_c} + {16 \k_c^2 \over x_c^2} + \k_c x_c \, )
\, \e^2 \xi^2
\cr\sp(4.0)
\, &+ \,
( \, {3 \k_c \over x_c} - {4 \k_c^2 \over x_c^2} - 2 \k_c x_c \, )
\, \e^2 c_0 \cc
\, + \, O(\e^3) \, \bigr\} .
\cr\sp(3.0)} \eqn\EqOmegaConI $$
{}From the naive calculation we find that
$\dim{\Phi^\dagger} = -1/2$ and $\dim{\H} = 1/2$,
which does not satisfy the condition {\EqConLimitCondition} again.
However, if the leading terms of $\omega (x,\k)$ is
higher than or equal to $\e^2$ order,
the condition {\EqConLimitCondition} is satisfied.
Therefore,
the coefficients of $\e^0$ and $\e^1$
in the right-hand side of eq.\ {\EqOmegaConI}
should be zero, i.e.,
$$ \eqalign{ \sp(2.0)
& 1 - {3 \k_c \over x_c} + {2 \k_c^2 \over x_c^2} + 2 \k_c x_c \ = \ 0 ,
\cr\sp(6.0)
& 2 - {9 \k_c \over x_c} + {8 \k_c^2 \over x_c^2} + 2 \k_c x_c \ = \ 0 .
\cr\sp(3.0)} \eqn\EqCriticalCond $$
Since $x_c$ and $\k_c$ are positive real,
we obtain the unique solution of eq.\ {\EqCriticalCond} as
$$ \eqalign{ \sp(2.0)
  x_c   \ = \  {3^{1/4} \, - \, 3^{-1/4} \over 2} ,
\qquad
  \k_c  \ = \  {3^{1/4} \over 6} .
\cr\sp(3.0)} \eqn\EqCriticalPoint $$
In the case of {\EqCriticalPoint},
the leading term of $\omega (x,\k)$ is proportional to $\e^2$.
Thus we find from {\EqOneDeformConI} that
$\dim{\Phi^\dagger} = -3/2$ and $\dim{\H} = -1/2$,
which satisfies the condition {\EqConLimitCondition}.
Then, we obtain
$$ \eqalign{ \sp(2.0)
  [ \, \H (G,\cc) \, , \, \Phi^\dagger (\xi,\cc) \, ] \, \cuum
  \ = \  \Bigl\{ \,
         - \, \omega (\xi,\cc)
         \, + \, {\der \over \der \xi} \, ( \Phi^\dagger (\xi,\cc) )^2
         \, \Bigr\} \, \cuum \, ,
\cr\sp(3.0)} \eqn\EqOneDeformConIII $$
where
$$ \eqalign{ \sp(2.0)
  \omega (\xi,\cc) \
& \define \  c_3 c_4 \, \lim_{\e \rightarrow 0}
             {1 \over \e^2} \, \omega ( x , \k )
\cr\sp(6.0)
& =       \  3 \xi^2 \, - \, {3 \over 4} \cc \, .
\cr\sp(3.0)} \eqn\EqOmegaConII $$
Here we have set
$c_0  = ( 3+\sqrt{3} )^2 / 16 $,
$c_3  = 2 / (1+\sqrt{3})^{5/2}$,
and
$c_4  = 2 \sqrt{3} / (1+\sqrt{3})^{1/2}$
in order to make the forms of
{\EqOneDeformConIII} and {\EqOmegaConII} simple.
\par
As a result, one can take the continuum limit
if and only if $x_c$ and $\k_c$ take the values {\EqCriticalPoint},
and
$$ \eqalign{ \sp(2.0)
& \dim{\Phi^\dagger}  \ = \  \dim{\Psi^\dagger}  \ = \  - \, {3 \over 2} ,
  \qquad
  \dim{\Psi}  \ = \  {5 \over 2} ,
\cr\sp(6.0)
& \dim{\T}  \ = \  - \, \dim{\H}  \ = \  {1 \over 2} .
\cr\sp(3.0)} \eqn\EqDimension $$
We have chosen the values of the coefficients,
$c_0$, $c_1$, $c_2$, $c_3$, and $c_4$, as
$$ \eqalign{ \sp(2.0)
 c_0  \, = \, \bigl( { 3+\sqrt{3} \over 4 } \bigr)^2 , \quad
 c_1  \, = \,  c_3  \, = \,  {1 \over c_2}
 \, = \,  {2 \over (1+\sqrt{3})^{5/2}} ,   \quad
 c_4  \, = \,  {2 \sqrt{3} \over (1+\sqrt{3})^{1/2}} ,
\cr\sp(3.0)} \eqn\EqCoefficientC $$
in order to make the forms of
{\EqDefPhiDaggerCon}, {\EqCommutePsiPhiCon} and {\EqOneDeformConIII}
simple.
Since $\dim{\Phi^\dagger} = \dim{\Psi^\dagger} = - 3 / 2$,
the $\lambda(\xi,\cc)$ defined in {\EqDefLambdaConI} becomes
$$ \eqalign{ \sp(2.0)
  \lambda (\xi)
  \ = \
  {1 \over \sqrt{3} (1+\sqrt{3})^{3/2}} \,
  \bigl( \, \e^{-3/2} - \sqrt{3} \e^{-1/2} \xi + O(\e^{1/2}) \, \bigr)  \, .
\cr\sp(3.0)} \eqn\EqDefLambdaConII $$
Not only $\lambda$ but also $\Phi^\dagger$ are independent of
the cosmological constant $\cc$
because the dependence of $\cc$ is neglected
in the continuum limit $\e \rightarrow 0$.
Thus, from now on, we use the notations,
$\lambda(\xi) = \lambda(\xi,\cc)$ and
$\Phi^\dagger(\xi) = \Phi^\dagger(\xi,\cc)$ for simplicity.
Therefore, we obtain the continuous Hamiltonian
which leads to {\EqOneDeformConIII} as
$$ \eqalign{ \sp(2.0)
  \H (G=0,\cc)  \ = \
  \intdzeta \, \bigl\{ \,
  - \, \omega (\zeta,\cc) \, \Psi (- \zeta)
  \, - \, ( \Phi^\dagger (\zeta) )^2
       \, {\der \over \der \zeta} \, \Psi (- \zeta)
  \, \bigr\} \, .
\cr\sp(3.0)} \eqn\EqHDiskCon $$
Then, we can calculate the disk amplitude by
$$ \eqalign{ \sp(2.0)
  F_1^{(0)} (\xi;\cc)
  \  &=  \
  \lim_{\T \rightarrow \infty}
  \vac e^{- \T \H (G=0,\cc)} \, \Psi^\dagger (\xi) \cuum
\cr\sp(6.0)
  \  &=  \
  \hat F_1^{(0)} (\xi;\cc)  \, + \, \lambda (\xi) ,
\cr\sp(3.0)} \eqn\EqAmpDiskConI $$
where $\hat F_1^{(0)}(\xi;\cc)$
is the universal part of disk amplitude defined by
$$ \eqalign{ \sp(2.0)
  \hat F_1^{(0)} (\xi;\cc) \  \define  \
  \lim_{\T \rightarrow \infty}
  \vac e^{- \T \H (G=0,\cc)} \, \Phi^\dagger (\xi) \cuum \, ,
\cr\sp(3.0)} \eqn\EqAmpDiskConHatF $$
while $\lambda(\xi)$ is the non-universal part of disk amplitude
because of the cut-off dependence.
In appendix D we have calculated the explicit form of the disk amplitude
by using the Schwinger-Dyson equation
which is derived from {\EqAmpDiskConHatF}.
See also ref.\ [{\IK}].
$\omega(\zeta,\cc)$ is related to $\hat F_1^{(0)}(\zeta;\cc)$ as
$\omega(\zeta,\cc) = {\der \over \der \zeta} ( \hat F_1^{(0)}(\zeta;\cc) )^2$.
\par
Secondly, we consider to take the continuum limit of {\EqOneDeformPsiII}.
By using {\EqConLimitxyk}, {\EqConLimitWave},
{\EqConLimitH} and {\EqConLimitG},
we obtain
$$ \eqalign{ \sp(2.0)
& [ \, [ \, \H (G,\cc) \, , \, \Phi^\dagger (\xi_1) \, ]
                     \, , \, \Phi^\dagger (\xi_2) \, ] \, \cuum
\cr\sp(8.0)
& = \
    - \, {2 x_c^2 c_3 c_4 \over c_5} G
    \, \e^{\dim{\H} + \dim{\Phi^\dagger} - \dim{G} -3}
\cr\sp(6.0)
& \phantom{= \ \ }
    \times \, {\der \over \der \xi_1} \, {\der \over \der \xi_2}
    \, \intdzeta \, \Phi^\dagger (\zeta)
                 \, \delta( -\zeta , \xi_1 ) \, \delta( -\zeta , \xi_2 )
    \, \cuum \, .
\cr\sp(3.0)} \eqn\EqOneDeformConIV $$
{}From {\EqDimension}, we find $\dim{G} = -5$,
which is consistent with the result by the matrix model.\NPrefmark{\Brezin}
In the matrix model the continuum limit is taken by the double scaling limit,
which remains
$(\k_c - \k)^{5/2} / g$
finite as
$g \rightarrow 0$ and $\k \rightarrow \k_c$
in the case of pure gravity.
If $\dim{G} = -5$,
the double scaling limit is derived by canceling $\e$
in the third eq.\ of {\EqConLimitxyk} and the eq.\ {\EqConLimitG}.
If we choose the values of $c_i$ ($i=0,\ldots,4$) as {\EqCoefficientC}
and $c_5 = x_c^2 c_3 c_4$,
we make the form of {\EqOneDeformConIV} simpler as
$$ \eqalign{ \sp(2.0)
& [ \, [ \, \H (G,\cc) \, , \, \Phi^\dagger (\xi_1) \, ]
                       \, , \, \Phi^\dagger (\xi_2) \, ] \, \cuum
\cr\sp(6.0)
& = \
  - \, 2 G \, {\der \over \der \xi_1} \, {\der \over \der \xi_2}
    \, \intdzeta \, \Phi^\dagger (\zeta)
                 \, \delta( -\zeta , \xi_1 ) \, \delta( -\zeta , \xi_2 )
    \, \cuum .
\cr\sp(3.0)} \eqn\EqOneDeformConV $$
{}From {\EqOneDeformConIII}, {\EqOneDeformConV} and $\hbox{otherwise = 0}$,
we obtain the Hamiltonian,
$$ \eqalign{ \sp(2.0)
  \H (G,\cc)  \ = \
  \intdzeta \, \bigl\{ \,
  - \,      \omega (\zeta,\cc) \, \Psi (- \zeta) \,
& - \,      ( \Phi^\dagger (\zeta) )^2
         \, {\der \over \der \zeta} \, \Psi (- \zeta)
\cr\sp(6.0)
& - \, G \, \Phi^\dagger (\zeta)
         \, ( {\der \over \der \zeta} \, \Psi (- \zeta) )^2 \, \bigr\} \, .
\cr\sp(3.0)} \eqn\EqHGenCon $$
One can also obtain the Hamiltonian {\EqHGenCon} directly
by taking the continuum limit of the Hamiltonian {\EqHDis},
though one needs one's careful treatment of the integral contour
on the complex plain.
The Hamiltonian {\EqHGenCon} is consistent with that of ref.\ [{\IK}].
Precisely speaking,
in the calculation of the amplitudes
the analytic continuation was necessary in ref.\ [{\IK}],
while one does not need the analytic continuation
if one uses the Hamiltonian {\EqHGenCon}.
\par
The continuum limit of the transfer matrix $T_{M,N}^{(h)}$ in {\EqGenTMDisIII}
is
$$ \eqalign{ \sp(2.0)
& \sum_{h=0}^\infty G^{h+N-1} \,
  T_{M,N}^{(h)} ( \eta_1, \ldots, \eta_M ; \xi_1, \ldots, \xi_N ; \cc ; \T )
\cr\sp(6.0)
& = \
  \vac \,  \Psi (\eta_1) \, \cdots \, \Psi (\eta_M) \,
           e^{- \T \H (G,\cc)} \,
           \Phi^\dagger (\xi_1) \, \cdots \, \Phi^\dagger (\xi_N) \,
           \cuum_{\rm connected}
\cr\sp(6.0)
&\phantom{ = \ } \
  +  \, \delta_{M,0} \, \delta_{N,1} \, \lambda (\xi_1) \, .
\cr\sp(3.0)} \eqn\EqGenTMCon $$
We also obtain the continuum limit of the amplitude $F_N^{(h)}$
in {\EqGenAmpDisIII} by
$$ \eqalign{ \sp(2.0)
& \sum_{h=0}^\infty G^{h+N-1} \,
  F_N^{(h)} ( \xi_1, \ldots, \xi_N ; \cc )
\cr\sp(6.0)
& = \
  \lim_{\T\rightarrow\infty}
  \vac \, e^{- \T \H (G,\cc)} \,
          \Phi^\dagger (\xi_1) \, \cdots \, \Phi^\dagger (\xi_N) \,
          \cuum_{\rm connected} \,
  + \, \delta_{N,1} \, \lambda (\xi_1) \, .
\cr\sp(3.0)} \eqn\EqGenAmpCon $$
Since $\lambda (\xi_1)$ depends on the cut-off $\e$,
$\lambda (\xi_1)$ is non-universal.
Thus,
only the amplitudes of disk topology,
$T_{0,1}^{(0)} (\void;\xi_1;\cc;\T)$ and $F_1^{(0)} (\xi_1;\cc)$,
have the non-universal part $\lambda (\xi_1)$.
Note that the transfer matrices as well as the amplitudes
are analytic in the region
$0 \le \Real(\xi_i)$, $0 \le \Real(\eta_j)$ and $0 < \Real(\cc)$,
because of the assumptions $a$), $b$), $c$), $d$) and $e$)
in section 2 and 3.
This analyticity is consistent with the statement that
$\lim_{L_i \rightarrow \infty} F_N^{(h)}(L_1,\ldots,L_N;\cc) = 0$
(for any $i$) in refs.\ [{\MSS},{\IK}].
\NextPage
\topskip 30pt
\chapter{\ Universality }
\vskip 10pt
\par
In this section we study
the universality of the $c=0$ non-critical string field theory.
We will show that
some modified discretized string field theories
always lead to the same Hamiltonian $\H (G,\cc)$ in {\EqHGenCon}
after taking the continuum limit.
\par
To begin with, let us regard
Figs.\ {{\FigPeelSkinI}, {\FigPeelSkinII} and {\FigPeelSkinIII}}
as $(\alpha_a/l)$, $(\alpha_b/l)$, and $(\alpha_c/l)$-step decompositions
respectively
instead of $(1/l)$-step ones,
where $\alpha_a$, $\alpha_b$, and $\alpha_c$ are
finite positive real number.
However, one can eliminate these parameters,
$\alpha_a$, $\alpha_b$, and $\alpha_c$,
by rescaling $\k$, $\Psi$, $\Psi^\dagger$, and $g$.
Therefore, we have obtained the same Hamiltonian {\EqH} and {\EqHDis}
at the discrete level.
\par
Next, let us consider to introduce
some new minimal-step \lq peeling decompositions'
illustrated in Fig.\ {\FigPeelSkinIV}
besides three fundamental decompositions
in Figs.\ {{\FigPeelSkinI}, {\FigPeelSkinII} and {\FigPeelSkinIII}}.
The general Hamiltonian for these decompositions is
$$ \eqalign{ \sp(2.0)
  \H (g,\k)  \ = \  \sum_{n=1}^\infty \H^{(n)} (g,\k)
\cr\sp(3.0)} \eqn\EqHGenGenDisI $$
with
$$ \eqalign{ \sp(2.0)
  \H^{(n)} (g,\k)  \
  = \  \alpha^{(n)} \H_\alpha^{(n)} (g,\k) \,
  + \, \beta^{(n)}  \H_\beta^{(n)}  (g,\k) \,
  + \, \gamma^{(n)} \H_\gamma^{(n)} (g,\k) \, ,
\cr\sp(3.0)} \eqn\EqHGenGenDisII $$
and
$$ \eqalign{ \sp(2.0)
& \H_\alpha^{(n)} (g,\k)  \ = \  - \, \intdz{x_c} \,
  \Psi^\dagger (z) \, ( \, g \, z^2 \, )^{n-1} \,
       ( - z {\der \over \der z} \, \Psi (\inv{z}) )^n \, ,
\cr\sp(6.0)
& \H_\beta^{(n)} (g,\k)  \ = \  - \, \intdz{x_c} \,
  {\k \over z} \, \Psi^\dagger (z) \, ( \, g \, z^2 \, )^{n-1} \,
  ( - z {\der \over \der z} \, \Psi (\inv{z}) )^n \, ,
\cr\sp(6.0)
& \H_\gamma^{(n)} (g,\k)  \ = \  - \, \intdz{x_c} \,
  z^2 \, ( \Psi^\dagger (z) )^2 \, ( \, g \, z^2 \, )^{n-1} \,
  ( - z {\der \over \der z} \, \Psi (\inv{z}) )^n \, ,
\cr\sp(3.0)} \eqn\EqHGenGenDisIII $$
where $\alpha^{(n)}$, $\beta^{(n)}$ and $\gamma^{(n)}$
are non-negative real numbers except for $\alpha^{(1)} = -1$.
The Hamiltonian in {\EqHDis} is expressed by using {\EqHGenGenDisIII} as
$\H = - \H_\alpha^{(1)} + \H_\beta^{(1)} + \H_\gamma^{(1)} + \H_\alpha^{(2)}$.
Thus, the Hamiltonian in {\EqHGenGenDisI} is the generalization of
that in {\EqHDis}.
\par
Now, let us consider to take the continuum limit of
the Hamiltonian {\EqHGenGenDisI}.
Firstly, we consider $\H^{(n=1)}$, i.e.,
$$ \eqalign{ \sp(2.0)
  \H^{(1)} (g,\k)  \ = \
  \intdz{x_c} \, \bigl\{ \,
     ( 1 - {\beta^{(1)} \k \over z} ) \Psi^\dagger (z) \,
       - \, \gamma^{(1)} z^2 ( \Psi^\dagger (z) )^2 \,
  \bigr\} \, ( - z {\der \over \der z} \Psi (\inv{z}) ) \, .
\cr\sp(3.0)} \eqn\EqHGenGenOneDis $$
After rescaling $\Psi$, $\Psi^\dagger$, $\k$ and $\t$,
we find that $\H^{(1)} (g,\k)$ becomes equal to $\H (g=0,\k)$ in {\EqHDis}.
As was discussed in section 4,
we can take the continuum limit of $\H(g=0,\k)$
if we choose the critical values as {\EqCriticalPoint} and
the canonical dimensions as {\EqDimension}.
Then, we obtain {\EqHDiskCon}.
\par
Secondly, we consider $\H^{(n=2)}$, i.e.,
$$ \eqalign{ \sp(2.0)
  \H^{(2)} (g,\k)  \
& = \  - \, g \, \intdz{x_c} \,
       \bigl\{ \,
            ( {\alpha^{(2)} \over 2} + {\beta^{(2)} \k \over 2 z} +
              {\gamma^{(2)} \over 4} (1-{\k \over z})
            ) \, ( 1 - {\k \over z} ) \,
\cr\sp(6.0)
& \phantom{= \  - \, g \, \intdz{x_c} \, \bigl\{ \, } \
       + \, ( \alpha^{(2)} z^2 + \beta^{(2)} \k z +
              \gamma^{(2)} z^2 (1-{\k \over z})
             ) \, \Phi^\dagger (z,\k) \,
\cr\sp(6.0)
& \phantom{= \  - \, g \, \intdz{x_c} \, \bigl\{ \, } \
       + \, \gamma^{(2)} z^4 \, ( \Phi^\dagger (z,\k) )^2 \,
       \bigr\} \, ( - z {\der \over \der z} \, \Psi (\inv{z}) )^2 \, ,
\cr\sp(3.0)} \eqn\EqHGenGenTwoDis $$
where we have used the field redefinition
{\EqDefPhiDagger} with {\EqDefLambda}.
{}From the naive dimensional analysis one finds that
$(\Phi^\dagger)^0$, $(\Phi^\dagger)^1$ and $(\Phi^\dagger)^2$ order terms
in the right-hand side of {\EqHGenGenTwoDis}
are proportional to
$\e^{-6-\dim{G}}$, $\e^{-9/2-\dim{G}}$ and $\e^{-3-\dim{G}}$ respectively,
in the $\e \rightarrow 0$ limit,
where we have used {\EqDimension}.
However, $(\Phi^\dagger)^0$ order term is found to be proportional to
$\e^{-4-\dim{G}}$ after the detail calculation,
which is performed by calculating, for example,
$$ \eqalign{ \sp(2.0)
  \vac \, g \intdz{x_c} \, \omega'(z,\k) \,
       (- z {\der \over \der z} \Psi (\inv{z}) )^2 \,
       \Phi^\dagger (x_1,\k) \Phi^\dagger (x_2,\k) \, \cuum \, ,
\cr\sp(3.0)} \eqn\EqHII $$
where $\omega'(z,\k)$ is supposed to be
an arbitrary function of $z$ and $\k$.
Thus,
in the $\e \rightarrow 0$ limit,
the leading term in the right-hand side of {\EqHGenGenTwoDis}
is $(\Phi^\dagger)^1$ order terms not $(\Phi^\dagger)^0$ order terms.
Thus, we find that $\dim{G} = -5$ in order to let
$\dim{\H^{(2)}} = -1/2$, i.e.,
$\H^{(2)} (g,\k) \propto \e^{1/2}$ in the $\e \rightarrow 0$ limit.
Therefore, we find that
$\H^{(2)} (g,\k)$ leads to
the $G^1$ order term of $\H (G,\cc)$ in {\EqHGenCon} in the continuum limit.
\par
Thirdly, we consider $\H^{(n=3)}$.
{}From {\EqDimension} and $\dim{G} = -5$,
one finds that the naive dimensional analysis leads to
$\H^{(3)} (g,\k) \propto \e^{1/2}$ in the $\e \rightarrow 0$ limit.
However,
by calculating
$$ \eqalign{ \sp(2.0)
  \vac \, g^2 \intdz{x_c} \, \omega''(z,\k) \,
       (- z {\der \over \der z} \Psi (\inv{z}) )^3 \,
       \Phi^\dagger(x_1,\k) \Phi^\dagger(x_2,\k) \Phi^\dagger(x_3,\k) \,
       \cuum \, ,
\cr\sp(3.0)} \eqn\EqHIII $$
we find that $\H^{(n=3)}$ vanishes in the continuum limit.
Since the leading term of other Hamiltonians $\H^{(n\ge4)}$
is less than $\e^{3n/2 - 4}$ order in the $\e \rightarrow 0$ limit,
$\H^{(n\ge4)}$ also vanishes in the continuum limit.
As a result,
the Hamiltonian $\H (g,\k)$ in {\EqHGenGenDisI} leads to
the Hamiltonian $\H (G,\cc)$ in {\EqHGenCon}
after taking the continuum limit.
\par
Next, let us analyze the dynamical triangulation
without introducing the two-folded parts
which have simplified the formulation.
Namely, we consider Fig.\ {\FigPeelTriI} instead of Fig.\ {\FigPeelTriII}
when we remove a triangle.
Precisely speaking,
instead of the three decompositions
in Figs.\ {\FigPeelSkinI} $\sim$ {\FigPeelSkinIII},
we consider the seven decompositions
in Figs.\ {\FigPeelTriA} $\sim$ {\FigPeelTriG}
as $(1/l)$-step fundamental \lq peeling decompositions'.
Figs.\ {\FigPeelTriD} and {\FigPeelTriE} are
tadpole diagrams as the string field theory.
The \lq peeling decompositions'
in Figs.\ {\FigPeelTriA} $\sim$ {\FigPeelTriG}
deform the wave function $\Psi^\dagger (l)$ as
$$ \eqalign{ \sp(2.0)
& \Psi^\dagger (l)  \ \longrightarrow \
  \k \Psi^\dagger (l+1) \,  ,
  \qquad \qquad \hskip 62pt \hbox{ ( for Fig.\ {\FigPeelTriA} ) }
\cr\sp(6.0)
& \Psi^\dagger (l)  \ \longrightarrow \
  \k \Psi^\dagger (l-1) \,  ,
  \qquad \qquad \hskip 62pt
  \hbox{ ( for Figs.\ {\FigPeelTriB} and {\FigPeelTriC} ) }
\cr\sp(6.0)
& \Psi^\dagger (3)  \ \longrightarrow \
  \k \,  ,
  \phantom{\Psi^\dagger (l-1)} \qquad \qquad \hskip 60pt
  \hbox{ ( for Fig.\ {\FigPeelTriD} ) }
\cr\sp(6.0)
& \Psi^\dagger (1)  \ \longrightarrow \
  \k \,  ,
  \phantom{\Psi^\dagger (l-1)} \qquad \qquad \hskip 60pt
  \hbox{ ( for Fig.\ {\FigPeelTriE} ) }
\cr\sp(6.0)
& \Psi^\dagger (l)  \ \longrightarrow \
  \k \sum_{l'=1}^l \Psi^\dagger (l') \Psi^\dagger (l-l'+1) \, ,
  \qquad \qquad \hskip -10pt \hbox{ ( for Fig.\ {\FigPeelTriF} ) }
\cr\sp(6.0)
& \Psi^\dagger (l)  \ \longrightarrow \
  \k \sum_{l'=1}^\infty l' \Psi^\dagger (l+l'+1) \Psi (l') \, .
  \qquad \qquad \hskip -10pt \hbox{ ( for Fig.\ {\FigPeelTriG} ) }
\cr\sp(3.0)} \eqn\EqOneOverLDeformNoSkinI $$
The factor $l'$ is necessary
in the right-hand side of the last deformation in {\EqOneOverLDeformNoSkinI}
because there are $l'$ types of figures for Fig.\ {\FigPeelTriG}
owing to the location of the marked link on $\Psi(l')$.
By summing up all kinds of fundamental decompositions
in {\EqOneOverLDeformNoSkinI},
we obtain the $(1/l)$-step deformed wave function,
$\Psi^\dagger(l) + \delta_{1/l} \Psi^\dagger(l)$.
Then, we find
$$ \eqalign{ \sp(2.0)
\delta_{1/l} \, \Psi^\dagger (l) \ = \
&  - \, \Psi^\dagger (l)
\, + \, \k \Psi^\dagger (l+1)
\, + \, 2 \k \, ( 1 - \delta_{l,1} ) \, \Psi^\dagger (l-1)
\, + \, \k \, \delta_{l,3} \, + \, \k \, \delta_{l,1}
\cr\sp(4.0)
&  + \, \k \, \sum_{l'=1}^l \Psi^\dagger (l') \Psi^\dagger (l-l'+1)
\, + \, 2 g \k \, \sum_{l'=1}^\infty \Psi^\dagger (l+l'+1) l' \Psi (l') \, .
\cr\sp(3.0)} \eqn\EqOneOverLDeformNoSkinIII $$
{}From {\EqOneOverLHamilton} and {\EqOneOverLDeformNoSkinIII}
we obtain the Hamiltonian,
$$ \eqalign{ \sp(2.0)
\H (g,\k)  \ = \
&\sum_{l=1}^\infty \bigl\{ \,
   \Psi^\dagger (l)
   \, - \, \k \Psi^\dagger (l+1) \,
 \bigr\} \, l \, \Psi (l)
\cr\sp(6.0)
&- \, 2 \, \sum_{l=2}^\infty \, \k \, \Psi^\dagger (l-1) \, l \, \Psi (l) \,
 - \, 3 \k \, \Psi (3) \, - \, \k \, \Psi (1)
\cr\sp(6.0)
&- \, \sum_{l=1}^\infty \, \sum_{l'=1}^l \,
 \k \, \Psi^\dagger (l') \Psi^\dagger (l-l'+1) \, l \, \Psi (l)
\cr\sp(6.0)
&- \, g \, \sum_{l=1}^\infty \sum_{l'=1}^\infty
 \k \, \Psi^\dagger (l+l'+1) \, l \, \Psi (l) \, l' \, \Psi (l') \, .
\cr\sp(3.0)} \eqn\EqHNoSkin $$
Note that we have not introduced neither $\Psi^\dagger(l=0)$ nor $\Psi(l=0)$
because it does not make the form of the Hamiltonian {\EqHNoSkin} simple.
The discrete Laplace transformation of the Hamiltonian {\EqHNoSkin} is
$$ \eqalign{ \sp(2.0)
  \H (g,\k)  \ = \  \intdz{x_c} \,
  \Bigl\{ \,
&    \bigl\{ \,
       - \, \k z^3 \, - \, \k z \,
       + \, ( 1 - {\k \over z} - 2 \k z ) \, \Psi^\dagger (z) \,
       - \, {\k \over z} \, ( \Psi^\dagger (z) )^2 \, \bigr\}
\cr\sp(4.0)
& \phantom{ \bigl\{ \, }
     \times \, ( - z {\der \over \der z} \, \Psi (\inv{z}) )
\cr\sp(6.0)
&    - \, {g \k \over z} \, \Psi^\dagger (z) \,
     ( - z {\der \over \der z} \, \Psi (\inv{z}) )^2 \,
  \Bigr\} \, .
\cr\sp(3.0)} \eqn\EqHNoSkinDis $$
As the same as before,
in order to remove the terms which are proportional to
$\Psi^\dagger (z) ( - z {\der \over \der z} \Psi (\inv{z}) )$
from the Hamiltonian {\EqHNoSkinDis},
we redefine the wave function as {\EqDefPhiDagger} with
$$ \eqalign{ \sp(2.0)
  \lambda ( x , \k )  \ \define \
  {x \over 2 \k} \bigl( 1 - {\k \over x} - 2 \k x \bigr) \, .
\cr\sp(3.0)} \eqn\EqDefLambdaNoSkin $$
The commutation relations of the wave function
are {\EqCommutePhiDis}.
Then, we obtain
$$ \eqalign{ \sp(2.0)
& [ \, \H (g,\k) \, , \, \Phi^\dagger (x,\k) \, ]  \, \cuum
\cr\sp(6.0)
& = \  \Bigl\{ \,
  - \, \omega ( x , \k ) \, - \,
  x {\der \over \der x}
  \bigl\{ \, {\k \over x} \, ( \Phi^\dagger (x,\k) )^2 \, \bigr\}
  \, \Bigr\} \, \cuum \, ,
\cr\sp(3.0)} \eqn\EqOneDeformPhiNoSkinI $$
$$ \eqalign{ \sp(2.0)
& [ \, [ \, \H (g,\k) \, , \, \Phi^\dagger (x_1,\k) \, ]
                      \, , \, \Phi^\dagger (x_2,\k) \, ] \, \cuum
\cr\sp(6.0)
& = \
  - \, 2 g \, x_1 x_2 \, {\der \over \der x_1} \, {\der \over \der x_2} \,
  \intdz{x_c} \, {\k \over z} \, \Phi^\dagger (z,\k)
         \, \delta( \inv{z} , x_1 ) \, \delta( \inv{z} , x_2 )
  \, \cuum \, ,
\cr\sp(3.0)} \eqn\EqOneDeformPhiNoSkinII $$
and $\hbox{otherwise = 0}$,
where
$$ \eqalign{ \sp(2.0)
\omega ( x , \k ) \ \define \
- \, x {\der \over \der x}
\bigl\{ \,
  {x \over 4 k} \bigl( 1 - {\k \over x} - 2 \k x \bigr)^2 \,
  - \, \k x^3 \, - \, \k x \,
\bigr\} \, .
\cr\sp(3.0)} \eqn\EqOmegaNoSkinDis $$
\par
The continuum limit is taken by {\EqConLimitxyk} $\sim$ {\EqDefLambdaConI}
and {\EqConLimitH} $\sim$ {\EqConLimitIV},
where the canonical dimensions and the values of $x_c$ and $\k_c$
are determined so as to satisfy the condition {\EqConLimitCondition}.
As the same as before, the condition {\EqConLimitCondition} is satisfied
only when the leading term of $\omega(x,\k)$ is
higher than or equal to $\e^2$ order in the $\e \rightarrow 0$ limit.
{}From {\EqOmegaNoSkinDis} we uniquely determine the critical values as
$x_c = 2^{-1} 3^{-1/4}$ and $\k_c = 2^{-1} 3^{-3/4}$.
Then, we obtain {\EqOmegaConII}, {\EqDimension} and $\dim{G} = -5$ again.
As a result, we obtain the Hamiltonian {\EqHGenCon}
from the continuum limit of {\EqOneDeformPhiNoSkinI},
{\EqOneDeformPhiNoSkinII} and $\hbox{otherwise = 0}$, i.e.,
both formulations with and without the two-folded parts
lead to the same non-critical string field theory at the continuous level.
\par
In the rest of this section,
we consider to replace regular triangles on lattice surfaces
with regular $n$-polygons.
We will show in the following that
the same Hamiltonian $\H (G,\cc)$ in {\EqHGenCon} is obtained
in the continuum limit
though the form of the Hamiltonian is different from {\EqHDis}
at the discrete level.
One of the \lq peeling decompositions' in Fig.\ {\FigPeelSkinI}
is modified by replacing a regular triangle with a regular $n$-polygon.
This modified decomposition changes the length of string
from $l$ to $l+n-2$, i.e.,
$$ \eqalign{ \sp(2.0)
  \Psi^\dagger (l)  \ \longrightarrow \
  \k_n \Psi^\dagger (l+n-2) \, .
  \qquad \hbox{ ( for Fig.\ {\FigPeelSkinI} ) }
\cr\sp(3.0)} \eqn\EqOneOverLDeformNI $$
Other \lq peeling decompositions'
in Figs.\ {{\FigPeelSkinII} and {\FigPeelSkinIII}}
are still unchanged because they remove a two-folded part.
Note that the case $n=3$ is included as a special case,
i.e., $\k_3 = \k$.
Then, we obtain the $(1/l)$-step deformed wave function,
$$ \eqalign{ \sp(2.0)
\delta_{1/l} \, \Psi^\dagger (l) \ = \
&  - \, \Psi^\dagger (l)
\, + \, \k_n \Psi^\dagger (l+n-2)
\, + \, (1-\delta_{l,1}) \,
        \sum_{l'=0}^{l-2} \Psi^\dagger (l') \Psi^\dagger (l-l'-2)
\cr\sp(4.0)
&  + \, 2 g \, \sum_{l'=1}^\infty \Psi^\dagger (l+l'-2) l' \Psi (l') \, .
\cr\sp(3.0)} \eqn\EqOneOverLDeformNII $$
Therefore, the Hamiltonian which satisfies
{\EqOneOverLHamilton} with {\EqOneOverLDeformNII} is
$$ \eqalign{ \sp(2.0)
\H (g,\k_n)  \, = \,
&\sum_{l=1}^\infty \bigl\{
 \Psi^\dagger (l) - \k_n \Psi^\dagger (l+n-2)
 \bigr\} \, l \, \Psi (l)
 \, - \,
 \sum_{l=2}^\infty
 \sum_{l'=0}^{l-2} \Psi^\dagger (l') \Psi^\dagger (l-l'-2)
 \, l \, \Psi (l)
\cr\sp(6.0)
&- \, g \, \sum_{l=1}^\infty \sum_{l'=1}^\infty
 \Psi^\dagger (l+l'-2) \, l \, \Psi (l) \, l' \, \Psi (l') \, ,
\cr\sp(3.0)} \eqn\EqHN $$
where we have introduce the wave function {\EqNormalize} again.
The discrete Laplace transformation of the Hamiltonian
$\H (g,\k_n)$ has the form,
$$ \eqalign{ \sp(2.0)
  \H (g,\k_n)  \ = \  \intdz{x_c}
  \, \Bigl\{ \,
&    \bigl\{ \,
     ( 1 - {\k_n \over z^{n-2}} ) \, \Psi^\dagger (z) \,
       - \, z^2 \, ( \Psi^\dagger (z) )^2 \,
     \bigr\} \, ( - z {\der \over \der z} \, \Psi (\inv{z}) )
\cr\sp(6.0)
&    - \, g \, z^2 \, \Psi^\dagger (z) \,
     ( - z {\der \over \der z} \, \Psi (\inv{z}) )^2
  \, \Bigr\} \, .
\cr\sp(3.0)} \eqn\EqHDisN $$
Thus, we obtain the Hamiltonian
of the discretized $c=0$ string field theory
by using regular $n$-polygons instead of regular triangles.
Any amplitudes as well as any transfer matrices are calculated by
{\EqGenTMDisII} and {\EqGenAmpDisII}
with the Hamiltonian {\EqHDisN}.
\par
Now, let us take the continuum limit
from the viewpoint of the time evolution of string states.
The form of $[ \H , \Psi^\dagger ] \cuum$ is
$$ \eqalign{ \sp(2.0)
& [ \, \H (g,\k_n) \, , \, \Psi^\dagger (x) \, ]  \, \cuum
\cr\sp(6.0)
& = \
  x {\der \over \der x} \bigl\{ \,
  \Psi^\dagger (x) \, - \, {\k_n \over x^{n-2}} \,
  ( \, \Psi^\dagger (x)
    \, - \, \sum_{i=0}^{n-3} {1 \over i!} \, x^i
            {\der^i \Psi^\dagger (x=0) \over \der x^i} \, )
  \, - \, x^2 ( \Psi^\dagger (x) )^2
  \, \bigr\} \, \cuum ,
\cr\sp(3.0)} \eqn\EqOneDeformPsiNI $$
while the form of $[ [ \H , \Psi^\dagger ] , \Psi^\dagger ] \cuum$ is
exactly the same as {\EqOneDeformPsiII},
and $\hbox{otherwise = 0}$.
In order to remove the linear terms of the wave function
in {\EqOneDeformPsiNI},
we introduce the following redefined wave function:
$$ \eqalign{ \sp(2.0)
\Phi^\dagger (x,\k_n) \ \define \
\Psi^\dagger (x) \, - \, \lambda ( x , \k_n ) ,
\cr\sp(3.0)} \eqn\EqDefPhiDaggerN $$
where
$$ \eqalign{ \sp(2.0)
\lambda ( x , \k_n )  \ \define \
{1 \over 2 x^2} \bigl( 1 - {\k_n \over x^{n-2}} \bigr)  .
\cr\sp(3.0)} \eqn\EqDefLambdaN $$
Substituting {\EqDefPhiDaggerN} and {\EqDefLambdaN} into {\EqOneDeformPsiNI},
we obtain
$$ \eqalign{ \sp(2.0)
& [ \, \H (g,\k_n) \, , \, \Phi^\dagger (x,\k_n) \, ]  \, \cuum
\cr\sp(6.0)
& = \  \Bigl\{ \,
  - \, \omega ( x , \k_n ) \, - \,
  x {\der \over \der x} \bigl\{ \, x^2 ( \Phi^\dagger (x,\k_n) )^2 \, \bigr\}
  \, \Bigr\} \, \cuum ,
\cr\sp(3.0)} \eqn\EqOneDeformPhiN $$
where
$$ \eqalign{ \sp(2.0)
 \omega ( x , \k_n )
 \ \define \
 - \, x {\der \over \der x}
 \bigl\{ \,
   {1 \over 4 x^2} \bigl( 1 - {\k_n \over x^{n-2}} \bigr)^2
   \, + \, {\k_n \over x^{n-2}} \,
           \sum_{i=0}^{n-3} {1 \over i!} \, x^i
           {\der^i \Psi^\dagger (x=0) \over \der x^i}
 \, \bigr\} .
\cr\sp(3.0)} \eqn\EqOmegaDisN $$
We also find that
$[ [ \H , \Phi^\dagger] , \Phi^\dagger] \cuum$
has the same form as {\EqOneDeformPhiII} and $\hbox{otherwise = 0}$.
\par
Now let us consider to take the continuum limit,
$\e \rightarrow 0$ with {\EqConLimitxyk} and {\EqConLimitWave}.
Here, note that $\omega (x,\k_n)$ in {\EqOmegaDisN} includes
the operator,
$\der^i \Psi^\dagger (x=0) / \der x^i$,
while $\omega (x,\k)$ in {\EqOmegaDis} does not.
Thus, in order to take the continuum limit of
{\EqOneDeformPhiN} with {\EqOmegaDisN},
we need to give the continuum limit of
$\der^i \Psi^\dagger (x=0) / \der x^i$,
which is a string with length $i$.
Since such states are almost vanishing states
in the sense of continuum limit,
they are replaced by the disk amplitude $F_1^{(0)}$
before taking the continuum limit as
$$ \eqalign{ \sp(2.0)
  {\der^i \Psi^\dagger (x=0) \over \der x^i}
  \ \longrightarrow \
  {\der^i F_1^{(0)} (x=0,\k_n) \over \der x^i} \, .
\cr\sp(3.0)} \eqn\EqConLimitVanishingStates $$
Therefore, one obtains the continuum limit of $\omega (x,\k_n)$
if one knows the explicit forms of
$\der^i F_1^{(0)} (x=0,\k_n) / \der x^i$ for $0 \le i \le n-3$.
As was shown in section 3,
in the $\e \rightarrow 0$ limit,
the coefficients of $\e^0$ and $\e^1$ in the right-hand side of {\EqOmegaDisN}
should be zero
so as to satisfy the condition {\EqConLimitCondition}.
This requirement will determine
the values of the critical points, $x_c$ and $\k_c$, uniquely.
Then, one will have
$$ \eqalign{ \sp(2.0)
  \omega (\xi,\cc)
  \ = \
  \hbox{const.} \times \,
  \lim_{\e \rightarrow 0}
  {1 \over \e^2} \, \omega ( x , \k_n )
  \ = \
  3 \xi^2 \, - \, {3 \over 4} \cc \, ,
\cr\sp(3.0)} \eqn\EqOmegaConN $$
after choosing the values of the coefficients $c_i$ properly.
Therefore, one will find the Hamiltonian {\EqHGenCon} again.
As a result, the same $c=0$ non-critical string field theory will be obtained
at the continuous level
from the discretized lattice surfaces by regular $n$-polygons.
\par
In the previous section we study the case of $n=3$.
Since $F_1^{(0)} (x=0,\k_3) = 1$,
which comes from {\EqNormalizeDis},
we have obtained the Hamiltonian {\EqHGenCon}.
In the case of $n=4$, we find that
$F_1^{(0)} (x=0,\k_4) = 1$ and
$\der F_1^{(0)} (x=0,\k_4) / \der x = 0$,
because the former comes from {\EqNormalizeDis}
and the latter comes from the fact that
the number of links on the boundary of disk is even.
The vanishing coefficients of $\e^0$ and $\e^1$
in the $\e \rightarrow 0$ limit of {\EqOmegaDisN}
requires that
$x_c = 1 / 2^{3/2}$ and $\k_c = 1 / 12$.
Then, we obtain {\EqOmegaConN}.
Therefore, we find the Hamiltonian {\EqHGenCon} again.
We have also checked
for the case of $n = 2i$ ($2 \le i \le 14$)
by using Mathematica
that the Hamiltonian {\EqHDisN} always leads to
the same Hamiltonian {\EqHGenCon} in the continuum limit.
\par
To summarize, we have always obtained the same Hamiltonian {\EqHGenCon}
after taking the continuum limit,
in spite of the modification of
\lq peeling decomposition' at the discrete level.
Therefore, the form of the Hamiltonian $\H(G,\cc)$ is universal.
Only $\lambda(\xi)$, which appears in $T_{0,1}^{(0)}$ and $F_1^{(0)}$,
depends on the cut-off parameter $\e$.
\NextPage
\topskip 30pt
\chapter{\ Fractal Structure of Surface }
\vskip 10pt
\par
As one of applications of the string field theory,
we study about the fractal structure of surface for pure gravity theory
in this section.
In section 3 we have defined
$v^{(n)}$ in {\EqUnivNoDisI} or {\EqUnivNoDisVI}
which makes us possible to investigate
the number of strings, the total length of strings and so on.
Since the operator $v^{(n)}$ is proportional to
$\e^{-n}$ in the $\e \rightarrow 0$ limit,
the continuum limit of $v^{(n)}$ is taken by
$$ \eqalign{ \sp(2.0)
  V^{(n)}  \ = \  \lim_{\e \rightarrow 0} \e^n \, v^{(n)} \, .
\cr\sp(3.0)} \eqn\EqUnivNoConI $$
By using the redefined wave function $\Phi^\dagger$, we find
$$ \eqalign{ \sp(2.0)
  V^{(n)}  \
& = \  \intdzeta \, \Psi^\dagger (\zeta)
       \bigl( {\der \over \der \zeta} \bigr)^n \Psi (- \zeta)
\cr\sp(6.0)
& = \  \intdzeta \, \Phi^\dagger (\zeta)
       \bigl( {\der \over \der \zeta} \bigr)^n \Psi (- \zeta)
       \, + \, \delta_{n,0} f_0 \, + \, \delta_{n,1} f_1 \, ,
\cr\sp(3.0)} \eqn\EqUnivNoConII $$
where
$$ \eqalign{ \sp(2.0)
& f_0  \ = \
  {1 \over \sqrt{3} (1+\sqrt{3})^{3/2}} \,
  \bigl( \, \e^{-3/2} \Psi (L=0) \,
       + \, \sqrt{3} \e^{-1/2} {\der \Psi (L=0) \over \der L}
  \, \bigr)  \, ,
\cr\sp(6.0)
& f_1  \ = \
  {1 \over (1+\sqrt{3})^{3/2}} \, \e^{-1/2} \Psi (L=0) \, .
\cr\sp(3.0)} \eqn\EqUnivNoConIII $$
$V^{(0)}$ and $V^{(1)}$ suffer from
the contribution of the non-universal part $f_0$ and $f_1$,
which come from $\lambda(\xi,\cc)$
in the field redefinition {\EqDefPhiDaggerCon}.
This fact is consistent with the result in ref.\ [{\KKMW}].
\par
According to ref.\ [{\KKMW}],
we investigate the fractal structure of a disk surface
by the expectation value of
$\Psi^\dagger(L) \Psi(L)$ instead of
$V^{(n)} = \int d L L^n \Psi^\dagger(L) \Psi(L)$.
The operator $\Psi^\dagger(L) \Psi(L)$ counts
the number of strings with length $L$.
At the discrete level,
the expectation value of $\Psi^\dagger(l) \Psi(l)$ for a disk surface is
$$ \eqalign{ \sp(2.0)
  \rho (l;l_0;\k;\t)
  \ &\define \
  { \lim_{\t' \rightarrow \infty}
    \vac e^{- \t' \H(g=0,\k)}
    \Psi^\dagger(l) \Psi(l)
    e^{- \t \H(g=0,\k)} \Psi^\dagger(l_0) \cuum
    \over
    \lim_{\t' \rightarrow \infty}
    \vac e^{- \t' \H(g=0,\k)}
    e^{- \t \H(g=0,\k)} \Psi^\dagger(l_0) \cuum
  }
\cr\sp(6.0)
    &= \
  { F_1^{(0)} (l;\k) \, \mbar T_{1,1}^{(0)} (l;l_0;\k;\t)
    \over
    F_1^{(0)} (l_0;\k)
  } \, ,
\cr\sp(3.0)} \eqn\EqRhoDis $$
where $\mbar T_{1,1}^{(0)} (l;l_0;\k;\t)$
is the propagator of one universe defined by
$$ \eqalign{ \sp(2.0)
  \mbar T_{1,1}^{(0)} ( y ; x ; \k ; \t ) \
& \define \
  \sum_{m=0}^\infty \, {1 \over m!}
  \oint\limits_{|z_1|=x_c} \! \! \cdots \! \! \oint\limits_{|z_m|=x_c} \! \!
  {d z_1 \over 2 \pi i z_1} \cdots {d z_m \over 2 \pi i z_m} \,
  F_1^{(0)} ( z_1 ; \k ) \cdots F_1^{(0)} ( z_m ; \k )
\cr\sp(6.0)
& \phantom{\define \  \sum_{m=0}^\infty \, } \
  \times T_{m+1,1}^{(0)} ( y , \inv{z_1} , \ldots, \inv{z_m} ; x ; \k ; \t )
  \, .
\cr\sp(3.0)} \eqn\EqPropaOneUnivDisI $$
In the above calculation, we have used the composition law {\EqCompoLaw}.
The $\rho (l;l_0;\k;\t)$ counts
the number of final string states with length $l$
after $\t$ step starting from an initial string with length $l_0$.
Note that this propagator satisfies the composition law:
$$ \eqalign{ \sp(2.0)
\mbar T_{1,1}^{(0)} ( y ; x ; \k ; \t_2 + \t_1 )  \ = \
\intdz{x_c} \, \mbar T_{1,1}^{(0)} ( y ; z ; \k ; \t_2 )
            \, \mbar T_{1,1}^{(0)} ( \inv{z} ; x ; \k ; \t_1 ) \, .
\cr\sp(3.0)} \eqn\EqPropaPileDis $$
According to ref.\ [{\IK}],
we introduce the Hamiltonian $\Hone$
which produces the propagator {\EqPropaOneUnivDisI}.
Since all final strings except for one string are capped by disks,
the propagator {\EqPropaOneUnivDisI} is rewritten by
$$ \eqalign{ \sp(2.0)
  \mbar T_{1,1}^{(0)} ( y ; x ; \k ; \t )
  \ = \
  \vac \Psi (y) \, e^{- \t \Hone (\k)} \, \Psi^\dagger (x) \cuum \, ,
\cr\sp(3.0)} \eqn\EqPropaOneUnivDisII $$
where the Hamiltonian $\Hone (\k)$ is modified
from {\EqHDis} as
$$ \eqalign{ \sp(2.0)
  \Hone (\k) \
& = \  \intdz{x_c} \,
  \Psi^\dagger (z)
  \bigl( \, 1 - {\k \over z} - 2 z^2 F_1^{(0)} (z;\k) \, \bigr) \,
  ( - z {\der \over \der z} \, \Psi (\inv{z}) )
\cr\sp(6.0)
& = \  \intdz{x_c} \,
  \Psi^\dagger (z)
  \bigl( \, - 2 z^2 \hat F_1^{(0)} (z;\k) \, \bigr) \,
  ( - z {\der \over \der z} \, \Psi (\inv{z}) ) \, .
\cr\sp(3.0)} \eqn\EqHOneUnivDis $$
Here we have used $\hat F_1^{(0)} (z;\k)$ defined in {\EqAmpDiskDisHatF}.
{}From {\EqPropaOneUnivDisII} and {\EqHOneUnivDis}
we find the differential equation,
$$ \eqalign{ \sp(2.0)
  {\der \over \der \t} \, \mbar T_{1,1}^{(0)} ( y ; x ; \k ; \t )
  \ = \
  x {\der \over \der x} \, \bigl\{ \,
  2 x^2 \hat F_1^{(0)} (x;\k) \, \mbar T_{1,1}^{(0)} ( y ; x ; \k ; \t )
  \, \bigr\} ,
\cr\sp(3.0)} \eqn\EqPropaDiffEqDis $$
which is also derived from the original definition {\EqPropaOneUnivDisI}
directly.
Substituting {\EqTMandDelta} into {\EqPropaOneUnivDisI}
for $\t \rightarrow 0$,
we find that
$\mbar T_{1,1}^{(0)} (y;x;\k;\t=0) = \delta (y,x)$.
\par
Next, let us consider to take the continuum limit.
Similarly to {\EqConLimitH},
the continuum limit of the Hamiltonian $\Hone (\k)$
is assumed to be
$$ \eqalign{ \sp(2.0)
\Hone (\cc)  \ = \  \lim_{\e \rightarrow 0} c_4 \, \e^\dim{\H} \, \Hone (\k) .
\cr\sp(3.0)} \eqn\EqConLimitVI $$
We also assume for the transfer matrix operator that
$$ \eqalign{ \sp(2.0)
e^{- \T \Hone (\cc)}  \ = \  \lim_{\e \rightarrow 0} e^{- \t \Hone (\k)}   .
\cr\sp(3.0)} \eqn\EqConLimitDeformII $$
Therefore, if we take {\EqDimension},
we obtain the continuum limit of the Hamiltonian $\Hone (\k)$ as
$$ \eqalign{ \sp(2.0)
\Hone (\cc)  \ = \
- 2 \, \intdzeta \,
\Psi^\dagger (\zeta)
\hat F_1^{(0)} (\zeta;\cc) \, {\der \over \der \zeta} \, \Psi (-\zeta) \, ,
\cr\sp(3.0)} \eqn\EqHOneUnivCon $$
where $\hat F_1^{(0)} (\zeta;\cc)$ is the universal part of disk amplitude
defined in {\EqAmpDiskConHatF}.
The continuum limit of the propagator of one universe is
$$ \eqalign{ \sp(2.0)
  \mbar T_{1,1}^{(0)} ( \eta ; \xi ; \cc ; \T )
  \ = \
  \vac \Psi (\eta) \, e^{- \T \Hone (\cc)} \, \Psi^\dagger (\xi) \cuum \, ,
\cr\sp(3.0)} \eqn\EqPropaOneUnivCon $$
which satisfies
$$ \eqalign{ \sp(2.0)
\mbar T_{1,1}^{(0)} ( \eta ; \xi ; \cc ; \T_2 + \T_1 )  \ = \
\intdzeta \, \mbar T_{1,1}^{(0)} ( \eta ; \zeta ; \cc ; \T_2 )
          \, \mbar T_{1,1}^{(0)} ( - \zeta ; \xi ; \cc ; \T_1 ) \, .
\cr\sp(3.0)} \eqn\EqPropaPileCon $$
Then, from {\EqPropaOneUnivCon} with {\EqHOneUnivCon},
or directly from the continuum limit of {\EqPropaDiffEqDis},
we obtain the differential equation at the continuous level,
$$ \eqalign{ \sp(2.0)
 {\der \over \der \T} \, \mbar T_{1,1}^{(0)} ( \eta ; \xi ; \cc ; \T )
 \ = \
 {\der \over \der \xi} \, \bigl\{ \,
 2 \, \hat F_1^{(0)} (\xi;\cc) \,
 \mbar T_{1,1}^{(0)} ( \eta ; \xi ; \cc ; \T )
 \, \bigr\}  ,
\cr\sp(3.0)} \eqn\EqPropaDiffEqCon $$
which is the same equation obtained by refs.\ [{\KKMW},{\IK}].
\foot{
$\mbar T_{1,1}^{(0)} (L';L;\cc;\T)$ is related to $N(L,L';\T;\cc)$,
which is used in refs.\ [{\KKMW},{\IK}],
\hfill\break
by $\mbar T_{1,1}^{(0)} (L';L;\cc;\T) = (L/L') N(L,L';\T;\cc)$.
}
The Hamiltonian $\Hone$ is useful,
though one can derive {\EqPropaDiffEqCon} directly
from the continuum limit of {\EqPropaOneUnivDisI}.
The initial condition of {\EqPropaDiffEqCon} is
$\mbar T_{1,1}^{(0)} (\eta;\xi;\cc;\T=0) = \delta (\eta,\xi)$.
Note that
$\mbar T_{1,1}^{(0)} (L';L;\cc;\T)$
has no non-universal contributions
because $\Hone (\cc)$ in {\EqHOneUnivCon} does not have them.
\par
The fractal structure
of a large enough space-time with disk topology
can be studied through the function {\EqRhoDis} at the continuous level, i.e.,
$$ \eqalign{ \sp(2.0)
  \rho (L;L_0;\cc;\T)
  \ = \
  { F_1^{(0)} (L;\cc) \, \mbar T_{1,1}^{(0)} (L;L_0;\cc;\T)
    \over
    F_1^{(0)} (L_0;\cc)
  } \, .
\cr\sp(3.0)} \eqn\EqRhoCon $$
In ref.\ [{\KKMW}] the authors have obtained
the explicit form of $\rho (L;L_0=0;A=\infty;\T)$,
by solving the differential equation {\EqPropaDiffEqCon}.
The $\rho (L;L_0=0;A=\infty;\T) d L$
is the number of loops belonging to the boundary of
$\Surf (p;\T)$ whose lengths lie between $L$ and $L+dL$,
where $\Surf (p;\T)$ is the set of points
whose geodesic distances from $p$ are less than or equal to $\T$.
\NextPage
\topskip 30pt
\chapter{\ Matter Fields on Surface }
\vskip 10pt
\par
In this section we extend our formalism to
the string field theory with matter fields on lattice surface,
i.e., the central charge of matter fields is $c \neq 0$.
\par
Firstly, we consider the non-critical string field theory
which corresponds to
$m$-th multicritical one matrix model ($m \ge 2$).\NPrefmark{\Kazakov}
The case $m=2$ corresponds to $c=0$ pure quantum gravity,
which has been studied in the previous sections.
For $m \ge 3$, matter fields are incorporated in this string theory
because the central charge is $c = - 2 (m-2) (6m-7) / (2m-1) < 0$.
In this model all kinds of regular $n$-polygons
are introduced at the same time.
Therefore, the Hamiltonian has the form,
$$ \eqalign{ \sp(2.0)
  \H (g,\k_3,\ldots)  \ = \  \intdz{x_c}
  \, \Bigl\{ \,
&    \bigl\{ \,
     ( 1 - \sum_{n \ge 3} {\k_n \over z^{n-2}} ) \, \Psi^\dagger (z) \,
       - \, z^2 \, ( \Psi^\dagger (z) )^2 \,
     \bigr\} \, ( - z {\der \over \der z} \, \Psi (\inv{z}) )
\cr\sp(6.0)
&    - \, g \, z^2 \, \Psi^\dagger (z) \,
     ( - z {\der \over \der z} \, \Psi (\inv{z}) )^2
  \, \Bigr\} \, ,
\cr\sp(3.0)} \eqn\EqHKDis $$
where the commutation relations are the same as {\EqCommutePsiDisII}.
Any amplitudes as well as any transfer matrices are calculated by
{\EqGenTMDisII} and {\EqGenAmpDisII}
with the Hamiltonian {\EqHKDis}.
Here we require the assumptions $a$), $b$), $c$), $d$) and $e$) again,
i.e., we assume that
$\Psi^\dagger(x)$ and $\Psi(y)$ are analytic in the region
$|x| \le x_c$ and $|y| \le y_c = 1/x_c$,
where $x_c$ and $y_c$ are the convergence radii.
\par
The time evolution of the string state,
$[ \H , \Psi^\dagger ] \cuum$, is
$$ \eqalign{ \sp(2.0)
& [ \, \H (g,\k_3,\ldots) \, , \, \Psi^\dagger (x) \, ]  \, \cuum
\cr\sp(6.0)
& = \
  x {\der \over \der x} \bigl\{ \,
  \Psi^\dagger (x) \,
  - \, \sum_{n \ge 3} {\k_n \over x^{n-2}} \,
  ( \, \Psi^\dagger (x)
    \, - \, \sum_{i=0}^{n-3} {1 \over i!} \, x^i
            {\der^i \Psi^\dagger (x=0) \over \der x^i} \, )
  \, - \, x^2 ( \Psi^\dagger (x) )^2
  \, \bigr\} \, \cuum \, .
\cr\sp(3.0)} \eqn\EqOneDeformPsiK $$
As the same as before,
we introduce the following redefined wave function
instead of {\EqDefPhiDagger}:
$$ \eqalign{ \sp(2.0)
\Phi^\dagger (x,\k_3,\ldots) \ \define \
\Psi^\dagger (x) \, - \, \lambda (x,\k_3,\ldots) \, ,
\cr\sp(3.0)} \eqn\EqDefPhiDaggerK $$
where
$$ \eqalign{ \sp(2.0)
\lambda (x,\k_3,\ldots)  \ \define \
{1 \over 2 x^2} \bigl( 1 - \sum_{n \ge 3} {\k_n \over x^{n-2}} \bigr)  .
\cr\sp(3.0)} \eqn\EqDefLambdaK $$
Substituting {\EqDefPhiDaggerK} and {\EqDefLambdaK}
into {\EqOneDeformPsiK},
we obtain
$$ \eqalign{ \sp(2.0)
& [ \, \H (g,\k_3,\ldots) \, , \, \Phi^\dagger (x,\k_3,\ldots) \, ]  \, \cuum
\cr\sp(6.0)
& = \  \Bigl\{ \,
  - \, \omega ( x , \k_3,\ldots ) \, - \,
  x {\der \over \der x} \bigl\{ \,
    x^2 ( \Phi^\dagger (x,\k_3,\ldots) )^2 \, \bigr\}
  \, \Bigr\} \, \cuum ,
\cr\sp(3.0)} \eqn\EqOneDeformPhiK $$
where the form of $\omega (x,\k_3,\ldots)$ is
$$ \eqalign{ \sp(2.0)
 \omega ( x , \k_3,\ldots )
 \ \define \
 - \, x {\der \over \der x}
 \bigl\{ \,
   {1 \over 4 x^2} \bigl( 1 - \sum_{n \ge 3} {\k_n \over x^{n-2}} \bigr)^2
   \, + \, \sum_{n \ge 3} {\k_n \over x^{n-2}} \,
           \sum_{i=0}^{n-3} {1 \over i!} \, x^i
           {\der^i \Psi^\dagger (x=0) \over \der x^i}
 \, \bigr\} .
\cr\sp(3.0)} \eqn\EqOmegaDisK $$
The time evolutions of other states are still unchanged as
{\EqOneDeformPhiII} and $\hbox{otherwise = 0}$
under the field redefinition {\EqDefPhiDaggerK} and {\EqDefLambdaK}.
\par
Now, let us consider to take the continuum limit, $\e \rightarrow 0$.
The continuum limit of
$x$, $y$, $\Psi^\dagger$, $\Psi$ and $\Phi^\dagger$
are taken as the same as {\EqConLimitxyk} and {\EqConLimitWave},
while that of $\k_p$ is
$$ \eqalign{ \sp(2.0)
  \k_p
  \ = \
  \k_{pc} \, \exp ( \, - \sum_{q=2}^m \e^q c_{pq} \cc_q \, ) \, ,
  \qquad \hbox{ ( for \ $3 \le p$ ) }
\cr\sp(3.0)} \eqn\EqConLimitkp $$
where $\k_{pc}$ is the convergence radius for $\k_p$ ($3 \le p$).
The continuum limit of area
$A_q = \e^q \sum_p c_{pq} a_p$ ($2 \le q \le m$)
has the dimension
$\dim{A_q} = q$,
where $a_p$ is the number of regular $p$-polygons on surface.
The relationship between $\cc_i$ ($i=2,\ldots,m$) in the above
and the cosmological constant of the Liouville theory
is discussed in refs.\ [{\MSS},{\Klebanov}].
Similarly to section 6,
$\omega (x,\k_3,\ldots)$ in {\EqOmegaDisK} includes the operator,
$\der^i \Psi^\dagger (x=0) / \der x^i$,
then, we take the continuum limit of
$\der^i \Psi^\dagger (x=0) / \der x^i$
by the replacement as {\EqConLimitVanishingStates}.
Thus, we obtain the continuum limit of $\omega (x,\k_3,\ldots)$
if we know the explicit forms of
$\der^i F_1^{(0)} (x=0,\k_3,\ldots) / \der x^i$
for $0 \le i \le n-3$.
In the $m$-th multicritical one matrix model,
one requires that
the coefficients of $\e^0$, $\ldots$, $\e^{2m-3}$
in the right-hand side of {\EqOmegaDisK}
are zero in the $\e \rightarrow 0$ limit.
This requirement restricts
the values of the critical points, $x_c$ and $\k_{pc}$.
As a result, we have
$$ \eqalign{ \sp(2.0)
  \omega (\xi,\cc_2,\ldots,\cc_m)
  \ = \
  \hbox{const.} \times \,
  \lim_{\e \rightarrow 0}
  {1 \over \e^{2m-2}} \, \omega ( x , \k_3,\ldots )
  \ = \
  (2m-1) \, \xi^{2m-2} \, + \, \ldots
  \, ,
\cr\sp(3.0)} \eqn\EqOmegaConK $$
after choosing the values of the coefficients $c_i$ properly.
The factor $(2m-1)$ in front of $\xi^{2m-2}$ in {\EqOmegaConK} is convention.
The dimensions of the wave functions and the Hamiltonian are
$$ \eqalign{ \sp(2.0)
& \dim{\Phi^\dagger}  \ = \
  \dim{\Psi^\dagger}  \ = \  - \, {2m - 1 \over 2} ,
  \qquad
  \dim{\Psi}  \ = \  {2m + 1 \over 2} ,
\cr\sp(6.0)
& \dim{\T}  \ = \  - \, \dim{\H}  \ = \  {2m - 3 \over 2} ,
\cr\sp(3.0)} \eqn\EqDimensionK $$
which satisfies the condition {\EqConLimitCondition}.
We also obtain $\dim{G} = - 2m - 1$.
Note that
$\Phi^\dagger$ depends on
not only $\xi$ but also $\cc_1$, $\ldots$, $\cc_{m-1}$,
because
$$ \eqalign{ \sp(2.0)
  \Phi^\dagger (\xi,\cc_2,\ldots,\cc_{m-1})
  \ = \
  \Psi^\dagger (\xi) \, - \, \lambda(\xi,\cc_2,\ldots,\cc_{m-1})
\cr\sp(3.0)} \eqn\EqDefPhiDaggerKCon $$
with
$$ \eqalign{ \sp(2.0)
  \lambda (\xi,\cc_2,\ldots,\cc_{m-1})
  \ \define \
  \lim_{\e \rightarrow 0} c_1 \, \e^\dim{\Psi^\dagger} \,
  \lambda (x,\k_3,\ldots)
  \, .
\cr\sp(3.0)} \eqn\EqDefLambdaKConI $$
{}From the dimensional analysis,
$\lambda$ and $\Phi^\dagger$ are found to be independent of $\cc_m$
in the continuum limit $\e \rightarrow 0$.
Therefore,
from the continuum limit of {\EqOneDeformPhiK},
$[ [ \H , \Phi^\dagger ] , \Phi^\dagger ] \cuum$ and so on,
we find the Hamiltonian at the continuous level,
$$ \eqalign{ \sp(2.0)
\H (G,\cc_2,\ldots,\cc_m)  \ = \
\intdzeta \, \bigl\{ \,
& - \,      \omega (\zeta,\cc_2,\ldots,\cc_m) \, \Psi (- \zeta)
\cr\sp(6.0)
& - \,      ( \Phi^\dagger (\zeta,\cc_2,\ldots,\cc_{m-1}) )^2 \,
            {\der \over \der \zeta} \, \Psi (- \zeta)
\cr\sp(6.0)
& - \, G \, \Phi^\dagger (\zeta,\cc_2,\ldots,\cc_{m-1}) \,
            ( {\der \over \der \zeta} \, \Psi (- \zeta) )^2
\, \bigr\} \, .
\cr\sp(3.0)} \eqn\EqHKCon $$
The transfer matrices as well as the amplitudes are calculated
by using {\EqGenTMCon} and {\EqGenAmpCon}.
Thus, we have obtained the non-critical string field theory
for $m$-th multicritical one matrix model.
For example, in the case of $m=3$, we find
$$ \eqalign{ \sp(2.0)
  \omega (\zeta,\cc_2,\cc_3)
  \ = \
  5 \zeta^4 \, - \, {15 \over 4} \cc_2 \zeta^2 \,
  + \, {5 \over 4} ( \, - p \cc_2 \, + \, p^3 \, ) \, \zeta \,
  + \, {5 \over 64} ( \, 5 \cc_2^2 \, + \, 2 p^2 \cc_2 \, - \, 3 p^4 \, ) \, ,
\cr\sp(3.0)} \eqn\EqOmegaConKII $$
where $p$ is a positive solution of
eq.\ $p^3 = p \cc_2 + \cc_3$.
We have also calculated $\omega$ for $m=4$ as
$$ \eqalign{ \sp(2.0)
  \omega (\zeta,\cc_2,\cc_3,\cc_4)
  \ = \
& 7 \zeta^6 \, - \, {175 \over 24} \cc_2 \zeta^4 \,
  + \, {35 \over 8} \cc_3 \zeta^3 \,
  + \, {105 \over 64}
    ( \, {35 \over 36} \cc_2^2 \, + \, p \cc_3 \,
    + \, p^2 \cc_2 \, - \, p^4 \, ) \, \zeta^2
\cr\sp(4.0)
& + \, {7 \over 16}
    ( \, - \, {175 \over 48} \cc_2 \cc_3 \, - \, {5 \over 8} p^2 \cc_3 \,
    - \, {5 \over 6} p^3 \cc_2 \, + \, p^5 \, ) \, \zeta \,
\cr\sp(4.0)
& + \, {35 \over 256}
    ( \, {35 \over 16} \cc_3^2 \, - \, {35 \over 12} p \cc_2 \cc_3 \,
    - \, {35 \over 12} p^2 \cc_2^2 \, + \, {1 \over 2} p^3 \cc_3 \,
    + \, {11 \over 3} p^4 \cc_2 \, - \, p^6 \, ) \,  ,
\cr\sp(3.0)} \eqn\EqOmegaConKIII $$
where $p$ is a positive solution of
eq.\ $p^4 = p^2 \cc_2 - \cc_2^2 / 12 + p \cc_3 + \cc_4$.
According to the discussion in appendix D,
$\omega$ is related to
the universal part of disk amplitude $\hat F_1^{(0)}$ as
$\omega(\zeta,\cc_2,\ldots,\cc_m) =
 {\der \over \der \zeta} ( \hat F_1^{(0)}(\zeta;\cc_2,\ldots,\cc_m) )^2$.
For any $m$, the disk amplitude calculated
by the above non-critical string field theory
is the same as that obtained from
$m$-th multicritical one-matrix model.
\par
We have also studied the one universe propagator,
which corresponds to {\EqPropaOneUnivDisI} or {\EqPropaOneUnivDisII},
for $m$-th multicritical non-critical string field theory.
The Hamiltonians at the discrete level and at the continuous level are
the same as {\EqHOneUnivDis} and {\EqHOneUnivCon} respectively,
where $\hat F_1^{(0)}$ is the universal part of disk amplitude for
$m$-th multicritical model.
Therefore, we find the same differential equations
{\EqPropaDiffEqDis} and {\EqPropaDiffEqCon},
the latter of which agrees with the result by ref.\ [{\Klebanov}].
\par
Next, let us consider to put a matter field $\phi$ naively
on each link of triangulated surface.
Since matter fields are put on each link,
the wave function depends on
not only the length of string $l$
but also the values of matter fields on string,
i.e.,
$\Psi^\dagger = \Psi^\dagger ( l ; \phi_1 , \ldots , \phi_l )$.
We here consider triangulated surfaces, for simplicity.
The extension to the case with $n$-polygons is trivial.
The amplitudes and the transfer matrices for the gravity theory
coupled to matter fields are
$$ \eqalign{ \sp(2.0)
& F_N^{(h)} ( l_1, \ldots, l_N ; \k )  \ = \
  \sum_{a=0}^\infty \, \sum_{\Surf_N^{(h)}} \, \k^a
  \, Z_{\rm matter}^{(c)} [ \Surf_N^{(h)} ] \, ,
\cr\sp(4.0)
& T_{M,N}^{(h)} ( l'_1, \ldots, l'_M ; l_1, \ldots, l_N ; \k ; \t )
  \ = \
  \sum_{a=0}^\infty \, \sum_{\Surf_{M,N}^{(h)}} \, \k^a
  \, Z_{\rm matter}^{(c)} [ \Surf_{M,N}^{(h)} ] \, ,
\cr\sp(3.0)} \eqn\EqAmpAndTMwithMatter $$
where
$\Surf_N^{(h)}$ and $\Surf_{M,N}^{(h)}$ are the triangulated surfaces
defined in section 2.
$Z_{\rm matter}^{(c)} [ \Surf ]$
is the partition function for matter fields with the central charge $c$
on the triangulated surface $\Surf$.
Then, the $(1/l)$-step deformations of wave function are
$$ \eqalign{ \sp(2.0)
& \Psi^\dagger ( l ; \phi_1 , \ldots , \phi_l ) \
  \longrightarrow \
  \int \! d \phi' d \phi''
  \k (\phi_1,\phi',\phi'') \,
  \Psi^\dagger ( l+1 ; \phi_2 , \ldots , \phi_l , \phi' , \phi'' ) \,  ,
\cr\sp(6.0)
& \hskip 324pt \hbox{ ( for Fig.\ {\FigPeelSkinI} ) }
\cr\sp(8.0)
& \Psi^\dagger ( l ; \phi_1 , \ldots , \phi_l ) \
  \longrightarrow \
  \sum_{l'=0}^{l-2} \delta ( \phi_1 - \phi_{l'+2} ) \,
  \Psi^\dagger ( l' ; \phi_2 , \ldots , \phi_{l'+1} ) \,
\cr\sp(6.0)
& \phantom{\Psi^\dagger ( l ; \phi_1 , \ldots , \phi_l ) \
           \longrightarrow \
           \sum_{l'=0}^{l-2} }
  \times \,
  \Psi^\dagger ( l-l'-2 ; \phi_{l'+3} , \ldots , \phi_l ) \,  ,
  \hskip 30pt \hbox{ ( for Fig.\ {\FigPeelSkinII} ) }
\cr\sp(8.0)
& \Psi^\dagger ( l ; \phi_1 , \ldots , \phi_l ) \
  \longrightarrow \
  \sum_{l'=1}^\infty \sum_{i=1}^{l'}
  \int \! d \tilde \phi_1 \cdots d \tilde \phi_{l'} \,
  \delta ( \phi_1 - \tilde \phi_i ) \,
\cr\sp(4.0)
& \phantom{\Psi^\dagger ( l ; \phi_1 , \ldots , \phi_l ) \
           \longrightarrow \  \sum_{l'=1}^\infty \sum_{i=1}^{l'} }
  \times \,
  \Psi^\dagger ( l+l'-2 ; \tilde \phi_1, \ldots, \tilde \phi_{i-1},
                          \phi_2,  \ldots, \phi_l,
                          \tilde \phi_{i+1}, \ldots, \tilde \phi_{l'} )
\cr\sp(4.0)
& \phantom{\Psi^\dagger ( l ; \phi_1 , \ldots , \phi_l ) \
           \longrightarrow \  \sum_{l'=1}^\infty \sum_{i=1}^{l'} }
  \times \,
  \Psi ( l' ; \tilde \phi_1, \ldots, \tilde \phi_{l'} ) \, ,
  \hskip 65pt \hbox{ ( for Fig.\ {\FigPeelSkinIII} ) }
\cr\sp(3.0)} \eqn\EqOneOverLDeformMatter $$
where $\k (\phi,\phi',\phi'')$ is represented by
the action of matter fields, $S (\phi,\phi')$, as
$$ \eqalign{ \sp(2.0)
  \k (\phi,\phi',\phi'')
  \ = \
  \k \, \exp \{ \, - \, S(\phi,\phi') \, - \, S(\phi,\phi'')
                \, - \, S(\phi',\phi'') \, \} \, .
\cr\sp(3.0)} \eqn\EqCosmoMatter $$
One may generalize $\k (\phi,\phi',\phi'')$ to
an arbitrary function of $\phi$, $\phi'$ and $\phi''$.
As the same as before, we have introduced
$\Psi^\dagger (l=0;\void) = 1$ again.
In the non-critical string theory with $c$ bosonic scalars,
the matter fields $\phi^\mu$ ($\mu = 1, \ldots, c$) are real valued.
The action $S(\phi,\phi')$ has the form,
$$ \eqalign{ \sp(2.0)
 S(\phi,\phi')
 \ = \
 {1 \over 4 \pi \alpha'} \, \sum_{\mu=1}^c \, ( \phi^\mu - {\phi'}^\mu )^2 ,
\cr\sp(3.0)} \eqn\EqActionMatterI $$
where $\alpha'$ is the slope parameter.
The continuum limit of the case $c=25$ will lead to
the light-cone string field theory
of the critical bosonic string.\NPrefmark{\KakuKikkawa}
In the Ising model,
the field $\phi$ takes the values, $+1$ or $-1$.
The form of action is
$$ \eqalign{ \sp(2.0)
 S(\phi,\phi')  \ = \
 - \, \beta \, \phi \, \phi' \, - \, H \, (\phi + \phi') \, ,
\cr\sp(3.0)} \eqn\EqActionMatterII $$
where $\beta$ is the inverse temperature and $H$ is the magnetic field.
\par
Since we have marked one of the links on each initial string
by technical reason,
the location of the marked link should be unphysical.
Namely, the wave function has the cyclic symmetry as
$$ \eqalign{ \sp(2.0)
 \Psi^\dagger ( l ; \phi_1, \phi_2, \ldots, \phi_l )
 \ = \
 \Psi^\dagger ( l ; \phi_2, \ldots, \phi_l, \phi_1 ) \, .
\cr\sp(3.0)} \eqn\EqWaveCyclicMatter $$
Then, the commutation relations of the wave function are
$$ \eqalign{ \sp(2.0)
& [ \, \Psi (l;\phi_1,\ldots,\phi_l) \, ,
    \, \Psi^\dagger (l';\tilde\phi_1,\ldots,\tilde\phi_{l'}) \, ]
\cr\sp(6.0)
& = \  {\delta_{l,l'} \over l} \,
       \bigl\{ \, \delta(\phi_1-\tilde\phi_1) \, \cdots \,
                  \delta(\phi_l-\tilde\phi_l)
\cr\sp(4.0)
& \phantom{= \  {\delta_{l,l'} \over l} \, \bigl\{ \, \, }
                  + \, \hbox{(cyclic permutation with respect to
                              $\tilde\phi_1, \ldots, \tilde\phi_l$)} \,
       \bigr\} \, ,
\cr\sp(3.0)} \eqn\EqCommutePsiDisMatter $$
and $\hbox{otherwise = 0}$.
Therefore, we can construct the Hamiltonian
which generates the $(1/l)$-step deformation {\EqOneOverLDeformMatter},
$$ \eqalign{ \sp(2.0)
& \H (g,\k)  \ = \
\cr\sp(6.0)
& \sum_{l=1}^\infty \int \! \prod_{i=1}^l d \phi_i \bigl\{ \,
  \Psi^\dagger ( l ; \phi_1, \ldots, \phi_l ) \,
  - \, \int \! d \phi' d \phi'' \,
  \k (\phi_1,\phi',\phi'') \,
  \Psi^\dagger ( l+1 ; \phi_2, \ldots, \phi_l, \phi', \phi'' )
\cr\sp(6.0)
& \phantom{\sum_{l=1}^\infty \int \! \prod_{i=1}^l d \phi_i \bigl\{ \, } \
  \bigr\} \, l \, \Psi ( l ; \phi_1 , \ldots , \phi_l )
\cr\sp(6.0)
& - \, \sum_{l=2}^\infty \int \! \prod_{i=1}^l d \phi_i \,
  \sum_{l'=0}^{l-2}
  \delta ( \phi_1 - \phi_{l'+2} ) \,
  \Psi^\dagger ( l' ; \phi_2, \ldots, \phi_{l'+1} ) \,
  \Psi^\dagger ( l-l'-2 ; \phi_{l'+3}, \ldots, \phi_{l} )
\cr\sp(6.0)
& \phantom{- \, \sum_{l=1}^\infty \int \! \prod_{i=1}^l d \phi_i \, } \
  \times \, l \, \Psi ( l ; \phi_1 , \ldots , \phi_l )
\cr\sp(6.0)
& - \, g \, \sum_{l=1}^\infty \sum_{l'=1}^\infty
  \int \! \prod_{i=1}^l d \phi_i \prod_{j=1}^{l'} d \tilde \phi_j \,
  \delta ( \phi_1 - \tilde \phi_1 ) \,
  \Psi^\dagger ( l+l'-2 ; \phi_2, \ldots, \phi_l,
                          \tilde \phi_2, \ldots, \tilde \phi_{l'} )
\cr\sp(4.0)
& \phantom{- \, g \, \sum_{l=1}^\infty \sum_{l'=1}^\infty
           \int \! \prod_{i=1}^l d \phi_i \prod_{j=1}^{l'} d \tilde \phi_j \, }
  \ \times \,
  l \,  \Psi ( l  ; \phi_1, \ldots, \phi_l ) \,
  l' \, \Psi ( l' ; \tilde \phi_1, \ldots, \tilde \phi_{l'} ) \, ,
\cr\sp(3.0)} \eqn\EqHMatter $$
where the integration $\int d \phi$
is replaced by the summation $\sum_\phi$
if the matter fields take the discrete values.
The extension to the case with some kinds of regular polygons
is straightforward.
However, it seems difficult
to take the discrete Laplace transformation of {\EqHMatter},
because of matter dependence.
Therefore, we leave the problem of taking the continuum limit
of {\EqHMatter} to the future study.
\NextPage
\topskip 30pt
\chapter{\ Conclusion }
\vskip 10pt
\par
In this paper we have proposed a new method
which analyzes the dynamical triangulation
from the viewpoint of the non-critical string field theory.
The \lq peeling decomposition' has played an important role
in the construction of the discretized non-critical string field theories.
As a simplest example, we have first constructed
the $c=0$ non-critical string field theory at the discrete level.
The assumptions $a$), $b$), $c$), $d$) and $e$) are indispensable
in order to construct the string field theory.
Namely, the wave function $\Psi^\dagger(x)$ and $\Psi(y)$,
which satisfy the commutation relations {\EqCommutePsiDisII},
are analytic in the region $|x| \le x_c$ and $|y| \le 1/x_c$,
where $x_c$ and $1/x_c$ are the convergence radii of
$\Psi^\dagger(x)$ and $\Psi(y)$ respectively.
The amplitude $F_N^{(h)}$ at the discrete level,
which has $h$ ($\ge 0$) handles and $N$ ($\ge 1$) boundaries,
is calculated by {\EqGenAmpDisII}.
The Hamiltonian at the discrete level has the form {\EqHDis}.
Note that the amplitude $F_N^{(h)} ( x_1, \ldots, x_N ; \k )$
is analytic in the region
$|x_i| \le x_c$ ($1 \le i \le N$) and $|\k| < \k_c$
because of the analyticity of $\Psi^\dagger(x)$.
\par
We have also succeeded in taking the continuum limit
and have obtained the $c=0$ continuous string field theory,
which is consistent with that of ref.\ [{\IK}].
The continuum limit is taken by {\EqConLimitxyk} $\sim$ {\EqDefLambdaConI}
and {\EqConLimitH} $\sim$ {\EqConLimitIV}
with {\EqCriticalPoint} and {\EqDimension}.
The field redefinition {\EqDefPhiDaggerCon} with {\EqDefLambdaConI}
is important in order to take the continuum limit.
The wave functions at the continuous level,
$\Psi^\dagger(\xi)$, $\Psi(\eta)$ and $\Phi^\dagger(\xi')$,
are found to be analytic in the region
$0 \le \Real(\xi)$, $0 \le \Real(\eta)$ and
$0 \le \Real(\xi') < \infty$.
The amplitude $F_N^{(h)}$ at the continuous level
is calculated by {\EqGenAmpCon}.
Only the disk amplitude $F_1^{(0)}$ has
the non-universal part $\lambda$ in {\EqDefLambdaConII}
which depends on a cut-off parameter.
The Hamiltonian at the continuous level has the form
{\EqHGenCon} with {\EqOmegaConII}.
$\cc$ is the cosmological constant.
Note that the amplitude $F_N^{(h)} ( \xi_1, \ldots, \xi_N ; \cc )$
is analytic in the region
$0 \le \Real(\xi_i)$ ($1 \le i \le N$) and $0 < \Real(\cc)$
because of the analyticity of $\Psi^\dagger(\xi)$.
For example, we have calculated the explicit forms of
$F_1^{(0)}$, $F_2^{(0)}$, $F_1^{(1)}$, $F_0^{(0)}$ and $F_0^{(1)}$
in appendix D.
$\omega(\zeta,\cc)$ in {\EqHGenCon} is found to be related to
the universal part of the disk amplitude, $\hat F_1^{(0)}$, by
$\omega(\zeta,\cc) = {\der \over \der \zeta} ( \hat F_1^{(0)}(\zeta;\cc) )^2$.
We have also studied the universality of the $c=0$ non-critical string theory
by showing that
some modified string field theories at the discrete level always lead to
the same $c=0$ continuous string field theory
after taking the continuum limit.
As an application of the string field theory,
we have studied about the fractal structure of disk surface.
We have derived the differential equations
{\EqPropaDiffEqDis} and {\EqPropaDiffEqCon},
the latter of which coincides with the result by refs.\ [{\KKMW},{\IK}].
\par
Moreover,
we have extended our formalism to the string field theory with matter fields.
As one of extensions we have obtained the non-critical string field theory
which corresponds to $m$-th multicritical one matrix model ($m = 3,4,\ldots$).
We have succeeded in taking the continuum limit
and have found the Hamiltonian {\EqHKCon},
where
$\omega(\zeta,\cc_2,\ldots,\cc_m) =
 {\der \over \der \zeta} ( \hat F_1^{(0)}(\zeta;\cc_2,\ldots,\cc_m) )^2$
and $\hat F_1^{(0)}(\zeta;\cc_2,\ldots,\cc_m)$
is the universal part of disk amplitude.
Here note that the redefined wave function $\Phi^\dagger$ depends on
not only $\zeta$ but also $\cc_2$, $\ldots$, $\cc_{m-1}$.
For $m$-th multicritical model
we have also obtained the differential equations
{\EqPropaDiffEqDis} and {\EqPropaDiffEqCon},
the latter of which agrees with the result by ref.\ [{\Klebanov}].
As another extension to the quantum gravity coupled to matter fields,
we have incorporated matter fields
by putting a matter field naively on each link of the triangulated surface.
In this case the wave function depends on
not only the length of string
but also the matter fields on string like
$\Psi^\dagger(l;\phi_1,\ldots,\phi_l)$.
However, we have not succeeded in taking the continuum limit,
though one expects the light-cone string field theory\NPrefmark{\KakuKikkawa}
for $c=25$ non-critical string theory.
The extension to the string field theories
which correspond to the two matrix models is now under study.
In the near future we hope that
the transfer matrix formalism in the dynamical triangulation will bring us
the non-critical string field theories for any value of $c$
(including $c>1$ cases) at the continuous level.
\NextPage
%
%
%
\ack
The author would like to thank
Dr.\ N.\ Ishibashi, Dr.\ H.\ Kawai, Dr.\ N.\ Kawamoto and Dr.\ T.\ Mogami
for useful discussions.
\NextPage
%
%
\topskip 30pt
\Appendix{A}
\vskip 10pt
\noindent{ \it Discrete Laplace Transformation and Its Continuum Limit}
\par
In this appendix
we give the definition and the properties of
the discrete Laplace transformation.
The usual continuous Laplace transformation
is obtained by taking the continuum limit.
In the following we apply the Laplace transformation
to the length of boundaries.
The application to the area of surface is exactly the same.
\par
Firstly, we give the definition of the discrete Laplace transformation.
Let introduce a function $f(l)$ which is defined for $l \ge l_0$,
where $l_0$ is an arbitrary integer.
The discrete Laplace transformation and its inverse transformation is
$$ \eqalign{ \sp(2.0)
  \tilde f (z)  \ = \  \sum_{l=l_0}^\infty \, z^l \, f(l) \, , \qquad
  f (l)  \ = \  \intdz{z_c} \, z^{-l} \, \tilde f(z) \, ,
\cr\sp(3.0)} \eqn\EqLapDis $$
where we suppose that
the convergence radius of $z$ of $\tilde f(z)$ is $z_c$
and the function $\tilde f(z)$ is analytic in the region
$|z| = z_c$ as well as $|z| < z_c$.
The continuum limit of the discrete Laplace transformation
is taken by $\e \rightarrow 0$ with
$$ \eqalign{ \sp(2.0)
L  \ = \  \e \, l \, ,  \quad
\hbox{ and } \quad
z  \ = \  z_c   \, e^{ - \e \zeta } \, .
\cr\sp(3.0)} \eqn\EqLapLimit $$
The continuum limit of $f(l)$ and $\tilde f(z)$ are
$$ \eqalign{ \sp(2.0)
  F (L)  \ = \
  \lim_{\e \rightarrow 0}
  c_f \, \e^{\dim{\tilde F}-1} \, z_c^l \, f (l)    \, ,
\cr\sp(3.0)} \eqn\EqLapLimitFII $$
and
$$ \eqalign{ \sp(2.0)
  \tilde F (\zeta)   \ = \
  \lim_{\e \rightarrow 0} c_f \, \e^\dim{\tilde F} \, \tilde f (z) \, ,
\cr\sp(3.0)} \eqn\EqLapLimitFI $$
where $c_f$ is a positive real number which one can choose arbitrarily.
$\dim{\tilde F}$ is the dimension of $\tilde F(\zeta)$
in the unit of $\dim{\e}=1$.
By using the formulae,
$$ \eqalign{ \sp(2.0)
  \sum_{l=l_0}^\infty  \ = \
  {1 \over \e} \int\limits_{\e l_0}^\infty d L \, ,
  \qquad
  \intdz{z_c}  \ = \  \e \int\limits_{- i \pi / \e}^{+ i \pi / \e}
                         {d \zeta \over 2 \pi i} \, ,
\cr\sp(3.0)} \eqn\EqLapLimitInt $$
we obtain the usual continuous Laplace transformation,
$$ \eqalign{ \sp(2.0)
  \tilde F (\zeta) \ = \
  \int\limits_0^\infty d L \, e^{- L \zeta} \, F (L) \, , \qquad
  F (L)     \ = \
  \intdzeta \, e^{L \zeta} \, \tilde F (\zeta) \, ,
\cr\sp(3.0)} \eqn\EqLapDis $$
in the continuum limit, $\e \rightarrow 0$.
The function $\tilde F(\zeta)$ is analytic, i.e., has no singularities
in the region $0 \le \Real(\zeta)$,
because $\tilde f(z)$ is analytic in the region $|z| \le z_c$.
\par
Next, let us consider the inner product.
In this case we introduce two functions, $p(l)$ and $q(l)$,
which have the different convergence radii, $x_c$ and $1/x_c$, respectively.
The discretized Laplace transformation of $p(l)$ and $q(l)$
and their continuum limit,
$\tilde p(x)$, $\tilde q(y)$, $P(L)$, $Q(L)$,
$\tilde P(\xi)$ and $\tilde Q(\eta)$,
are defined as mentioned above.
The functions $\tilde p(x)$ and $\tilde q(y)$
are analytic in the region,
$|x| \le x_c$ and $|y| \le 1/x_c$,
while the functions $\tilde P(\xi)$ and $\tilde Q(\eta)$ are analytic
in the region, $0 \le \Real(\xi)$ and $0 \le \Real(\eta)$.
The inner product of $p$ and $q$ is defined by
$$ \eqalign{ \sp(2.0)
  \sum_{l=l_0}^\infty \, p(l) \, q(l)
  \ = \
  \intdz{x_c} \, \tilde p(z) \, \tilde q(\inv{z}) \, .
\cr\sp(3.0)} \eqn\EqLapInternalDis $$
The discretized Laplace transformation of the $\delta$-function is
$$ \eqalign{ \sp(2.0)
  \delta ( y , x )
  \ = \
  \sum_{l'=l_0}^\infty \sum_{l=l_0}^\infty \, y^{l'} \, x^l \, \delta_{l',l}
  \ = \
  { ( y x )^{l_0} \over 1 \, - \, y x } \, ,
\cr\sp(3.0)} \eqn\EqDeltaDis $$
which has the following properties:
$$ \eqalign{ \sp(2.0)
 \intdz{x_c} \, \tilde p(z) \, \delta ( \inv{z} , x )
 \ &= \  \tilde p(x) \, ,
\cr\sp(6.0)
 \intdz{x_c} \, \delta ( y , z ) \, \tilde q( \inv{z} )
 \ &= \  \tilde q(y) \, .
\cr\sp(3.0)} \eqn\EqDeltaDisII $$
The continuum limit of the inner product {\EqLapInternalDis} becomes
$$ \eqalign{ \sp(2.0)
 \int\limits_0^\infty d L \, P (L) \, Q (L)
 \ = \
 \intdzeta \, \tilde P (\zeta) \, \tilde Q (- \zeta) \, .
\cr\sp(3.0)} \eqn\EqLapInternalCon $$
The continuum limit of the $\delta$-function, {\EqDeltaDis}, is
$$ \eqalign{ \sp(2.0)
  \delta ( \eta , \xi )
  \  \define \  \lim_{\e \rightarrow 0} \e \, \delta ( y , x )
  \  = \        {1 \over \eta \, + \, \xi} \, ,
\cr\sp(3.0)} \eqn\EqDeltaCon $$
which is also obtained directly
from the continuous Laplace transformation of $\delta ( L - L' )$.
The $\delta$-function {\EqDeltaCon} satisfies
$$ \eqalign{ \sp(2.0)
& \intdzeta \,  \tilde P (\zeta) \, \delta ( - \zeta , \xi )    \ = \
  \tilde P (\xi) \, ,
\cr\sp(6.0)
& \intdzeta \,  \delta ( \eta , \zeta ) \, \tilde Q( - \zeta )  \ = \
  \tilde Q (\eta) \, ,
\cr\sp(3.0)} \eqn\EqDeltaConII $$
which are the continuum limit of {\EqDeltaDisII}.
\NextPage
\Appendix{B}
\vskip 10pt
\noindent{ \it Notations and Properties about Transfer Matrices and Amplitudes}
\par
In this appendix we summarize the notations and the properties
of the transfer matrices as well as the amplitudes
in the dynamical triangulation for pure gravity.
The extension to other non-critical string field theories
with matter fields is straightforward.
The Laplace transformations of the transfer matrices and the amplitudes
are defined by
$$ \eqalign{ \sp(2.0)
& T_{M,N}^{(h)} ( y_1, \ldots, y_M ; x_1, \ldots, x_N ; \k ; \t )
\cr\sp(6.0)
& = \
  \sum_{l'_1,\ldots,l'_M,l_1,\ldots,l_N = 1}^\infty
  y_1^{l'_1} \cdots y_M^{l'_M} \,
  x_1^{l_1}  \cdots x_N^{l_N} \,
  T_{M,N}^{(h)} ( l'_1, \ldots, l'_M ; l_1, \ldots, l_N ; \k ; \t ) \, ,
\cr\sp(6.0)
& F_N^{(h)} ( x_1, \ldots, x_N ; \k ) \
  = \
  \sum_{l_1,\ldots,l_N = 1}^\infty
  x_1^{l_1}  \cdots x_N^{l_N} \,
  F_N^{(h)} ( l_1, \ldots, l_N ; \k ) \, ,
\cr\sp(3.0)} \eqn\EqLTofTMAmpDis $$
and
$$ \eqalign{ \sp(2.0)
& T_{M,N}^{(h)} ( l'_1, \ldots, l'_M ; l_1, \ldots, l_N ; \k ; \t ) \
  = \
  \sum_{a = 0}^\infty \, \k^a \,
  T_{M,N}^{(h)} ( l'_1, \ldots, l'_M ; l_1, \ldots, l_N ; a ; \t ) \, ,
\cr\sp(6.0)
& F_N^{(h)} ( l_1, \ldots, l_N ; \k ) \
  = \
  \sum_{a = 0}^\infty \, \k^a \,
  F_N^{(h)} ( l_1, \ldots, l_N ; a ) \, ,
\cr\sp(3.0)} \eqn\EqLTofTMAmpDisII $$
at the discrete level, and
$$ \eqalign{ \sp(2.0)
& T_{M,N}^{(h)} ( \eta_1, \ldots, \eta_M ; \xi_1, \ldots, \xi_N ; \cc ; \T )
\cr\sp(6.0)
& = \
  \int\limits_0^\infty \! d L'_1 \cdots d L'_M d L_1 \cdots d L_N \,
  e^{ - L'_1 \eta_1 - \ldots - L'_M \eta_M
      - L_1  \xi_1  - \ldots - L_N  \xi_N  } \,
\cr\sp(3.0)
& \phantom{= \
  \int\limits_0^\infty \! d L'_1 \cdots d L'_M d L_1 \cdots d L_N \,
  } \ \times \,
  T_{M,N}^{(h)} ( L'_1, \ldots, L'_M ; L_1, \ldots, L_N ; \cc ; \T ) \, ,
\cr\sp(6.0)
& F_N^{(h)} ( \xi_1, \ldots, \xi_N ; \cc ) \
  = \
  \int\limits_0^\infty \! d L_1 \cdots d L_N \,
  e^{ - L_1  \xi_1  - \ldots - L_N  \xi_N  } \,
  F_N^{(h)} ( L_1, \ldots, L_N ; \cc ) \, ,
\cr\sp(3.0)} \eqn\EqLTofTMAmpCon $$
and
$$ \eqalign{ \sp(2.0)
& T_{M,N}^{(h)} ( L'_1, \ldots, L'_M ; L_1, \ldots, L_N ; \cc ; \T )
\cr\sp(6.0)
& = \
  \int\limits_0^\infty \! d A \, e^{ - A \cc } \,
  T_{M,N}^{(h)} ( L'_1, \ldots, L'_M ; L_1, \ldots, L_N ; A ; \T ) \, ,
\cr\sp(6.0)
& F_N^{(h)} ( L_1, \ldots, L_N ; \cc ) \
  = \
  \int\limits_0^\infty \! d A \, e^{ - A \cc } \,
  F_N^{(h)} ( L_1, \ldots, L_N ; A ) \, ,
\cr\sp(3.0)} \eqn\EqLTofTMAmpConII $$
at the continuous level.
The transfer matrices and the amplitudes are analytic in the region
$|x_i| \le x_c$, $|y_j| \le 1/x_c$ and $|\k| < \k_c$
at the discrete level,
and
$0 \le \Real(\xi_i)$, $0 \le \Real(\eta_j)$ and $0 < \Real(\cc)$
at the continuous level.
They are related each other as
$$ \eqalign{ \sp(2.0)
& T_{M,N}^{(h)} ( \eta_1, \ldots, \eta_M ; \xi_1, \ldots, \xi_N ; \cc ; \T )
\cr\sp(6.0)
& = \
  \lim_{\e \rightarrow 0}
  { c_1^N c_2^M \over c_5^{h+N-1} } \,
  \e^{N \dim{\Psi^\dagger} + M \dim{\Psi} - (h+N-1) \dim{G} } \,
\cr\sp(6.0)
& \phantom{= \  \lim_{\e \rightarrow 0} } \  \times \,
  T_{M,N}^{(h)} ( y_1, \ldots, y_M ; x_1, \ldots, x_N ; \k ; \t )  \, ,
\cr\sp(3.0)} \eqn\EqGenTMConII $$
and
$$ \eqalign{ \sp(2.0)
& F_N^{(h)} ( \xi_1, \ldots, \xi_N ; \cc )
\cr\sp(6.0)
& = \
  \lim_{\e \rightarrow 0}
  { c_1^N \over c_5^{h+N-1} } \,
  \e^{N \dim{\Psi^\dagger} - (h+N-1) \dim{G} } \,
  F_N^{(h)} ( x_1, \ldots, x_N ; \k )  \, ,
\cr\sp(3.0)} \eqn\EqGenAmpConII $$
where $\dim{\Psi^\dagger}$, $\dim{\Psi}$ and $\dim{G}$
are $-3/2$, $5/2$ and $-5$, respectively, for pure gravity.
The relationship of $\mbar T_{1,1}^{(0)}$
between at the discrete level and at the continuous level is
the same as that of $T_{1,1}^{(0)}$.
\par
As is manifest from the definition given in section 2,
the transfer matrices and the amplitudes are invariant
under the exchange of two initial strings or two final strings, i.e.,
$$ \eqalign{ \sp(2.0)
& T_{M,N}^{(h)} ( \ldots, y_i, \ldots, y_j, \ldots ; \ldots ; \k ; \t )
  \ = \
  T_{M,N}^{(h)} ( \ldots, y_j, \ldots, y_i, \ldots ; \ldots ; \k ; \t ) \, ,
\cr\sp(6.0)
& T_{M,N}^{(h)} ( \ldots ; \ldots, x_i, \ldots, x_j, \ldots ; \k ; \t )
  \ = \
  T_{M,N}^{(h)} ( \ldots ; \ldots, x_j, \ldots, x_i, \ldots ; \k ; \t ) \, ,
\cr\sp(6.0)
& F_N^{(h)} ( \ldots, x_i, \ldots, x_j, \ldots ; \k )
  \ = \
  F_N^{(h)} ( \ldots, x_j, \ldots, x_i, \ldots ; \k ) \, ,
\cr\sp(3.0)} \eqn\EqTMAmpDisI $$
and
$$ \eqalign{ \sp(2.0)
& T_{M,N}^{(h)} ( \ldots, \eta_i, \ldots, \eta_j, \ldots ; \ldots ; \cc ; \T )
  \ = \
  T_{M,N}^{(h)} ( \ldots, \eta_j, \ldots, \eta_i, \ldots ; \ldots ; \cc ; \T )
  \, ,
\cr\sp(6.0)
& T_{M,N}^{(h)} ( \ldots ; \ldots, \xi_i, \ldots, \xi_j, \ldots ; \cc ; \T )
  \ = \
  T_{M,N}^{(h)} ( \ldots ; \ldots, \xi_j, \ldots, \xi_i, \ldots ; \cc ; \T )
  \, ,
\cr\sp(6.0)
& F_N^{(h)} ( \ldots, \xi_i, \ldots, \xi_j, \ldots ; \cc )
  \ = \
  F_N^{(h)} ( \ldots, \xi_j, \ldots, \xi_i, \ldots ; \cc ) \, ,
\cr\sp(3.0)} \eqn\EqTMAmpConI $$
which correspond to
$[\Psi,\Psi] = [\Psi^\dagger,\Psi^\dagger] = 0$
from the viewpoint of the string field theory.
{}From {\EqAmpAndTM} and {\EqTMandDelta}, we find for $N \ge 1$ that
$$ \eqalign{ \sp(2.0)
& \lim_{\t \rightarrow \infty}
  T_{0,N}^{(h)} ( \void ; x_1, \ldots, x_N ; \k ; \t )
  \ = \
  F_N^{(h)} ( x_1, \ldots, x_N ; \k ) \, ,
\cr\sp(6.0)
& \lim_{\t \rightarrow \infty}
  T_{M>0,N}^{(h)} ( y_1, \ldots, y_M ; x_1, \ldots, x_N ; \k ; \t )
  \ = \
  0 \, ,
\cr\sp(6.0)
& \lim_{\t \rightarrow 0}
  T_{M,N}^{(h)} ( y_1, \ldots, y_M ; x_1, \ldots, x_N ; \k ; \t )
  \ = \
  \delta_{h,0} \, \delta_{M,1} \, \delta_{N,1} \, \delta (y_1,x_1) \, ,
\cr\sp(3.0)} \eqn\EqTMAmpDisII $$
and
$$ \eqalign{ \sp(2.0)
& \lim_{\T \rightarrow \infty}
  T_{0,N}^{(h)} ( \void ; \xi_1, \ldots, \xi_N ; \cc ; \T )
  \ = \
  F_N^{(h)} ( \xi_1, \ldots, \xi_N ; \cc ) \, ,
\cr\sp(6.0)
& \lim_{\T \rightarrow \infty}
  T_{M>0,N}^{(h)} ( \eta_1, \ldots, \eta_M ; \xi_1, \ldots, \xi_N ; \cc ; \T )
  \ = \
  0 \, ,
\cr\sp(6.0)
& \lim_{\T \rightarrow 0}
  T_{M,N}^{(h)} ( \eta_1, \ldots, \eta_M ; \xi_1, \ldots, \xi_N ; \cc ; \T )
  \ = \
  \delta_{h,0} \, \delta_{M,1} \, \delta_{N,1} \, \delta (\eta_1,\xi_1) \, .
\cr\sp(3.0)} \eqn\EqTMAmpConII $$
\par
At the discrete level,
the transfer matrices and the amplitudes are calculated by
$$ \eqalign{ \sp(2.0)
& T_{M,N} ( g ; y_1, \ldots, y_M ; x_1, \ldots, x_N ; \k ; \t )
\cr\sp(6.0)
& = \
  \vac \,  \Psi (y_1) \, \cdots \, \Psi (y_M) \,
           e^{- \t \H (g,\k)} \,
           \Psi^\dagger (x_1) \, \cdots \, \Psi^\dagger (x_N) \,
           \cuum_{\rm connected} \, ,
\cr\sp(6.0)
& F_N ( g ; x_1, \ldots, x_N ; \k )
\cr\sp(6.0)
& = \
  \lim_{\t\rightarrow\infty}
  \vac \, e^{- \t \H (g,\k)} \,
          \Psi^\dagger (x_1) \, \cdots \, \Psi^\dagger (x_N) \,
          \cuum_{\rm connected} \, ,
\cr\sp(3.0)} \eqn\EqTMAmpDisIII $$
where
$$ \eqalign{ \sp(2.0)
& T_{M,N} ( g ; y_1, \ldots, y_M ; x_1, \ldots, x_N ; \k ; \t )
\cr\sp(6.0)
& \define \
  \sum_{h=0}^\infty g^{h+N-1} \,
  T_{M,N}^{(h)} ( y_1, \ldots, y_M ; x_1, \ldots, x_N ; \k ; \t ) \, ,
\cr\sp(6.0)
& F_N ( g ; x_1, \ldots, x_N ; \k )
  \ \define \
  \sum_{h=0}^\infty g^{h+N-1} \,
  F_N^{(h)} ( x_1, \ldots, x_N ; \k )  \, .
\cr\sp(3.0)} \eqn\EqTMAmpDisIV $$
Since we have set $\Psi^\dagger(l=0) = 1$,
the operator $\Psi(l=0)$ must not be used in our formalism.
By introducing the new wave function $\Phi^\dagger$,
{\EqTMAmpDisIII} and {\EqTMAmpDisIV} are rewritten as
$$ \eqalign{ \sp(2.0)
& T_{M,N} ( g ; y_1, \ldots, y_M ; x_1, \ldots, x_N ; \k ; \t )
\cr\sp(6.0)
& = \
  \hat T_{M,N} ( g ; y_1, \ldots, y_M ; x_1, \ldots, x_N ; \k ; \t ) \,
  +  \, \delta_{M,0} \, \delta_{N,1} \, \lambda (x_1,\k) \, ,
\cr\sp(6.0)
& F_N ( g ; x_1, \ldots, x_N ; \k )
\cr\sp(6.0)
& = \
  \hat F_N ( g ; x_1, \ldots, x_N ; \k ) \,
  +  \, \delta_{N,1} \, \lambda (x_1,\k) \, ,
\cr\sp(3.0)} \eqn\EqTMAmpDisV $$
where
$$ \eqalign{ \sp(2.0)
& \hat T_{M,N} ( g ; y_1, \ldots, y_M ; x_1, \ldots, x_N ; \k ; \t )
\cr\sp(6.0)
& = \
  \vac \,  \Psi (y_1) \, \cdots \, \Psi (y_M) \,
           e^{- \t \H (g,\k)} \,
           \Phi^\dagger (x_1,\k) \, \cdots \, \Phi^\dagger (x_N,\k) \,
           \cuum_{\rm connected} \, ,
\cr\sp(6.0)
& \hat F_N ( g ; x_1, \ldots, x_N ; \k )
\cr\sp(6.0)
& = \
  \lim_{\t\rightarrow\infty}
  \vac \, e^{- \t \H (g,\k)} \,
          \Phi^\dagger (x_1,\k) \, \cdots \, \Phi^\dagger (x_N,\k) \,
          \cuum_{\rm connected} \, .
\cr\sp(3.0)} \eqn\EqTMAmpDisVI $$
The continuum limit of
{\EqTMAmpDisV}, {\EqTMAmpDisVI} and {\EqTMAmpDisIV} are
$$ \eqalign{ \sp(2.0)
& T_{M,N} ( G ; \eta_1, \ldots, \eta_M ; \xi_1, \ldots, \xi_N ; \cc ; \T )
\cr\sp(6.0)
& = \
  \hat T_{M,N} ( G ; \eta_1, \ldots, \eta_M ;
                     \xi_1, \ldots, \xi_N ; \cc ; \T ) \,
  +  \, \delta_{M,0} \, \delta_{N,1} \, \lambda (\xi_1) \, ,
\cr\sp(6.0)
& F_N ( G ; \xi_1, \ldots, \xi_N ; \cc )
\cr\sp(6.0)
& = \
  \hat F_N ( G ; \xi_1, \ldots, \xi_N ; \cc ) \,
  + \, \delta_{N,1} \, \lambda (\xi_1) \, ,
\cr\sp(3.0)} \eqn\EqTMAmpConV $$
where
$$ \eqalign{ \sp(2.0)
& \hat T_{M,N} ( G ; \eta_1, \ldots, \eta_M ;
                     \xi_1, \ldots, \xi_N ; \cc ; \T )
\cr\sp(6.0)
& = \
  \vac \,  \Psi (\eta_1) \, \cdots \, \Psi (\eta_M) \,
           e^{- \T \H (G,\cc)} \,
           \Phi^\dagger (\xi_1) \, \cdots \, \Phi^\dagger (\xi_N) \,
           \cuum_{\rm connected} \, ,
\cr\sp(6.0)
& \hat F_N ( G ; \xi_1, \ldots, \xi_N ; \cc )
\cr\sp(6.0)
& = \
  \lim_{\T\rightarrow\infty}
  \vac \, e^{- \T \H (G,\cc)} \,
          \Phi^\dagger (\xi_1) \, \cdots \, \Phi^\dagger (\xi_N) \,
          \cuum_{\rm connected} \, ,
\cr\sp(3.0)} \eqn\EqTMAmpConVI $$
and
$$ \eqalign{ \sp(2.0)
& T_{M,N} ( G ; \eta_1, \ldots, \eta_M ; \xi_1, \ldots, \xi_N ; \cc ; \T )
\cr\sp(6.0)
& \define \
  \sum_{h=0}^\infty G^{h+N-1} \,
  T_{M,N}^{(h)} ( \eta_1, \ldots, \eta_M ; \xi_1, \ldots, \xi_N ; \cc ; \T )
  \, ,
\cr\sp(6.0)
& F_N ( G ; \xi_1, \ldots, \xi_N ; \cc )
  \ \define \
  \sum_{h=0}^\infty G^{h+N-1} \,
  F_N^{(h)} ( \xi_1, \ldots, \xi_N ; \cc )  \, .
\cr\sp(3.0)} \eqn\EqTMAmpConIV $$
Expanding {\EqTMAmpDisV} and {\EqTMAmpConV}
with respect to $g$ and $G$ respectively, we find
$$ \eqalign{ \sp(2.0)
& T_{M,N}^{(h)} ( y_1, \ldots, y_M ; x_1, \ldots, x_N ; \k ; \t )
  \ = \
  \hat T_{M,N}^{(h)} ( y_1, \ldots, y_M ; x_1, \ldots, x_N ; \k ; \t )
\cr\sp(3.0)
& \phantom{T_{M,N}^{(h)} ( y_1, \ldots, y_M ; x_1, \ldots, x_N ; \k ; \t )
           \ = \  }
  + \, \delta_{h,0} \, \delta_{M,0} \, \delta_{N,1} \, \lambda (x_1,\k) \, ,
\cr\sp(6.0)
& F_N^{(h)} ( x_1, \ldots, x_N ; \k )
  \ = \
  \hat F_N^{(h)} ( x_1, \ldots, x_N ; \k ) \,
  + \, \delta_{h,0} \, \delta_{N,1} \, \lambda (x_1,\k) \, ,
\cr\sp(3.0)} \eqn\EqTMAmpDisVII $$
at the discrete level and
$$ \eqalign{ \sp(2.0)
& T_{M,N}^{(h)} ( \eta_1, \ldots, \eta_M ; \xi_1, \ldots, \xi_N ; \cc ; \T )
  \ = \
  \hat T_{M,N}^{(h)} ( \eta_1, \ldots, \eta_M ;
                       \xi_1, \ldots, \xi_N ; \cc ; \T )
\cr\sp(3.0)
& \phantom{T_{M,N}^{(h)}
           ( \eta_1, \ldots, \eta_M ; \xi_1, \ldots, \xi_N ; \cc ; \T )
           \ = \  }
  + \, \delta_{h,0} \, \delta_{M,0} \, \delta_{N,1} \, \lambda (\xi_1) \, ,
\cr\sp(6.0)
& F_N^{(h)} ( \xi_1, \ldots, \xi_N ; \cc )
  \ = \
  \hat F_N^{(h)} ( \xi_1, \ldots, \xi_N ; \cc ) \,
  + \, \delta_{h,0} \, \delta_{N,1} \, \lambda (\xi_1) \, ,
\cr\sp(3.0)} \eqn\EqTMAmpConVII $$
at the continuous level.
The non-universal part $\lambda(\xi_1)$,
which depends on the cut-off parameter,
contributes only to the disk topology, $T_{0,1}^{(0)}$ and $F_1^{(0)}$.
{}From {\EqNormalizeDis} we find at the discrete level that
$$ \eqalign{ \sp(2.0)
& T_{M,N}^{(h)} ( y_1=0, y_2, \ldots, y_M ; x_1, \ldots, x_N ; \k ; \t )
  \ = \  0 \, ,
\cr\sp(6.0)
& T_{M,N}^{(h)} ( y_1, \ldots, y_M ; x_1=0, x_2, \ldots, x_N ; \k ; \t )
  \ = \  \delta_{h,0} \, \delta_{M,0} \, \delta_{N,1} \, ,
\cr\sp(6.0)
& F_N^{(h)} ( x_1=0, x_2, \ldots, x_N ; \k )
  \ = \  \delta_{h,0} \, \delta_{N,1} \, ,
\cr\sp(3.0)} \eqn\EqTMAmpDisVIII $$
which lead to
$$ \eqalign{ \sp(2.0)
& \lim_{\eta_j \rightarrow \infty}
  T_{M,N}^{(h)} ( \eta_1, \ldots, \eta_M ; \xi_1, \ldots, \xi_N ; \cc ; \T )
  \ = \  0 \, ,
\cr\sp(6.0)
& \lim_{\xi_i \rightarrow \infty}
  T_{M,N}^{(h)} ( \eta_1, \ldots, \eta_M ; \xi_1, \ldots, \xi_N ; \cc ; \T )
  \ = \  c_1 \, \e^{\dim{\Psi^\dagger}} \,
         \delta_{h,0} \, \delta_{M,0} \, \delta_{N,1} \, ,
\cr\sp(6.0)
& \lim_{\xi_i \rightarrow \infty}
  F_N^{(h)} ( \xi_1, \ldots, \xi_N ; \cc )
  \ = \  c_1 \, \e^{\dim{\Psi^\dagger}} \,
         \delta_{h,0} \, \delta_{N,1} \, ,
\cr\sp(3.0)} \eqn\EqTMAmpConVIII $$
at the continuous level.
Since the dimensions of
$\xi_i$, $\eta_j$, $\cc$, $D$, $\Phi^\dagger$, $\Psi$, and $G$ are
$-1$, $-1$, $-2$, $1/2$, $-3/2$, $5/2$, and $-5$, respectively,
the dimensional analysis leads to
$$ \eqalign{ \sp(2.0)
& \bigl\{ \,
       \sum_{i=1}^N \xi_i {\der \over \der \xi_i} \,
  + \, \sum_{j=1}^M \eta_j {\der \over \der \eta_j} \,
  + \, 2 \cc {\der \over \der \cc} \,
  - \, {1 \over 2} \T {\der \over \der \T} \,
\cr\sp(4.0)
& \phantom{\bigl\{ \, } \,
  - \, {3 \over 2} N \, + \, {5 \over 2} M \, + \, 5 ( h + N - 1 )
  \, \bigr\}
  \hat T_{M,N}^{(h)} ( \eta_1, \ldots, \eta_M ;
                       \xi_1, \ldots, \xi_N ; \cc ; \T )
  \  =  \  0  \, ,
\cr\sp(6.0)
& \bigl\{ \,
       \sum_{i=1}^N \xi_i {\der \over \der \xi_i} \,
  + \, 2 \cc {\der \over \der \cc} \,
  - \, {3 \over 2} N \, + \, 5 ( h + N - 1 )
  \, \bigr\}
  \hat F_N^{(h)} ( \xi_1, \ldots, \xi_N ; \cc )
  \  =  \  0  \, .
\cr\sp(3.0)} \eqn\EqTMAmpConIX $$
%
\NextPage
\Appendix{C}
\vskip 10pt
\noindent{ \it Schwinger-Dyson Equations at the Discrete Level}
\par
In this appendix we derive the Schwinger-Dyson equations
or Wheeler de Witt equations
in the discrete level
by using the properties {\EqAmpAndTM},
which mean that
any transfer matrix is convergent to an amplitude or zero
for $\t \rightarrow \infty$.
\par
Firstly, we obtain the Schwinger-Dyson equation
for the disk amplitude $F_1^{(0)}$.
Since
$T_{0,1}^{(0)}$ is convergent to $F_1^{(0)}$ for $\t \rightarrow \infty$,
i.e., {\EqAmpAndTM}, or equivalently, {\EqTMAmpDisII},
we derive the Schwinger-Dyson equation from
$$ \eqalign{ \sp(2.0)
0  \
& = \  \lim_{\t \rightarrow \infty}
       {\der \over \der \t} \,
       T_{0,1}^{(0)} ( \void ; x ; \k ; \t )
\cr\sp(6.0)
& = \  \lim_{\t \rightarrow \infty}
       {\der \over \der \t} \,
       \vac \, e^{- \t \H (g=0,\k)} \, \Psi^\dagger (x) \, \cuum
\, .
\cr\sp(3.0)} \eqn\EqSDDiskI $$
By using $\H (g=0,\k) \cuum = 0$,
eq.\ {\EqSDDiskI} becomes
$$ \eqalign{ \sp(2.0)
  \lim_{\t\rightarrow\infty} \,
  \vac \, e^{- \t \H (g=0,\k)} \, [ \, \H (g=0,\k) , \Psi^\dagger (x) \, ]
  \, \cuum
  \ = \ 0  .
\cr\sp(3.0)} \eqn\EqSDDiskII $$
Substituting {\EqOneDeformPsiI} into {\EqSDDiskII},
we obtain
$$ \eqalign{ \sp(2.0)
  \lim_{\t\rightarrow\infty} \,
  \vac \, e^{- \t \H (g=0,\k)} \,
  x {\der \over \der x} \bigl\{ \,
    \Psi^\dagger (x) \, - \, {\k \over x} ( \Psi^\dagger (x) - 1 )
    \, - \, x^2 ( \Psi^\dagger (x) )^2
  \, \bigr\} \, \cuum
  \ = \ 0  .
\cr\sp(3.0)} \eqn\EqSDDiskIII $$
By using {\EqAmpDiskDisII} and
$\lim_{\t\rightarrow\infty} \vac e^{- \t \H (g=0,\k)}
 \Psi^\dagger (x) \Psi^\dagger (x') \cuum
 = F_1^{(0)} (x;\k) F_1^{(0)} (x';\k)$,
we find
$$ \eqalign{ \sp(2.0)
  x {\der \over \der x} \bigl\{ \,
  F_1^{(0)} (x;\k) \, - \, {\k \over x} \, \bigl( F_1^{(0)} (x;\k) - 1 \bigr)
  \, - \, x^2 \, \bigl( F_1^{(0)} (x;\k) \bigr)^2
  \, \bigr\}
  \ = \ 0  .
\cr\sp(3.0)} \eqn\EqSDDiskIV $$
This eq.\ {\EqSDDiskIV} is also written as
$$ \eqalign{ \sp(2.0)
& F_1^{(0)} (x;\k) \, - \, 1
  \, - \, {\k \over x} \,
          \bigl( F_1^{(0)} (x;\k) - 1
                 - x {\der F_1^{(0)} (x=0;\k) \over \der x} \bigr)
\cr\sp(6.0)
& - \, x^2 \, \bigl( F_1^{(0)} (x;\k) \bigr)^2
  \ = \ 0  \, .
\cr\sp(3.0)} \eqn\EqSDDiskV $$
The solution of the equation {\EqSDDiskIV} or {\EqSDDiskV}
with $F_1^{(0)} (x=0;\k) = 1$
is already obtained in ref.\ [{\BIPZ}].
We do not solve the equation explicitly in this paper,
because we need some help of the matrix model calculation.
As we will show in the next appendix,
the explicit form of the disk amplitude is calculable
at the continuous level without any help of the matrix model calculation.
\par
Next, we calculate the Schwinger-Dyson equation
for the amplitudes of general genus topologies, $F_N^{(h)}$.
We here introduce the generating functional $Z [J;g,\k]$,
$$ \eqalign{ \sp(2.0)
   Z [J;g,\k]  \ \define \
   \lim_{\t\rightarrow\infty} \vac \,
   e^{- \t \H (g,\k)} \, \exp \bigl\{ \intdz{x_c} \,
   \Psi^\dagger (z) J (\inv{z}) \bigr\} \, \cuum  \, ,
\cr\sp(3.0)} \eqn\EqGenFunDis $$
which generates the connected amplitudes as
$$ \eqalign{ \sp(2.0)
  F_N ( g ; x_1, \ldots, x_N ; \k )
  \ = \
  { \delta^N \over \delta J (\inv{x_1}) \, \cdots \, \delta J (\inv{x_N}) } \,
  \bigl\{ \ln Z [J;g,\k] \bigr\} \Big|_{J=0} \, ,
\cr\sp(3.0)} \eqn\EqGenAmpDis $$
where $F_N$ is the amplitude defined in {\EqTMAmpDisIII}.
As the same as before for the disk amplitude,
we obtain the Schwinger-Dyson equation
from {\EqAmpAndTM} or {\EqTMAmpDisII},
i.e.,
$$ \eqalign{ \sp(2.0)
  \lim_{\t \rightarrow \infty} \,
  {\der \over \der \t} \,
  \vac \,
  e^{- \t \H (g,\k)} \,
  \exp \bigl\{ \intdz{x_c} \,
  \Psi^\dagger (z) J (\inv{z}) \bigr\} \, \cuum
  \ = \ 0  \, .
\cr\sp(3.0)} \eqn\EqSDGenDisI $$
By using
$\H (g=0,\k) \cuum = 0$, {\EqOneDeformPsiI} and {\EqOneDeformPsiII},
we obtain
$$ \eqalign{ \sp(2.0)
\intdz{x_c} \, \Bigl[ \,
& ( 1 - {\k \over z} ) \, \bigl( - z {\der \over \der z} J (\inv{z}) \bigr) \,
  {\delta \over \delta J(\inv{z})} \, \bigl\{ \ln Z [J;g,\k] \bigr\}
\cr\sp(6.0)
& - \, z^2 \, \bigl( - z {\der \over \der z} J (\inv{z}) \bigr)
  \Bigl( \bigl( {\delta \over \delta J(\inv{z})} \bigr)^2
         \bigl\{ \ln Z [J;g,\k] \bigr\}
    \, + \, \bigl( {\delta \over \delta J(\inv{z})}
                   \bigl\{ \ln Z [J;g,\k] \bigr\} \bigr)^2 \Bigr)
\cr\sp(6.0)
& - \, g \, z^2 \, \bigl( - z {\der \over \der z} J (\inv{z}) \bigr)^2 \,
  {\delta \over \delta J(\inv{z})} \, \bigl\{ \ln Z [J;g,\k] \bigr\} \,
\Bigr]
\ = \ 0  \, ,
\cr\sp(3.0)} \eqn\EqSDGenDisII $$
from eq.\ {\EqSDGenDisI}.
\par
Especially,
the linear term with respect to $J$ in {\EqSDGenDisII} is
$$ \eqalign{ \sp(2.0)
  0 \
& = \
  {\delta \over \delta J (\inv{x})} \,
  \bigl\{ \hbox{L.H.S. of {\EqSDGenDisII}} \bigr\} \Big|_{J=0}
\cr\sp(8.0)
& = \
  x {\der \over \der x} \, \bigl\{ \,
  F_1 (g;x;\k)
  \, - \, {\k \over x} \, ( F_1 (g;x;\k) - F_1 (g;x=0;\k) )
\cr\sp(6.0)
& \phantom{= \  x {\der \over \der x} \, \bigl\{ \, } \
  - \, x^2 \, F_2 (g;x,x;\k) \, - \, x^2 \, (F_1 (g;x;\k))^2
  \, \bigr\}
  \, .
\cr\sp(3.0)} \eqn\EqSDMIGenDisI $$
We obtain eq.\ {\EqSDDiskIV} again
from the $g^0$ order terms in {\EqSDMIGenDisI}.
Eq.\ {\EqSDMIGenDisI} is also written as
$$ \eqalign{ \sp(2.0)
  \omega (x,\k) \, + \,
  x {\der \over \der x} \, \bigl\{ \,
  x^2 \, \hat F_2 (g;x,x;\k) \, + \, x^2 \, (\hat F_1 (g;x;\k))^2
  \, \bigr\}
  \ = \ 0  \, ,
\cr\sp(3.0)} \eqn\EqSDMIGenDisII $$
where we have used $\hat F_N$ defined in {\EqTMAmpDisVI}.
\par
The quadratic term with respect to $J$ in {\EqSDGenDisII} is
$$ \eqalign{ \sp(2.0)
  0 \
& = \
  {\delta \over \delta J (\inv{x_1})} {\delta \over \delta J (\inv{x_2})} \,
  \bigl\{ \hbox{L.H.S. of {\EqSDGenDisII}} \bigr\} \Big|_{J=0}
\cr\sp(8.0)
& = \
  x_1 {\der \over \der x_1} \, \bigl\{ \,
  F_2 (g;x_1,x_2;\k)
  \, - \, {\k \over x_1} \, ( F_2 (g;x_1,x_2;\k) - F_2 (g;x_1=0,x_2;\k) )
\cr\sp(6.0)
& \phantom{= \  x_1 {\der \over \der x_1} \, \bigl\{ \, } \
  - \, x_1^2 \, F_3 (g;x_1,x_1,x_2;\k)
  - \, 2 x_1^2 \, F_1 (g;x_1;\k) F_2 (g;x_1,x_2;\k) )
  \, \bigr\}
\cr\sp(6.0)
& \phantom{= \ } \
  + \, ( \, x_1 \, \leftrightarrow \, x_2 \, )
\cr\sp(6.0)
& \phantom{= \ } \
  - \, 2 g x_1 {\der \over \der x_1} x_2 {\der \over \der x_2}
  \intdz{x_c} \delta ( \inv{z} , x_1 ) \, \delta ( \inv{z} , x_2 ) \,
  z^2 F_1 (g;z;\k)
  \, .
\cr\sp(3.0)} \eqn\EqSDMIGenDisIII $$
Since
$F_2 (g;x_1=0,x_2;\k) = 0$,
by using $\hat F_N$ we obtain
$$ \eqalign{ \sp(2.0)
& x_1 {\der \over \der x_1} \, \bigl\{ \,
  - \, x_1^2 \, \hat F_3 (g;x_1,x_1,x_2;\k)
  - \, 2 x_1^2 \, \hat F_1 (g;x_1;\k) \hat F_2 (g;x_1,x_2;\k) )
  \, \bigr\}
\cr\sp(6.0)
& + \, ( \, x_1 \, \leftrightarrow \, x_2 \, )
\cr\sp(6.0)
& - \, 2 g x_1 {\der \over \der x_1} x_2 {\der \over \der x_2}
  \intdz{x_c} \delta ( \inv{z} , x_1 ) \, \delta ( \inv{z} , x_2 ) \,
  z^2 \hat F_1 (g;z;\k) \,
  \ = \ 0  \, .
\cr\sp(3.0)} \eqn\EqSDMIGenDisIV $$
For example,
the $g^1$ order of eq.\ {\EqSDMIGenDisIV}
is the Schwinger-Dyson equation which determines the cylinder amplitude.
\NextPage
\Appendix{D}
\vskip 10pt
\noindent{ \it Calculation of Amplitudes at the Continuous Level
               by using Schwinger-Dyson Equations}
\par
In this appendix we calculate
the explicit forms of some amplitudes at the continuous level
by using the Schwinger-Dyson equations.
According to ref.\ [{\IK}], we also use the fact that
the amplitude
$F_N^{(h)} ( \xi_1 , \ldots , \xi_N ; \cc )$
is analytic, i.e., has no singularities in the region
$0 \le \Real(\xi_i)$ ($1 \le i \le N$) and $0 < \Real(\cc)$.
\par
Firstly, let us calculate the explicit form of the disk amplitude
at the continuous level.
Similarly to the discrete level in appendix C,
we obtain the Schwinger-Dyson equation from {\EqAmpDiskConHatF} as
$$ \eqalign{ \sp(2.0)
  \lim_{\T \rightarrow \infty} \,
  {\der \over \der \T} \,
  \vac \, e^{- \T \H (G=0,\cc)} \, \Phi^\dagger (\xi) \, \cuum
  \ = \ 0  .
\cr\sp(3.0)} \eqn\EqSDDiskConI $$
By using
$\H (G=0,\cc) \cuum = 0$ and {\EqOneDeformConIII},
we find
$$ \eqalign{ \sp(2.0)
  {\der \over \der \xi} \, ( \hat F_1^{(0)} (\xi;\cc) )^2
  \ = \
  \omega (\xi,\cc) \, .
\cr\sp(3.0)} \eqn\EqSDDiskConII $$
The solution of eq.\ {\EqSDDiskConII} is
$$ \eqalign{ \sp(2.0)
  \hat F_1^{(0)} (\xi;\cc)
  \ = \
  \sqrt{ \xi^3 - {3 \over 4} \cc \xi + c(\cc) }  \, ,
\cr\sp(3.0)} \eqn\EqSDDiskConIII $$
where $c(\cc)$ is an integral constant and may depend on $\cc$.
If and only if $c(\cc) = \cc^{3/2} / 4$,
the amplitude, $\hat F_1^{(0)} (\xi;\cc)$,
has no poles and no cuts
in the region $0 \le \Real(\xi) < \infty$ on the complex plain.
Then, we obtain
$$ \eqalign{ \sp(2.0)
  \hat F_1^{(0)} (\xi;\cc)
  \ = \
  ( \xi - {\sqrt{\cc} \over 2} ) \, \sqrt{ \xi + \sqrt{\cc} } \, .
\cr\sp(3.0)} \eqn\EqSDDiskConIII $$
Therefore from eq.\ {\EqAmpDiskConI} the disk amplitude is found to be
$$ \eqalign{ \sp(2.0)
  F_1^{(0)} (\xi;\cc)
  \ = \
  \lambda (\xi) \, + \,
  ( \xi - {\sqrt{\cc} \over 2} ) \, \sqrt{ \xi + \sqrt{\cc} } \, .
\cr\sp(3.0)} \eqn\EqSDDiskConIV $$
The inverse Laplace transformation of {\EqSDDiskConIV}
is
$$ \eqalign{ \sp(2.0)
  F_1^{(0)} (L;\cc)
  \ = \
  {3 \over 4 \sqrt{\pi}} \,
  {1 \over L^{5/2}} \,
  ( \, 1 \, + \, \sqrt{\cc} L \, ) \, e^{- \sqrt{\cc} L}  \, ,
\cr\sp(3.0)} \eqn\EqAmpDiskConL $$
where we have introduced the proper regularization
into the inverse Laplace transformation
so as to absorb the divergent part $\lambda (\xi)$.
In other words, from the Laplace transformation of {\EqAmpDiskConL},
one obtains the divergent part $\lambda (\xi)$
because of the cut-off for small $L$.
The $\e$ in $\lambda (\xi)$ is proportional to
the cut-off parameter for small $L$.
\par
Next, let us calculate other amplitudes.
The continuum limit of the generating functional $Z [J;G,\cc]$ is
$$ \eqalign{ \sp(2.0)
  Z [J;G,\cc]
  \ \define \
  \lim_{\T\rightarrow\infty} \,
  \vac \,
  e^{- \T \H (G,\cc)} \,
  \exp \bigl\{ \intdzeta \,
               \Phi^\dagger (\zeta) J (-\zeta) \bigr\} \, \cuum  \, .
\cr\sp(3.0)} \eqn\EqGenFunCon $$
The connected amplitudes for general genus topologies are obtained by
$$ \eqalign{ \sp(2.0)
  \hat F_N ( G ; \xi_1, \ldots, \xi_N ; \cc )
  \ = \
  { \delta^N \over \delta J (-\xi_1) \, \cdots \, \delta J (-\xi_N) } \,
  \bigl\{ \ln Z [J;G,\cc] \bigr\} \Big|_{J=0} \, .
\cr\sp(3.0)} \eqn\EqGenAmpCon $$
where $\hat F_N$ is defined in {\EqTMAmpConVI}.
The Schwinger-Dyson equation is derived from
$$ \eqalign{ \sp(2.0)
  \lim_{\T \rightarrow \infty} {\der \over \der \T} \,
  \vac \,
  e^{- \T \H (G,\cc)} \,
  \exp \bigl\{ \intdzeta \,
               \Phi^\dagger (\zeta) J (-\zeta) \bigr\} \, \cuum
  \ = \ 0  \, .
\cr\sp(3.0)} \eqn\EqSDGenConI $$
Then we have
$$ \eqalign{ \sp(2.0)
\intdzeta \, \Bigl[ \,
& - \, \omega (\zeta,\cc) \, J (-\zeta)
\cr\sp(6.0)
& - \, \bigl( {\der \over \der \zeta} J (-\zeta) \bigr) \,
  \Bigl( \,    \bigl( {\delta \over \delta J(-\zeta)} \bigr)^2 \,
               \bigl\{ \ln Z [J;G,\cc] \bigr\} \,
          + \, \bigl( {\delta \over \delta J(-\zeta)} \,
               \bigl\{ \ln Z [J;G,\cc] \bigr\} \, \bigr)^2  \, \Bigr)
\cr\sp(6.0)
& - \, G \, \bigl( {\der \over \der \zeta} J (-\zeta) \bigr)^2 \,
    {\delta \over \delta J(-\zeta)} \, \bigl\{ \ln Z [J;G,\cc] \bigr\} \,
\Bigr]
\ = \ 0  \, .
\cr\sp(3.0)} \eqn\EqSDGenConII $$
\par
Especially,
the linear term with respect to $J$ in {\EqSDGenConII} is
$$ \eqalign{ \sp(2.0)
  0 \
& = \
    {\delta \over \delta J (-\xi)} \,
    \bigl\{ \hbox{L.H.S. of {\EqSDGenConII}} \bigr\} \Big|_{J=0}
\cr\sp(8.0)
& = \
    - \, \omega (\xi,\cc) \,
    + \, {\der \over \der \xi} \, \bigl\{ \,
            \hat F_2 (G;\xi,\xi;\cc) \, + \, (\hat F_1 (G;\xi;\cc))^2
         \, \bigr\}
    \, .
\cr\sp(3.0)} \eqn\EqSDMIGenConI $$
Expanding eq.\ {\EqSDMIGenConI} with respect to $G$,
we obtain eq.\ {\EqSDDiskConII} again from the $G^0$ order terms.
The $G^1$ order terms in {\EqSDMIGenConI} lead to
$$ \eqalign{ \sp(2.0)
  {\der \over \der \xi} \, \bigl\{ \,
   \hat F_2^{(0)} (\xi,\xi;\cc) \,
   + \, 2 \hat F_1^{(0)} (\xi;\cc) \hat F_1^{(1)} (\xi;\cc)
  \, \bigr\}
  \ = \ 0  \, ,
\cr\sp(3.0)} \eqn\EqSDDiskHandleCon $$
which determines the form of $\hat F_1^{(1)}$
from $\hat F_1^{(0)}$ and $\hat F_2^{(0)}$.
\par
The quadratic term with respect to $J$ in {\EqSDGenConII} is
$$ \eqalign{ \sp(2.0)
  0 \
& = \
  {\delta \over \delta J (-\xi_1)} {\delta \over \delta J (-\xi_2)} \,
  \bigl\{ \hbox{L.H.S. of {\EqSDGenConII}} \bigr\} \Big|_{J=0}
\cr\sp(8.0)
& = \
  {\der \over \der \xi_1} \, \bigl\{ \,
  \hat F_3 (G;\xi_1,\xi_1,\xi_2;\cc) \,
  + \, 2 \hat F_1 (G;\xi_1;\cc) \hat F_2 (G;\xi_1,\xi_2;\cc)
  \, \bigr\}
\cr\sp(6.0)
& \phantom{= \ } \
  + \, ( \, \xi_1 \, \leftrightarrow \, \xi_2 \, )
\cr\sp(6.0)
& \phantom{= \ } \
  - \, 2 G {\der \over \der \xi_1} {\der \over \der \xi_2} \,
  \intdzeta \delta ( - \zeta , \xi_1 ) \, \delta ( - \zeta , \xi_2 ) \,
  \hat F_1 (G;\zeta;\cc)
  \, .
\cr\sp(3.0)} \eqn\EqSDMIGenConII $$
The $G^1$ order terms in eq.\ {\EqSDMIGenConII} are
$$ \eqalign{ \sp(2.0)
& {\der \over \der \xi_1} \, \bigl\{ \,
    \hat F_1^{(0)} (\xi_1;\cc) \hat F_2^{(0)} (\xi_1,\xi_2;\cc)
  \, \bigr\} \,
  + \, ( \, \xi_1 \, \leftrightarrow \, \xi_2 \, )
\cr\sp(6.0)
& = \  {\der \over \der \xi_1} {\der \over \der \xi_2} \,
  \intdzeta \, \delta ( - \zeta , \xi_1 ) \, \delta ( - \zeta , \xi_2 ) \,
  \hat F_1^{(0)} (\zeta;\cc)
  \, .
\cr\sp(3.0)} \eqn\EqSDMICylinderCon $$
By using the explicit form of $\hat F_1^{(0)} (\xi;\cc)$ in {\EqSDDiskConIII},
we find the solution of {\EqSDMICylinderCon} as
$$ \eqalign{ \sp(2.0)
  F_2^{(0)} (\xi_1,\xi_2;\cc)  \
  = \  \hat F_2^{(0)} (\xi_1,\xi_2;\cc) \
  = \
  { 1 \over 2 ( \xi_1 - \xi_2 )^2 } \,
  \bigl( \, { \xi_1 + \xi_2 + 2 \sqrt{\cc}
              \over
              2 \sqrt{\xi_1 + \sqrt{\cc}} \sqrt{\xi_2 + \sqrt{\cc}} }
         \, - \, 1 \,
  \bigr) \, ,
\cr\sp(3.0)} \eqn\EqAmpCylinderCon $$
where we have determined the integral constant
so as to vanish poles and cuts
in the region $0 \le \Real(\xi_1)$ and $0 \le \Real(\xi_2)$.
The inverse Laplace transformation of {\EqAmpCylinderCon} is
$$ \eqalign{ \sp(2.0)
  F_2^{(0)} (L_1,L_2;\cc)
  \ = \
  { 1 \over 2 \pi } \,
  { \sqrt{ L_1 L_2 } \over L_1 + L_2 } \,
  e^{- \sqrt{\cc} ( L_1 + L_2 )} \, ,
\cr\sp(3.0)} \eqn\EqAmpCylinderConL $$
which agrees with the result by ref.\ [{\MSS}].
Substituting {\EqSDDiskConIII} and {\EqAmpCylinderCon}
into {\EqSDDiskHandleCon},
we also find
$$ \eqalign{ \sp(2.0)
  F_1^{(1)} (\xi;\cc)  \
  = \  \hat F_1^{(1)} (\xi;\cc) \
  = \
  { 1 \over 72 \cc } \,
  { \xi + {5 \over 2} \sqrt{\cc}
    \over
    ( \xi + \sqrt{\cc} )^{5/2} } \, ,
\cr\sp(3.0)} \eqn\EqAmpDiskIICon $$
which leads to
$$ \eqalign{ \sp(2.0)
  F_1^{(1)} (L;\cc)
  \ = \
  {1 \over 36 \sqrt{\pi}} \,
  {L^{1/2} \over \cc} \,
  ( 1 + \sqrt{\cc} L ) \, e^{- \sqrt{\cc} L}  \, .
\cr\sp(3.0)} \eqn\EqAmpDiskIIConL $$
\par
Notice that the amplitudes for the closed surface,
$F_{N=0}^{(h)}$, are obtained from $F_{N=1}^{(h)}$ by
$$ \eqalign{ \sp(2.0)
  F_{N=0}^{(h)} ( \void ; A )
  \ \propto \
  \hbox{the leading term of} \
  {1 \over A L} F_{N=1}^{(h)} ( L ; A )  \, ,
  \qquad \hbox{ for $L \rightarrow 0$ } \, .
\cr\sp(3.0)} \eqn\EqAmpClosedSurface $$
Therefore, by using {\EqAmpClosedSurface}
the amplitudes of sphere and torus are calculated
from {\EqAmpDiskConL} and {\EqAmpDiskIIConL}
as
$$ \eqalign{ \sp(2.0)
& F_0^{(0)} ( \void ; A )  \ \propto \  A^{-7/2} \, ,
\cr\sp(6.0)
& F_0^{(1)} ( \void ; A )  \ \propto \  A^{-1}  \, ,
\cr\sp(3.0)} \eqn\EqAmpSphereTorus $$
which agree with the well-known results of the matrix model.
\NextPage
%
%
\def\Figures{
\centerline{\BIGr FIGURE CAPTIONS}
\vskip 8pt
\item{\rm Fig.~1}{
  A triangulated surface
  $\Surf_N^{(h)} (l_1,\ldots,l_N;a)$
  with $h$ handles and $N$ boundaries (denoted by $\Loop_i$).
  The surface has $a$ triangles and
  each boundary $\Loop_i$ has $l_i$ links.
  The point on each $\Loop_i$ denotes a marked link.
  We have omitted triangles and links for simplicity.
  }
\item{\rm Fig.~2}{
  A triangulated surface
  $\Surf_{M,N}^{(h)} (l'_1,\ldots,l'_M;l_1,\ldots,l_N;a;\t)$
  with $h$ handles,
  $N$ initial boundaries (denoted by $\Loop_i$) and
  $M$ final boundaries (denoted by $\Loop'_j$).
  The surface has $a$ triangles and $\t$ height.
  The number of links on $\Loop_i$ and $\Loop'_j$ is
  $l_i$ and $l'_j$, respectively.
  The point on each $\Loop_i$ denotes a marked link.
  We have omitted triangles and links for simplicity.
  }
\item{\rm Fig.~3}{
  Decomposition of
  the transfer matrix of disk topology with height $\t_1 + \t_2$
  into transfer matrices with height $\t_1$ and $\t_2$.
  }
\item{\rm Fig.~4}{
  Decomposition of a surface by slicing (Fig.~4a) and peeling (Fig.~4b)
  }
\item{\rm Fig.~5}{
  Fig.~5a shows three different decompositions
  when one removes a triangle with a marked link.
  After introducing two-folded parts like $\alpha$ and $\beta$,
  these three different decompositions are identified with one decomposition
  like Fig.~5b.
  }
\item{\rm Fig.~6}{
  Three basic minimal-step \lq peeling decompositions'
  where a solid line and a broken line represent
  an initial string and a final string with a marked link, respectively.
  Fig.~6a shows removing a triangle
  while Figs.~6b and 6c show removing a two-folded part.
  }
\item{\rm Fig.~7}{
  Other possible minimal-step \lq peeling decompositions'
  besides those in Fig.~6.
  }
\item{\rm Fig.~8}{
  Seven basic minimal-step \lq peeling decompositions'
  without introducing the two-folded parts.
  A solid line and a broken line represent
  an initial string and a final string with a marked link, respectively.
  In all Figures one triangle is removed.
  }
          }
%
%

%
%
\refout
\NextPage
\Figures
\vfill
%
%
\bye